\pdfoutput=1
\documentclass[12pt,a4paper]{article}

\usepackage{geometry}
\usepackage[english]{babel}
\usepackage[utf8]{inputenc}

\usepackage{amsmath,amssymb,amsfonts,amsxtra,mathrsfs,mathtools}
\usepackage[bbgreekl]{mathbbol}
\usepackage{dsfont}
\usepackage{fixmath}
\usepackage{textgreek}
\usepackage{bm}

\usepackage{graphicx,graphics,epsfig}
\usepackage{caption}
\usepackage{subcaption}
\usepackage{float}
\usepackage{booktabs}
\restylefloat{table}

\usepackage{tikz}
\usetikzlibrary{decorations.pathmorphing,decorations.markings}
\usetikzlibrary{quotes,arrows.meta,arrows,calc,fadings,patterns,positioning,decorations.pathreplacing}
\usepackage{tikz-cd}

\usepackage{xcolor}
\usepackage[inline]{enumitem}
\usepackage{xpatch}
\usepackage{lipsum}
\usepackage{cite}

\usepackage[
    linkbordercolor=blue!50,
    citebordercolor=red!50,
    urlbordercolor={0 0.7 0.7}
]{hyperref}

\makeatletter
\g@addto@macro\bfseries{\boldmath}
\makeatother

\newcommand{\romannum}[1]{\romannumeral #1}
\newcommand{\Romannum}[1]{\MakeUppercase{\romannumeral #1}}

\numberwithin{equation}{section}

\newcommand{\ii}{\mathrm{i}}

\newcommand{\nn}{\nonumber}

\newcommand{\be}{\begin{equation}} \newcommand{\ee}{\end{equation}}
\newcommand{\bea}{\begin{equation} \begin{aligned}} \newcommand{\eea}{\end{aligned} \end{equation}}

\def\U{\mathrm{U}}

\def\SU{\mathrm{SU}}

\newcommand{\rd}{\mathrm{d}}

\newcommand{\Vol}{\mathrm{Vol}}

\newcommand{\wt}{\widetilde}
\newcommand{\wh}{\widehat}

\DeclareMathOperator{\Tr}{Tr}

\newcommand{\pd}{\partial}

\newcommand{\cC}{\mathcal{C}} 
 \newcommand{\cF}{\mathcal{F}}
 
\newcommand{\cI}{\mathcal{I}} 
\newcommand{\cK}{\mathcal{K}} \newcommand{\cL}{\mathcal{L}}
\newcommand{\cM}{\mathcal{M}} \newcommand{\cN}{\mathcal{N}}
 \newcommand{\cP}{\mathcal{P}}
\newcommand{\cQ}{\mathcal{Q}} 
\newcommand{\cS}{\mathcal{S}} 
 \newcommand{\cV}{\mathcal{V}}
\newcommand{\cW}{\mathcal{W}} 
 \newcommand{\cZ}{\mathcal{Z}}

\newcommand{\bC}{\mathbb{C}}

 \newcommand{\bR}{\mathbb{R}}
\newcommand{\bV}{\mathbb{V}} \newcommand{\bZ}{\mathbb{Z}}

\newcommand{\fg}{\mathfrak{g}}

 \newcommand{\fn}{\mathfrak{n}}
 
 \newcommand{\fp}{\mathfrak{p}}
\newcommand{\fq}{\mathfrak{q}}

\begin{document}

\begin{titlepage}

\vskip 2.5cm

\begin{center}

\today

\vskip 2.5cm

{\Large \bf $\mathcal{I}$-Extremization for AdS$_4$ Black Holes:\\[0.5em]
Master Volume, Free Energy, and Baryonic Charges}

\vskip 2cm
{Seyed Morteza Hosseini$^{\mathrm{a}}$ and Alberto Zaffaroni$^{\mathrm{b,c}}$}

\vskip 0.8cm

${}^{\mathrm{a}}$\textit{Centre for Theoretical Physics, Department of Physics and Astronomy,\\
Queen Mary University of London, London E1 4NS, UK}

\vskip 0.2cm

${}^{\mathrm{b}}$\textit{Dipartimento di Fisica, Universit\`a di Milano-Bicocca\\
Piazza della Scienza 3, 20126 Milano, Italy}

\vskip 0.2cm

${}^{\mathrm{c}}$\textit{INFN, sezione di Milano-Bicocca, Piazza della Scienza 3, 20126 Milano, Italy}

\end{center}

\vskip 2 cm

\begin{abstract}
\noindent  
In a previous paper, we proposed an entropy function for AdS$_4$ BPS black holes in M-theory with general magnetic charges, resolving a long-standing puzzle about baryonic charges in three-dimensional holography and offering a prediction for the large-$N$ limit of several partition functions whose saddle points have yet to be found. The entropy function is constructed from the master volume of the internal manifold. In this paper, we prove that the entropy of a general class of black holes based on toric geometry can indeed be reformulated as an $\mathcal{I}$-extremization problem, and we provide a set of examples. As an aside, we also simplify existing proofs of the equivalence between $a$-, $c$-, and $F$-extremizations and their gravitational duals.
\end{abstract}

\end{titlepage}

\tableofcontents

\section{Introduction}
\label{sec:intro}

The field theory derivation of the entropy of supersymmetric black holes in anti de Sitter (AdS) spacetime is a remarkable success of holography. In recent years, the large-$N$ limit of many supersymmetric partition functions has been used to match the entropy of BPS black holes and black strings that are asymptotically AdS. The relevant field theory quantity of interest is the logarithm of the partition function $\log \cZ ( \Delta_a , \epsilon_i , \fn_a )$, which depends on chemical potentials $\Delta_a$ and $\epsilon_i$ for the global symmetries of the theory and the rotational isometries of the background, respectively, as well as magnetic fluxes $\fn_a$ associated with the global symmetries (if present). The entropy of the black object, with electric charges $\fq_a$ and angular momenta $J_i$, is then obtained via a Legendre transform, or equivalently by extremizing the entropy function
\bea
 \label{ef0}
 \cI ( \Delta_a , \epsilon_i , \fn_a ) = \log \cZ ( \Delta_a  , \epsilon_i , \fn_a ) + 2 \pi \ii \biggl( \epsilon_i J_i + \sum_{a} \Delta_a \fq_a \biggr) \, .
\eea 
There is also strong evidence that $\log \cZ(\Delta_a,\epsilon_i,\fn_a)$ exhibits characteristic factorization properties at large $N$, and can be obtained by gluing together gravitational blocks \cite{Hosseini:2019iad}. For example, for rotating black objects with horizon geometry $S^2$ (or more generally, a spindle), with corresponding rotation parameter $\epsilon$, the partition function factorizes into two contributions from the poles:%
\footnote{See \cite{Hosseini:2021fge, Faedo:2021nub, Hosseini:2023ewi} for explicit examples of gluing of gravitational blocks.}
\bea
 \label{id0}
 \log \cZ  ( \Delta_a, \epsilon  , \fn_a ) = \frac{1}{2\epsilon} \left ( \cF (\Delta_a^+) \pm \cF (\Delta_a^-) \right ) \, ,
\eea
where the gravitational block $\cF ( \Delta_a )$ is a universal quantity associated with the dual superconformal field theory (SCFT), while the gluing variables $\Delta_a^\pm = \Delta_a^\pm ( \fn_a , \epsilon )$ and the relative sign depend on the choice of supersymmetric index used to count microstates---or equivalently, on how supersymmetry is preserved by the black object. The factorization arise naturally from the holomorphic block factorization of supersymmetric partition functions \cite{Pasquetti:2011fj,Beem:2012mb}.

The gravitational block $\cF(\Delta_a)$ for four-dimensional $\cN=1$ SCFTs is given by the large-$N$ trial central charge
\bea
 \label{cubic0}
 a_{(4)}(\Delta_a) =\frac{9}{32} \Tr R(\Delta_a)^3
 \, ,
\eea
where $R(\Delta_a)$ denotes the matrix of $R$-charges of the fermionic fields, expressed as a function of all the global symmetries of the theory. The entropy function of known BPS black holes and black strings in AdS$_5 \times \text{SE}_5$, where $\text{SE}_5$ is a Sasaki-Einstein manifold, can indeed be constructed by gluing together copies of $a_{(4)} ( \Delta_a )$ (see, for a non-exhaustive list, \cite{Hosseini:2017mds,Cabo-Bizet:2018ehj, Hosseini:2019iad,Choi:2018hmj,Benini:2018ywd,Cabo-Bizet:2019osg,Amariti:2019mgp,Lanir:2019abx,Hosseini:2019use,Hosseini:2020vgl,Benini:2020gjh,Boido:2022mbe,Martelli:2023oqk}). 

The situation in three dimensions is more subtle. A natural candidate for the gravitational block $\cF(\Delta_a)$ in three-dimensional $\cN=2$ SCFTs is the large-$N$ partition function on $S^3$. However, a long-standing puzzle arises in this context. The global symmetries of theories dual to AdS$_4 \times \text{SE}_7$ solutions in M-theory, where $\text{SE}_7$ is a seven-dimensional Sasaki-Einstein manifold, can be divided into \emph{mesonic}, associated with the isometries of the internal manifold, and \emph{baryonic}, associated with gauge fields arising from the reduction of antisymmetric forms on nontrivial cycles. So far, the large-$N$ limit of three-dimensional supersymmetric partition functions has been computed only for a subset of dual SCFTs, and, when available, it is independent of baryonic fugacities and fluxes \cite{Jafferis:2011zi,Hosseini:2016tor,Hosseini:2016ume,Choi:2019dfu}. When restricting to purely mesonic variables---in the so-called \emph{mesonic twist}--- it is possible to show that the quantum field theory predictions match the entropy of the corresponding black holes in a large class of examples \cite{Hosseini:2019ddy,Gauntlett:2019roi,Choi:2019dfu,Hosseini:2022vho,Colombo:2024mts}. However, the entropy of the most general black hole depends on \emph{all} fluxes, including the baryonic ones. This is confirmed by the existence of explicit black hole solutions with baryonic charges \cite{Halmagyi:2013sla,Azzurli:2017kxo,Hong:2019wyi,Kim:2020qec}.

In \cite{Hosseini:2025ggq}, we proposed a resolution to this puzzle, along with a natural candidate for the universal gravitational block $\cF(\Delta_a)$ for three-dimensional SCFTs that depends on \emph{all} chemical potentials. In this paper, we provide further evidence supporting this proposal, prove the main results for all toric $\text{SE}_7$, and present explicit examples.

Our strategy is to consider a large class of M-theory black hole horizons with topology AdS$_2 \times \Sigma \times \text{SE}_7$, where $\Sigma$ is a two-sphere or a spindle. This configuration is general enough to encompass rotating and magnetically charged twisted black holes in AdS$_4$, Kerr-Newman black holes, and rotating and accelerating black holes with spindle horizons \cite{Ferrero:2020twa}. These solutions can be analyzed with the methods of \cite{Couzens:2018wnk,Gauntlett:2018dpc}, which involve defining an entropy function $\cS ( b , \lambda_a )$ that depends on the supersymmetric Killing vector $b$ and a set of K\"ahler parameters $\lambda_a$ of the internal geometry. The conditions of supersymmetry allows to determine the parameters $\lambda_a$, reducing the problem to a function of the vector field $b$, which must be extremized to obtain the entropy. The method of \cite{Couzens:2018wnk,Gauntlett:2018dpc} can be applied for arbitrary choices of magnetic charges and predicts the entropy even when there are no field theory results to compare with.

In this paper, we show that the full set of supersymmetry conditions of \cite{Couzens:2018wnk,Gauntlett:2018dpc} for a general toric $\text{SE}_7$ can be recast in the form of $\cI$-extremization \eqref{id0}, for an arbitrary choice of charges and fluxes. In the special case of a mesonic twist, the equivalence between the gravitational extremization and $\cI$-extremization with $\cF(\Delta_a) = F_{S^3}(\Delta_a)$ was proven in \cite{Hosseini:2019ddy,Gauntlett:2019roi} for a large class of examples of static, magnetically charged black holes. Here, we extend the definition of $\cF ( \Delta_a )$ to include cases where no field theory results are available, as well as to incorporate all charges, including the baryonic ones.

In our formalism, the gravitational block $\cF(\Delta_a)$ is defined as a constrained Legendre transform of the master volume $\cV(b, \lambda_a)$ \cite{Gauntlett:2018dpc} of the internal geometry---an intrinsic, topological object that depends only on the toric data. This quantity agrees with the large-$N$ free energy on $S^3$ in cases where the latter has been computed. The equivalence between the gravitational extremization of \cite{Couzens:2018wnk,Gauntlett:2018dpc}, which is performed only along the mesonic directions parameterized by the supersymmetric Killing vector $b$, and  $\cI$-extremization, which is carried out with respect to all chemical potentials $\Delta_a$, is possible because the extremization of $\cS ( \Delta_a , \fn_a )$ along the baryonic directions is automatically enforced by the supersymmetry conditions---and is, in fact, equivalent to them. Indeed, we find
\bea
 \label{barextr0}
 \sum_{a = 1}^{d} B_a^{(r)} \frac{\pd \cI ( \Delta_a , \epsilon , \fn_a )}{\pd \Delta_a}= 0 \, , 
\eea
where the charge vectors $B^{(r)}$ specify the baryonic directions.

The main ingredient of our analysis is the generalized free energy $\cF(\Delta_a)$, defined as a constrained Legendre transform of the master volume. This quantity has a two-fold interpretation. On the gravitational side, it provides the universal gravitational block that allows one to construct entropy functions for all asymptotically AdS$_4\times \text{SE}_7$ black holes. We will prove this for a particular class of solutions, those that can be analyzed with the methods of \cite{Couzens:2018wnk,Gauntlett:2018dpc}, but we believe that the result applies to \emph{all} possible black objects, in the spirit of gravitational blocks \cite{Hosseini:2019iad}. On the field theory side, $\cF(\Delta_a)$ should be regarded as a generalization of the large-$N$ free energy $F_{S^3}(\Delta_a)$, reproducing known results when available and offering predictions where they are not. In particular, the extremization of $\cF(\Delta_a)$ with respect to $\Delta_a$ determines the exact R-charges for all the fields in the SCFT, providing predictions for the spectrum of both mesonic and baryonic operators. Note that the extremization of the existing $F_{S^3}(\Delta_a)$ yields information only about mesonic operators \cite{Jafferis:2011zi}, and indirect methods must be employed to extract the dimensions of baryons.

The master volume of the internal geometry depends only on the toric data. It can be extracted from the equivariant volume of the Calabi-Yau (CY) cone over the $\text{SE}_7$ \cite{Martelli:2023oqk}. Both the equivariant volume and the master volume can be computed from purely topological data, and they can be defined even when the explicit supergravity solution is not known. This is, in fact, what we expect. The field theory free energy $\cF(\Delta_a)$ is invariant under small variations of the couplings in the superpotential and in the Lagrangian, and is therefore a renormalization group invariant. The central charge $a_{(4)}(\Delta_a)$ is constructed purely from anomalies, and the $S^3$ free energy $F_{S^3}(\Delta_a)$ can be computed by localizing a supersymmetric partition function; the result is independent of the gauge couplings and the superpotential, unless some term is turned on or off. It thus makes sense that the gravitational counterpart of $\cF(\Delta_a)$ is topological in nature. It was argued in \cite{Martelli:2023oqk} that the equivariant volume is the natural object to appear in all gravitational extremization problems involving branes, and it would be interesting to extend the analysis of this paper to more general supergravity backgrounds involving, for example, M5- or D4-branes in M-theory or massive type IIA. The master volume has also appeared in a recent investigation of the giant graviton expansion of the superconformal index \cite{Chen:2024erz}.  

Although our main interest in this paper is M2-branes and $\cI$-extremization, our results can also be applied in other contests. In particular, our methods provide an alternative and considerably simpler derivations of the equivalence between $a$-maximization and volume minimization \cite{Martelli:2005tp,Martelli:2006yb}, which was originally proven for all toric CY cones in \cite{Butti:2005vn,Butti:2005ps}. They also offer a streamlined derivation of the equivalence between $c$-extremization with its gravitational dual \cite{Couzens:2018wnk,Gauntlett:2018dpc}, as established in \cite{Hosseini:2019use,Boido:2022mbe}, and, of the equivalence between $F$-maximization and volume minimization, checked in numerous examples in \cite{Martelli:2005tp,Martelli:2006yb,Jafferis:2011zi}. For this reason, in section \ref{sec:extr} we consider in parallel the case of M2- and D3-branes.

The paper is organized as follows. In section \ref{sec:extr}, we introduce the equivariant volume and the master volume of the internal geometry for systems of M2- and D3-branes probing CY singularities in M-theory or type IIB string theory, respectively. We define the generalized free energy $\cF(\Delta_a)$ as the contrained Legendre transform of the master volume, and show that the supersymmetry conditions of \cite{Couzens:2018wnk,Gauntlett:2018dpc} can be recast in the form of an $\cI$-extremization. In the process, we also provide alternative proofs of the equivalence between $a$-, $c$- and $F$-extremizations and their gravitational duals. In section \ref{sec:examples}, we present a set of examples and explicit computation of the free energy $\cF(\Delta_a)$ based on toric CY four-folds. In section \ref{sec:BH}, as an application of our formalism, we reproduce the entropy for a class of known black holes with purely baryonic magnetic charges, asymptotic to AdS$_4\times Q^{1,1,1}$ and AdS$_4\times M^{1,1,1}$. We end with conclusions and discussions. An appendix contains technical details.

\section{The extremization problem}
\label{sec:extr}

In this section we introduce the basic object of our story, the constrained Legendre transform $\cV ( \Delta )$ of the master volume of the internal geometry, which allows us to write the extremal functions for all extremization problems involving M2-branes at conical singularities. We will extend the existing partial results for M2-branes to the most general assignment of mesonic and baryonic charges. Although our main interest in this paper is M2-branes and the case of D3-branes is well understood, in this section we will treat both cases together, since the formalism is common. Along the way, we will also recover and simplify the existing results for D3-branes. 

\subsection{The equivariant volume}
\label{sec:equiv}
 
We start by introducing the geometrical objects that enter into the supersymmetric conditions for branes at conical singularities. Consider a toric Calabi-Yau (CY) manifold of complex dimension $m$ (CY$_m$), which is a cone over a $(2m - 1)$-dimensional Sasaki-Einstein manifold $\text{SE}_{2m - 1}$. This can be the conical singularity CY$_n$ that the D3- and M2-branes are probing, but it can also represent the topology of the total internal geometry when the branes are compactified on a two-dimensional variety or orbifold (in which case $m = n + 1$). The CY can be defined by specifying its fan, a set of integer vectors $v^a$, $a = 1, \ldots, d$, in $\bR^m$. The CY condition requires that all the $v^a$ lie on a plane, and in this paper we adopt the convention that one component---always taken to be $i = 1$---of each vector is equal to one, i.e. $v_1^a = 1$. It is well known from toric geometry that the cone is a $\U(1)^m$ fibration over a polytope $\cP_0 = \{ v^a_i y_i \geq 0 \}$ in $\bR^m$. At the $a$-th facet of $\cP_0$, the vector $\sum_{i=1}^m v^a_i \pd_{\phi_i}$ degenerates, where the $\phi_i$ are a basis of $2\pi$-periodic angles for the torus $\U(1)^m$.
The torus fibration over the $a$-th facet then defines a conical torus-invariant divisor $D_a$ in CY$_m$. 

We equip the geometry with two sets of data: a vector
\bea
 b = \sum_{i=1}^m b_i \pd_{\phi_i} \, ,
\eea
in $\U(1)^m$, which we call the Reeb vector, and a set of $d$ K\"ahler parameters $\lambda_a$. The latter can be incorporated by deforming the cone into a toric fibration $M_{2m}$ over the deformed polytope
$$
 \cP = \{ v^a_i y_i - \lambda_a \geq 0 \} \, .
$$
It turns out that all the conditions for supersymmetry for the brane systems of interest can be expressed in terms of the equivariant volume \cite{Martelli:2023oqk}
\bea
 \bV (\lambda_a , b_i) = \frac{1}{(2 \pi)^m} \int_{M_{2m}} e^{ - b_i y_i} \frac{\omega^m}{m!}   \, ,
\eea
where $\omega = \rd y_i \wedge \rd \phi_i$ is a symplectic form on $M_{2m}$. The equivariant volume can be written as the integral of the equivariantly closed form $\omega^b = \omega - b_i y_i$, satisfying $\rd \omega^b - \iota_b \omega^b = 0$, with the components of the Reeb vector $b_i$ playing the role of equivariant parameters:
\bea
 \bV (\lambda_a , b_i)  = \, \frac{1}{(2 \pi)^m} \int_{M_{2m}} e^{ \omega^b} \, .
\eea
It follows that the equivariant volume  can be computed using equivariant localization \cite{Duistermaat,berline1982classes,ATIYAH19841} as a sum over fixed points of the $\U(1)^m$ action and it depends only on topological data. The fixed points correspond to the vertices of the polytope $\cP$. In order to apply the localization formula, we need to have at most orbifold singularities at the fixed points, which correspond to a vertex of $\cP$ where precisely $m$ facets meet. For the original conical CY$_m$, the polytope $\cP_0$ has only one vertex at the origin, where all the facets meet, and cannot be used for localization. We can arrange, on the other hand,  that the deformed polytope $\cP$ has only orbifold singularities. As familiar from toric geometry, this is equivalent to resolving the CY conical singularity by decomposing the fan into a union of simplices $\{ v^{a_1}, v^{a_2}, \ldots , v^{a_m}\}$. Each simplex corresponds to $m$ facets intersecting in $\cP$, and thus to a fixed point. The localization formula then reads \cite{Martelli:2023oqk}
\bea
 \label{fixedpoint}
 \bV ( \lambda_a , b_i ) = \sum_{A=\{a_1 , \ldots , a_{m}\}} \frac{e^{-\frac{\phi_A}{d_A}}}{ d_A \prod_{i=1}^{m} \epsilon^{(A)}_i} \, ,
\eea
where $d_A = |(v^{a_1}, \ldots, v^{a_m})|$, and the weights of the $\U(1)^m$ action are given by
\bea
 \epsilon^{(A)}_i = \frac{(b, v^{a_{j_2}}, \ldots, v^{a_{j_m}})}{(v^{a_{i}}, v^{a_{j_2}} , \ldots, v^{a_{j_m}})}
 \quad \text{ for } \quad i \neq j_2 \neq \ldots \neq j_m \, ,
\eea
and the exponent is
\bea
 \frac{\phi_A}{d_A} = \sum_{i=1}^{m} \epsilon^{(A)}_i \lambda_{a_i} \,   .
\eea
Here and in the rest of the paper  $( v^a , v^b , \ldots , v^d)$ denotes the determinant of the vectors $v^a$, $v^b$, $\ldots , v^d$. 
 
The equivariant volume can be expanded in a formal power series in the K\"ahler parameters $\lambda_a$. The terms of order $m$ encode information about the intersection numbers of the geometry \cite{Nekrasov:2021ked,Cassia:2022lfj}. The terms of order less than $m$ are rational functions of the $b_i$ that contain interesting geometric information about the base of the cone $\text{SE}_{2 m - 1}$, as shown in \cite{Martelli:2023oqk}. In particular, the restriction of the equivariant volume to $\lambda_a = 0$ yields the Sasakian volume of $\text{SE}_{2 m - 1}$, as defined in \cite{Martelli:2005tp,Martelli:2006yb},
\bea
 \label{SE}
 \bV(0, b_i) = \frac{( m - 1)!}{2\pi^m} \Vol_{\text{SE}_{2m-1}}(b_i) \, ,
\eea
which is a function of the Reeb vector only. Extremizing the Sasakian volume with respect to $b_i$ determines the Reeb vector of the Sasaki-Einstein metric and the exact R-symmetry of the dual CFT \cite{Martelli:2005tp,Martelli:2006yb}. In the following, we will also need a formula for the Sasakian volume of the $(2 n - 3)$-cycles $S_a$, obtained by intersecting the divisor $D_a$ with $\text{SE}_{2 m - 1}$:%
\footnote{This formula easily follows from the formalism in \cite{Martelli:2023oqk}.}
\bea
 \label{Sa}
 \frac{\pd \bV(\lambda_a, b_i)}{\pd \lambda_a}  \biggl|_{\lambda_a = 0} = -\frac{( m - 2)!}{2\pi^{m-1}} \Vol_{S_a}(b_i) \, .
\eea
Another interesting object is the $(m - 1)$-th term in the formal Taylor expansion in $\lambda_a$. As shown in \cite{Martelli:2023oqk}, this is equal to the master volume $\cV_{2 m - 1} (\lambda_a, b_i)$, defined in \cite{Gauntlett:2018dpc}:
\bea
 \label{master}
 \cV_{2m - 1}(\lambda_a,b_i) = (2 \pi)^m \bV^{(m-1)}(\lambda_a,b_i) \, .
\eea
The master volume plays an important role in the study of AdS$_2$ and AdS$_3$ supersymmetric solutions \cite{Gauntlett:2018dpc}.  It is a homogeneous function of degree $m-1$ in $\lambda_a$ and degree minus one in $b_i$. In the  approach of \cite{Gauntlett:2018dpc}, the value of $b_1$ is fixed by supersymmetry, but we will keep it arbitrary for convenience.

The equivariant volume depends on the choice of resolution, but the Sasakian volume does not \cite{Martelli:2005tp,Martelli:2006yb}. The master volume is independent of the resolution only when $\text{SE}_{2 m - 1}$ is smooth. Note that only $d - m+1$ parameters of the $\lambda_a$ are independent. Indeed, one can check that \cite{Martelli:2023oqk}
\bea
 \label{shift}
 \bV \Bigl( \lambda_a + \sum_{j=1}^m\beta_j v_j^a,b_i \Bigr) = e^{-\sum_{i=1}^m \beta_i b_i} \bV(\lambda_a , b_i) \, ,
\eea
for arbitrary $\beta_i \in \bR$, which also implies
\bea
 \label{idder}
 \sum_{a=1}^d v_i^a \frac{\pd \bV(\lambda_a , b_i)}{\pd \lambda_a} = -b_i \bV(\lambda_a , b_i) \, .
\eea
It follows that the $\phi_A$, and therefore $\bV ( \lambda_a, b_i)$ and $\cV (\lambda_a, b_i)$, are invariant under the gauge transformation
\bea
 \label{gauge}
 \lambda_a \to \lambda_a + \sum_{i=1}^m \gamma_i (b_1 v_i^a - b_i) \, ,
\eea
with $\gamma_i \in \bR$. In particular,
\bea
 \label{conb}
 b_1 \sum_{a = 1}^d v_i^a \frac{\pd \cV(\lambda_a , b_i)}{\pd \lambda_a}
 = b_i \sum_{a=1}^d \frac{\pd \cV(\lambda_a , b_i)}{\pd \lambda_a}\, .
\eea
 
\subsection{Four- and three-dimensional SCFTs}
\label{sec:free}

We now consider a system of $N$ branes probing a CY singularity CY$_n$, which is a cone over a Sasaki-Einstein manifold $\text{SE}_{2 n-1}$. The case $n = 3$ corresponds to D3-branes in type IIB string theory, which support a four-dimensional SCFT and have a dual AdS$_5 \times \text{SE}_5$ geometry. The case $n = 4$ corresponds to M2-branes in M-theory, which support a three-dimensional SCFT and have a dual AdS$_4\times \text{SE}_7$ geometry.   Given the similarity of the formalism, we will treat both cases in parallel, using the complex dimension $n$ of the CY to distinguish them. 

We also assume that $CY_n$ is toric and specified by a fan with vectors $v^a$, $a = 1, \ldots, d$, whose first component is equal to one. Let us now consider the holographic description of the theory. According to the rules of the AdS/CFT correspondence, every massless gauge field in the AdS bulk corresponds to a global symmetry of the dual SCFT. Massless gauge fields arise from the isometries of the internal manifold $\text{SE}_{2 n-1}$, and therefore we find $n$ abelian global symmetries of the dual SCFT associated with the $\U(1)^n$ toric action. We refer to these symmetries as \emph{mesonic}. In addition, the reduction of antisymmetric forms along non-trivial cycles in the internal geometry $\text{SE}_{2 n-1}$ give rise to extra abelian symmetries. For D3-branes, the relevant field is the RR four-form of type IIB reduced over three-cycles. For M2-branes, the relevant field is the M-theory three-form reduced over two-cycles. By Poincar\`e duality, the relevant cohomology group is $H^{2 n - 3}( \text{SE}_{2 n-1} )$, accommodating both cases in a single expression. For toric $\text{SE}_{2 n-1}$, the dimension of $H^{2 n - 3}(\text{SE}_{2 n-1})$ is $d - n$. We then have extra $d - n$ extra abelian symmetries, which we dub \emph{baryonic}. The name refers to the fact that the charged objects are branes wrapped over $( 2 n - 3 )$-cycles, which give rise to baryonic operators in the SCFT. 

One geometric way to see the splitting between mesonic and baryonic symmetries is to use the symplectic reduction description of the CY. In this description, the CY can be written as the quotient of $\bC^d$ with respect to the subgroup $( \bC^*)^{d-n}$, whose generators $B^{(r)}$ satisfy
\bea
 \label{baryonic}
 \sum_{a = 1}^{d} B_a^{(r)} v^a = 0 \, , \quad \text{ for } \quad r = 1, \dots, d - n \, .
\eea
From this realization, we see that the global symmetry of the theory is the isometry group $\U(1)^d$ of $\bC^d$, which can be split into a $\U(1)^n$ mesonic subgroup corresponding to the isometries of $\text{SE}_{2 n-1}$, and $d - n$ baryonic $\U(1)$'s coming from the ambient space and associated with the charges $B_a^{(r)}$ in \eqref{baryonic}. Notice that the distinction between mesonic and baryonic directions is by no means canonical, as it depends on a choice of splitting of the exact sequence of tori embeddings:
\bea
 0 ~ \rightarrow ~ (\bC^\star)^n ~ \rightarrow ~ (\bC^\star)^d ~\rightarrow ~ (\bC^\star)^{d-n} ~ \rightarrow ~ 0 \, .
\eea
One general piece of information coming  from the symplectic description is that the $\U(1)$ symmetries of the theory can be naturally labeled by the index $a=1, \ldots , d$, and are associated with vectors in the fan.

The quantum field theory quantities of interest for holographic comparisons are the trial central charge $a_{(4)}$ for the four-dimensional SCFTs and the three-sphere free-energy, $F_{S^3}$, for three-dimensional SCFTs. These are functions of the R-symmetry of the theory. Recall that the R-symmetry in a SCFT is not necessarily unique: given an R-symmetry $R$ and a global symmetry $J$, the combination $R + J$ is also an $R$-symmetry. In our setting, we therefore have a family of R-symmetries parameterized by $d$ real parameters. We can introduce a basis of generators $T_a$, each with integer spectrum, associated with the vectors $v_a$ of the fan, and write the general R-symmetry as $\sum_{a = 1}^d \Delta_a T_a$. We normalize the R-symmetry such that the superpotential has charge two, and we adopt conventions where this constraint corresponds to
\bea
 \label{cons}
 \sum_{a=1}^d \Delta_a=2 \, .
\eea
Every chiral multiplet in the SCFT has an $R$-charge that is a linear combinations of the $\Delta_a$. For D3-branes probing toric CY$_3$, there is an explicit parameterization of the field theory $R$-charges in terms of the $\Delta_a$, based on the combinatorial data of the fan \cite{Butti:2005vn,Butti:2005ps}. For M2-branes, on the other hand, this mapping has mostly been derived on a case-by-case basis \cite{Aharony:2008ug,Hanany:2008cd,Hanany:2008fj,Martelli:2008si,Benini:2009qs,Jafferis:2009th}. We use the $d$ charges $\Delta_a$ to parameterize the trial central charge $a_{(4)} (\Delta)$ of the four-dimensional SCFTs and the $S^3$ free-energy  $F_{S^3} ( \Delta )$ of the three-dimensional SCFTs. With some abuse of language, we will refer to these quantities generically as \emph{free energies}, and we will denote them by the common symbol $\cF ( \Delta )$ when needed. The extremization of the free energy $\cF ( \Delta )$ with respect to the $\Delta_a$, subject to the constraint \eqref{cons}, identifies the exact $R$-symmetry that sits in the superconformal algebra \cite{Intriligator:2003jj, Jafferis:2010un}.

We now want to make contact with the gravitational description and relate the free energy $\cF ( \Delta )$ to the master volume of the Sasaki-Einstein manifold $\text{SE}_{2 n - 1}$. More precisely, we will see that, up to a numerical factor, $\cF ( \Delta )$ is a \emph{constrained Legendre transform} of the master volume.

\subsubsection{The gravitational interpretation of the free energy}
\label{freegrav}
 
Recall that the master volume $\cV ( \lambda_a , b_i)$ of $\text{SE}_{2 n-1}$ is a function of both the K\"ahler moduli $\lambda_a$ and the components of the Reeb vector $b_i$. We now want to perform a Legendre transform with respect to the $\lambda_a$ and define the conjugate variables:
\bea
 \label{conj}
 \eta_a  \equiv \frac{\pd \cV_{2 n - 1} ( \lambda_a , b_i )}{\pd \lambda_a} \, .
\eea
The components of the Reeb vector $b_i$ can also be expressed in terms of the dual variables $\eta_a$ and $b_1$ using \eqref{conb}. Since the master volume is a homogeneous function of the $\lambda_a$, we have
\bea
 \label{leg}
 \cV_{2 n - 1} ( \lambda_a , b_i ) \equiv \alpha_n \biggl( \sum_{a = 1}^{d} \lambda_a \frac{\pd \cV_{2 n - 1} ( \lambda_a , b_i )}{\pd \lambda_a} - \cV_{2 n - 1} ( \lambda_a , b_i ) \biggr) \, ,
\eea
with $\alpha_3 = 1$ and $\alpha_4 = \frac12$. The Legendre transform, up to a constant, thus coincides with the master volume itself, now viewed as a function of the dual variables.

Let us now discuss in some detail how the Legendre transform is carried out. We associate the variables $\eta_a$ with the field theory quantities $( \Delta_a , N )$, which satisfy the constraint $\sum_{a=1}^d \Delta_a = 2$, through the identification
\bea
 - 2 \eta_a = N \Delta_a \, .
\eea
More explicitly, \eqref{conj} becomes
\bea
 \label{def:N}
 N \equiv - \sum_{a=1}^d \frac{\pd \cV_{2 n - 1} ( \lambda_a , b_i )}{\pd \lambda_a} \, ,
\eea
and \cite{Gauntlett:2018dpc,Hosseini:2019use}
\bea
 \label{def:R-charge}
 \Delta_a  \equiv - \frac{2}{N} \frac{\pd \cV_{2 n - 1} ( \lambda_a , b_i )}{\pd \lambda_a} \, .
\eea
We can now invert the relations \eqref{def:N} and \eqref{def:R-charge}, and express the variables $( \lambda_a, b_i)$ in terms of $( N , \Delta_a ,  b_1)$.
The $d$ relations \eqref{def:N}, \eqref{def:R-charge} are subject to the $n - 1$ constraints \eqref{conb}, which can be rewritten as
\bea
 \label{reeb:id}
 b_i = \frac{b_1}{2} \sum_{a = 1}^d v_i^a \Delta_a \, , \quad \text{ for } \quad i = 1 , \ldots, n \, .
\eea
These identities allow us to express $b_i$ as a function of $\Delta_a$ and $b_1$. Note that \eqref{conb} imposes this relation only for $i \ne 1$, but the $i = 1$ case follows automatically from the constraint \eqref{cons}. As a result, we are left with $d - n + 1$  independent relations \eqref{def:N} and \eqref{def:R-charge} to solve. Fortunately, due to the gauge invariance \eqref{gauge}, only $d - n + 1$ of the $\lambda_a$ are independent, and the system of equations is solvable. Using also \eqref{reeb:id}, we can solve for $\lambda_a$ as functions of $\Delta_a$, $N$ and $b_1$. The solution is not unique and is defined up to a gauge transformation with $n - 1$ free parameters:
\bea
 \lambda_a (N, \Delta, b_1) \rightarrow \lambda_a(N, \Delta ,b_1) + \sum_{i = 1}^n \beta_i v_i^a \, ,
 \qquad \sum_{i=1}^n \beta_i b_i = 0 \, .
\eea 
The parameters $\beta_i$ drop out when $\lambda_a$ is inserted into any gauge invariant quantity, in particular, the master volume and its derivatives. We can finally compute the Legendre transform by re-expressing the master volume as a function of  $(N, \Delta, b_1)$:
\bea
 \label{free}
 \boxed{ \cV_{2 n-1}(N, \Delta_a, b_1) \equiv \cV_{2 n-1}(\lambda_a(N, \Delta, b_1), b_i(\Delta,b_1))} \, .
\eea

Let us discuss some general properties of the functional $\cV_{2 n-1} (N, \Delta_a,  b_1)$.  A simple scaling argument implies that
\bea
 \cV_{5} ( N , \Delta_a , b_1 ) = N^{2} b_1 \cW_{5} ( \Delta_a ) \, , \qquad
 \cV_{7} ( N , \Delta_a , b_1 ) = N^{3/2} \sqrt{b_1} \sqrt{\cW_{7} ( \Delta_a )} \, ,
\eea
where $\cW_{2 n - 1} ( \Delta_a )$ is a homogenous function of degree $n$ in $\Delta_a$. The scaling behavior above follows from the fact that $\cV_{2n-1}$ is homogeneous of degree $n - 1$ in $\lambda_a$ and of degree $-1$ in $b_i$, which implies the following scaling in $(\Delta, N, b_1)$ 
\be
\label{scaling}
\begin{array}{c|ccc|ccc}
 & \multicolumn{3}{c|}{n = 4} & \multicolumn{3}{c}{n = 3} \\
\cline{2-7}
\rule{0pt}{1.em} & N & \Delta_a & b_1 & N & \Delta_a & b_1 \\
\hline
\rule{0pt}{1.5em} \lambda_a & \dfrac{1}{2} & 1 & \dfrac{1}{2} & 1 & 2 & 1 \\
\rule{0pt}{1.em} b_i & 0 & 1 & 1 & 0 & 1 & 1 \\
\rule{0pt}{1.5em} \mathcal{V}_{2 n - 1} & \dfrac{3}{2} & 2 & \dfrac{1}{2} & 2 & 3 & 1 \\[0.6em]
\hline
\end{array}
\ee

Moreover, the Legendre transform can be inverted to give the variables $( b_i , \lambda_a )$. Due to the gauge invariance \eqref{gauge}, we can only determine the gauge invariant quantities $b_i$ and $X^{(r)} \equiv \sum_{a=1}^d B_a^{(r)} \lambda_a$. The former correspond to the mesonic moduli of the problem and are given by \eqref{reeb:id}, which we repeat here for completeness:
\bea
 \label{reeb:id2}
 \boxed {b_i = \frac{b_1}{2} \sum_{a = 1}^d v_i^a \Delta_a \, , \quad \text{ for } \quad i = 1 , \ldots, n \, .}
\eea
The latter correspond to the baryonic moduli and are given by
\bea
\boxed{
 \label{bar}
 X^{(r) } \equiv \sum_{a=1}^d B_a^{(r)} \lambda_a
 = - \frac{2}{\alpha_n N} \sum_{a = 1}^{d} B_a^{(r)} \frac{\pd \cV_{2 n - 1} (N , \Delta  , b_1 )}{\pd \Delta_a} \, , \quad \text{ for } \quad r = 1, \ldots , d - n \, .}
\eea

The previous equation can be proved as follows. From \eqref{def:R-charge} and \eqref{leg}, we obtain
\bea
 \cV_{2 n - 1} ( N , \Delta ,  b_1 ) = \alpha_n \biggl[ - \frac{N}{2}\sum_{c = 1}^{d} \lambda_c ( \Delta ) \Delta_c  - \cV_{2 n - 1} \left( \lambda_b (\Delta) , b_i ( \Delta) \right) \biggr] \, .
\eea
Therefore, using \eqref{reeb:id}, we find that
\bea
 - \frac{2}{\alpha_n N} \sum_{a = 1}^{d} B^{(r)}_a \frac{\pd \cV_{2 n - 1} ( N , \Delta , b_1 )}{\pd \Delta_a} & =
 \sum_{a = 1}^{d} B^{(r)}_{a} \biggl( \lambda_a ( \Delta) + \sum_{c = 1}^{d} \frac{\pd \lambda_c ( \Delta)}{\pd \Delta_a} \Delta_c \biggr) \\
 & + \frac{2}{N} \sum_{a , c = 1}^{d} B^{(r)}_a \frac{\pd \cV_{2 n - 1} \left( \lambda_b ( \Delta) , b_i ( \Delta) \right)}{\pd \lambda_c}
 \frac{\pd \lambda_c (\Delta)}{\pd \Delta_a} \\
 & + \frac{b_1}{N} \sum_{a = 1}^{d} \sum_{i = 1}^{n} B^{(r)}_a v_i^a \frac{\pd \cV_{2 n - 1} \left( \lambda_b ( \Delta) , b_i ( \Delta) \right)}{\pd b_i} \, .
\eea
Using \eqref{baryonic} and  \eqref{def:R-charge}, we indeed obtain
\bea
 - \frac{2}{\alpha_n N} \sum_{a = 1}^{d} B^{(r)}_a \frac{\pd \cV_{2 n - 1} ( N , \Delta , b_1 )}{\pd \Delta_a} & =
 \sum_{a=1}^d B_a^{(r)} \lambda_a ( \Delta) \, ,
\eea
ending the proof. 

The previous equation determines $\lambda_a$
\bea
 \label{po}
 \lambda_a (\Delta) = - \frac{2}{\alpha_n N}  \frac{\pd \cV_{2 n - 1} ( N , \Delta , b_1 )}{\pd \Delta_a} +\sum_{i=1}^n \wt \beta_i v_i^a \, ,
\eea
up to $n$ free parameters $\wt\beta_i$. Using the gauge freedom \eqref{gauge}, we can eliminate $n - 1$ of them by setting $\wt \beta_i = 0$ for $i > 1$. In this gauge, the remaining parameter $\wt \beta_1$ can be fixed by multiplying \eqref{po} by $\Delta_a$, summing over $a$, and using the homogeneity properties of the master volume. We thus find
\bea
 N \lambda_a(\Delta) = \cV_{2 n - 1} (N, \Delta, b_1) - \frac{2}{\alpha_n} \frac{\pd \cV_{2 n - 1} (N, \Delta , b_1 )}{\pd \Delta_a} \, ,
\eea
which holds in this particular gauge.

\subsubsection{Comparing with the quantum field theory free energy}
\label{freeext}

We now compare $\cV_{2 n - 1} (N, \Delta, b_1)$ with the the trial central charge $a_{(4)}(\Delta)$ of four-dimensional SCFTs for $n = 3$, and with the $S^3$ free-energy $F_{S^3} ( \Delta )$ of three-dimensional SCFTs for $n = 4$.

In the four-dimensional case, the vectors of the fan lie on a two-dimensional plane and can be ordered anticlockwise, with the convention that $v^{d+1}=v^1$ and $v^0=v^d$. There is a closed-form expression for the master volume \cite{Gauntlett:2018dpc}%
\footnote{For a derivation from the fixed point formula \eqref{fixedpoint}, see \cite{Martelli:2023oqk}.}
\bea
 \label{mastervol}
 {\cal V}=4 \pi^3 \sum_{a=1}^d \lambda_a  \frac{ \lambda_{a-1} (v_a,v_{a+1},b) - \lambda_{a} (v_{a-1},v_{a+1},b) + \lambda_{a+1} (v_{a-1},v_{a},b)}{  (v_{a-1},v_{a},b) (v_{a},v_{a+1},b)} \, ,
\eea
which allows one to solve \eqref{def:R-charge} for $\lambda_a$ very easily. In the gauge $\lambda_1 = \lambda_2=0$, one finds by induction \cite[(B.17)]{Hosseini:2019use}
 \bea
 \label{inversion}
 \lambda_a = -\frac{N}{16\pi^3}  \sum_{c=2}^{a} (v_c,v_a,b) \Delta_c \, ,\qquad a=3,\ldots ,d \, .
\eea 
Then one can easily show that \cite[Sect.\,B.4]{Hosseini:2019use}%
\footnote{See also \cite{Chen:2024erz} for a recent application.}
\bea
 \label{4da}
 \cV_{5} ( \Delta_a ) = \frac{b_1}{4 ( 3 \pi )^3} a_{(4)} ( \Delta_a ) \, ,
\eea
where the trial central charge is given, in the large-$N$ limit, by
\bea
 \label{cubic}
 a_{(4)}(\Delta_a) =\frac{9}{32} \Tr R(\Delta_a)^3 = \frac{9}{32} \sum_{a,b,c=1}^d c_{abc} \Delta_a \Delta_b \Delta_c \, ,
\eea
and t'Hooft anomaly coefficients are given in terms of toric data by \cite{Benvenuti:2006xg}
\bea
 c_{abc} = \frac{N^2}{2}|(v^a,v^b,v^c)| \, .
\eea 

The situation is more complicated in three dimensions. The origin of these complications lies in our incomplete understanding of the large-$N$ limit of the $S^3$ free energy for quiver gauge theories with a holographic dual. The large-$N$ computation was carried out using supersymmetric localization in \cite{Herzog:2010hf,Jafferis:2011zi,Martelli:2011qj}, but it only works for quivers with vector-like matter fields.%
\footnote{More precisely, the bi-fundamental fields must transform in a real representation of the gauge group, and the total number of fundamentals must be equal to the total number of anti-fundamentals.}
Moreover, there are accidental flat directions at large $N$, and the R-charges parameterizing the baryonic symmetries disappear from the free energy functional $F_{S^3}(\Delta_a)$ \cite{Jafferis:2011zi}. Correspondingly, the master volume of the associated CY$_4$ has been mostly investigated for the so-called mesonic twist \cite{Hosseini:2019ddy,Gauntlett:2019roi}, which corresponds to setting  the baryonic moduli $X^{(r) } \equiv \sum_{a=1}^d B_a^{(r)} \lambda_a$ to zero. Under this restriction, it was shown in \cite{Hosseini:2019ddy} for a large class of examples that
\bea
 \label{free3d}
 F_{S^3}(\Delta_a) \equiv \frac{(4 \pi)^3}{\sqrt{b_1}} \cV_7(\Delta_a) \, .
\eea
Our functional $\cV_7(\Delta_a)$ can  be defined also for CY$_4$ associated with chiral quivers and explicitly depends on the baryonic variable $X^{(r)}$. It is tempting to speculate that the existing restrictions on $ F_{S^3}(\Delta_a)$ are due to our ignorance about more general saddle points in the large-$N$ limit. In this case, \eqref{free3d} stands as a \emph{conjecture} for the relation between the large-$N$ free energy---still to be found---and the master volume of the internal geometry.

As a non-trivial check, we now show that the extremization of  $\cV_{2 n-1}(\Delta_a)$ reproduces the correct exact R-symmetry and is consistent with what we know about $a$- and $F$-maximization \cite{Intriligator:2003jj,Jafferis:2010un} and their equivalence with volume minimization \cite{Martelli:2005tp,Martelli:2006yb}. We will consider the two cases $n = 4$ and $n = 3$ in parallel. The functional $\cV_{2 n-1}(\Delta_a)$ needs to be extremized with respect to all $\Delta_a$, subject to the constraint \eqref{cons}; this includes extremizing with respect to both mesonic and baryonic directions. From \eqref{bar}, we see that extremization along the baryonic directions imposes $X^{(r)} = 0$ for $r = 1, \ldots, d - n$. Since the baryonic K\"ahler moduli vanish, we are left with functions of the mesonic moduli only, which can be parameterized by the components of the Reeb vector $b_i$. More explicitly, the general solution to $X^{(r) } \equiv \sum_{a=1}^d B_a^{(r)} \lambda_a = 0$ is given by $\lambda_a = \sum_{i=1}^n \beta_i v_i^a$. 
It then follows from \eqref{shift} that all the exponents in the fixed point formula \eqref{fixedpoint} are equal:
\bea
 \frac{\phi_A}{d_A} = \sum_{j = 1}^{n} \beta_j b_j \equiv - \Phi\, .
\eea  
Indeed, from \eqref{shift} and \eqref{SE} we have
\bea
 \label{shift2}
 \bV \Bigl( \sum_{j=1}^n\beta_j v_j^a,b_i \Bigr) = e^{-\sum_{i=1}^n \beta_i b_i} \bV(0 , b_i) =  \frac{( n - 1)!}{2\pi^n} e^{\Phi} \Vol_{\text{SE}_{2n-1}}(b_i) \, .
\eea
The master volume \eqref{master} then reads
\bea
 \label{shift3}
 \cV_{2 n-1} ( \Phi , b_i) =  (2 \Phi)^{n-1} \Vol_{\text{SE}_{2n-1}}(b_i) \, .
\eea
Moreover, we find: 
\bea
 \label{rell}
 \sum_{a = 1}^d \frac{\pd \Phi}{\pd \lambda_a} =\Phi  \bigl|_{\lambda_a = 1} = \Phi \bigl|_{\beta_i = \delta_{i1}} = - b_1 \, ,
\eea
and from \eqref{def:N}
\bea
 N = (n-1)2^{n-1} b_1 \Phi^{n-2} \Vol_{\text{SE}_{2 n-1}}( b_i ) \, .
\eea 
By eliminating $\Phi$, we finally obtain
\bea
 \cV_{2 n-1} (b_i) = \left( \frac{N}{2 b_1 (n-1)} \right)^{\frac{n-1}{n-2}}\frac{1}{\bigl( \Vol_{\text{SE}_{2 n-1}}(b_i)\bigr)^{\frac{1}{n-2}}}\, .
\eea
This expression must still be extremized with respect to the mesonic moduli $b_i$, and we see that the extremization of the free energy is equivalent to the minimization of the Sasakian volume \cite{Martelli:2005tp,Martelli:2006yb,Jafferis:2011zi}.

Moreover, differentiating \eqref{shift} with respect to $\lambda_a$ and using \eqref{Sa}, we find
  \bea
 \label{shift2}
\frac{\pd}{\pd \lambda_a}  \bV \Bigl( \lambda_a + \sum_{j=1}^n\beta_j v_j^a,b_i \Bigr) \biggl|_{\lambda_a = 0}
= e^{-\sum_{i=1}^n \beta_i b_i} \frac{\pd}{\pd \lambda_a}  \bV(\lambda_a , b_i) \biggl|_{\lambda_a = 0} = - \frac{( n - 2)!}{2\pi^{n-1}} e^{\Phi} \Vol_{S_a}(b_i) \, ,
\eea
which, when restricted to the terms of degree $n-2$, gives
\bea
 \frac{\pd \cV_{2 n-1} ( \Phi , b_i)}{\pd \lambda_a} =- 2 \pi (2\Phi)^{n-2}  \Vol_{S_a}(b_i) \, ,
\eea
and finally, from \eqref{def:R-charge}, we obtain
\bea
 \label{BZ}
 \Delta_a(b_i) =\frac{2\pi}{ b_1(n-1)} \frac{\Vol_{S_a}(b_i)}{\Vol_{\text{SE}_{2 n-1}}(b_i)}\, .
\eea
On the mesonic locus \eqref{BZ}, the quantities \eqref{4da} and \eqref{free3d} read 
\bea
 \label{aFv}
 a_{(4)} ( \Delta ( b_i ) ) = \frac{\pi^3 N^2}{4  \Vol_{\text{SE}_{5}}(b_i)} \, , \qquad
 F_{S^3} ( \Delta ( b_i ) ) = N^{3/2}\sqrt{\frac{2 \pi^6}{27 \Vol_{\text{SE}_{7}}(b_i)}} \, ,
\eea
where $b_1 = n$ has been fixed by supersymmetry. The formulae \eqref{aFv} reproduce well-established results in the literature  \cite{Martelli:2005tp,Martelli:2006yb,Butti:2005vn,Butti:2005ps,Jafferis:2011zi}. In particular, \eqref{BZ} is precisely the parameterization of the R-charges used in \cite{Butti:2005vn,Butti:2005ps} to prove the equivalence of $a$-maximization and volume minimization for general toric CY$_3$, as well as the one used in \cite{Jafferis:2011zi} to check, in many examples, the equivalence of $F$-maximization and volume minimization for CY$_4$. We see that this is a geometric consequence of imposing extremization with respect to the baryonic symmetries. 

Let us summarize the results of this section. For four-dimensional theories, we have simplified the proof of the equivalence of $a$-maximization and volume minimization provided in \cite{Butti:2005vn,Butti:2005ps}. For three-dimensional theories, we re-derived the gravitational expression \eqref{aFv} for the $S^3$ free energy. Moreover, we have defined a functional that naturally extends the  $S^3$ free energy at large $N$ to include CY$_4$ with chiral duals and general baryonic R-charges. Our proposal is a function of all the variables $\Delta_a$ and reduces to \eqref{aFv} when the baryonic variables $X^{(r)}$ vanish. It can be seen as the three-dimensional analogue of the cubic formula \eqref{cubic} for the large-$N$ trial $a_{(4)}$ central charge. Our formula generalizes and completes previous attempts to find a formula based on a quartic expression \cite{Amariti:2011uw,Amariti:2012tj} by incorporating the baryonic directions. In particular, we can now formulate an extremization problem with respect to all mesonic and baryonic directions. Since the extremization with respect to the baryonic directions \eqref{bar} imposes $X^{(r)}=0$, the new extremization problem is essentially equivalent to the old one. However, it provides the extremal value of all the R-charges $\Delta_a$ and can be used to predict the R-charges of both mesonic and baryonic operators. In contrast, the R-charges of baryonic operators can only be extracted through indirect methods in the approach of \cite{Jafferis:2011zi}.

\subsection{Compactifying on a spindle}
\label{spindle}

We want to argue that the Legendre transform $\cV ( \Delta )$ of the master volume is the \emph{gravitational block} that allows one to build extremal functions for the most general black hole or black string that can be embedded in  AdS$_4\times \text{SE}_{7}$ and  AdS$_5\times \text{SE}_{5}$.
 
To this purpose, we consider M-theory black hole horizons with topology AdS$_2 \times \Sigma \times \text{SE}_7$, which can be realized by further wrapping the stack of $N$ M2-branes on a spindle $\Sigma = \mathbb{WP}^1_{[n_+,n_-]}$, a two-sphere with conical deficits characterized by $n_+$ and $n_-$ at the poles, with a topological twist ($\sigma=1$) or an anti-twist ($\sigma=-1$) \cite{Ferrero:2020twa,Ferrero:2021ovq,Ferrero:2021etw,Boido:2022mbe}. This configuration is general enough to accommodate the general rotating and magnetically charged twisted black holes in AdS$_4$ for $\sigma = n_\pm = 1$, the Kerr-Newman black holes for $\sigma = - n_\pm = - 1$, and the general rotating and accelerating black holes for coprime $n_+$ and $n_-$. We will write an extremal function whose Legendre transform yields the entropy of these black holes. Similarly, we consider type IIB solutions with topology AdS$_3 \times \Sigma \times \text{SE}_5$, which are dual to general rotating black strings in AdS$_5$ \cite{Ferrero:2020laf,Hosseini:2021fge,Boido:2021szx}. The corresponding extremal function reproduces the trial central charge of the dual two-dimensional SCFT.

The geometry can be modeled as follows \cite{Boido:2022mbe}. In the examples that can be realized in the approach of \cite{Gauntlett:2018dpc}, the cone over $\Sigma \times \text{SE}_{2 n - 1}$ is topologically still a CY. It can be realized as a fibration of the CY$_n = C ( \text{SE}_{2 n-1} )$ over the spindle $\Sigma$
\bea
 \label{fibre}
 {\rm CY}_n \hookrightarrow {\rm CY}_{n+1} \to \Sigma \, ,
\eea
and described by the fan
\bea
 V^a = ( 0 , v^a ) \, , \qquad V^+ = ( n_+ , w^+ ) \, , \qquad V^- = ( - \sigma n_- , w^- ) \, ,
\eea
where we use capital letters $V_I^A$ with indices $I = 0 , \ldots , n$ to describe the $(n+1)$-dimensional complex geometry. The fibration is specified by the $n$-dimensional vectors $w^\pm$. We will assume that the first component of the vectors $w^\pm$ is one, $w^{\pm}_1 = 1$, so that the $(n+1)$-dimensional geometry is still a CY. Notice also that, in the anti-twist case $\sigma = - 1$, the toric diagram is not convex and it does not define a toric geometry. We will study the anti-twist by analytically continuing the results from the twist case. We now have an extra $\U(1)$ isometry associated with the rotation along the spindle. Correspondingly, we  introduce $n + 1$  equivariant parameters $b = ( b_0 , b_1 , \ldots , b_n)$, where we use conventions in which $b_i$, $i = 1 , \ldots , n$ refer to the CY$_n$ and $b_0$ to the spindle. Using the fixed point formula \eqref{fixedpoint} with $m = n + 1$, by resolving the CY$_{n+1}$ with polyhedra $( V^{\pm} , V^{a_1} , \ldots , V^{a_n})$, it is easy to show that \cite{Martelli:2023oqk}
\bea
 \label{FIN} 
 \bV_{\text{CY}_{n+1}}(\lambda_A,b_I) = \frac{1}{b_0}\bV_{\text{CY}_{n}} \left(\lambda^+_a \, , b^+_i  \right) -\frac{1}{b_0}\bV_{\text{CY}_{n}} \left(\lambda^-_a \, , b^-_i  \right) \, ,
\eea
where
\bea
 & b_i^\pm \equiv b_i -\frac{b_0 w_i^\pm}{V_0^\pm} \, , \qquad
 & \lambda_a^\pm \equiv \lambda_a +\frac{b_0}{V_0^\pm b_1^\pm} \lambda_\pm \, .
\eea
We see that $\bV_{\text{CY}_{n+1}} $ can be obtained by gluing two copies of $\bV_{\text{CY}_{n}}$, This generalizes an identity for the master volume derived  in \cite{Boido:2022mbe} by an explicit calculation. 

The condition of supersymmetry, as written in \cite{Gauntlett:2019roi}, reads%
\footnote{Compared with \cite{Gauntlett:2019roi}, we set $\nu_n =1$ for simplicity.}
\bea
 \label{BPS}
 \frac{\pd \cS}{\pd \lambda_A} = - 2 \pi M_A \, ,
\eea
where%
\footnote{The supersymmetric action is defined in \cite{Couzens:2018wnk,Gauntlett:2018dpc} as $\cS = -\sum_{A=1}^D \frac{\pd \cV_{2 n+1}}{\pd \lambda_A}$, where $\cV_{2n+1}$ is the master volume of the $(n+1)$-dimensional geometry. The equation in the text follows from \eqref{master} and \eqref{idder}.}
\bea
 \label{susu:action}
 \cS ( \lambda_A , b_I ) =  b_1( 2 \pi )^{n + 1} \bV^{(n-1)}_{\text{CY}_{n+1}} \, ,
\eea
is the \emph{supersymmetric action} introduced in \cite{Couzens:2018wnk,Gauntlett:2018dpc}, and the $M_A$ are the integrally quantized fluxes of the antisymmetric $( 2 n - 1 )$-curvature form through the torus-invariant $( 2 n - 1 )$-cycles of the internal geometry. These cycles arise by restricting the torus-invariant divisors $D_A$ of the CY$_{n+1}$, to the base. Each such cycle corresponds to a vector $V^A$ in the toric data. We set $D = d + 2$, and adopt the convention that the index $A$ runs over $\pm$, and $a = 1 , \ldots , d$, in that order. The fluxes satisfy the toric geometry conditions
\bea
 \label{lin}
 \sum_{A=1}^{D} V_I^A M_A = 0 \, , \quad \text{ for } \quad I = 0, \ldots , n \, ,
\eea
which reflect the fact that only $D - n - 1$ of the $( 2 n - 1 )$-cycles are homologically independent. The components $I = 0$ and $I=1$ of these relations allow us to parameterize the fluxes as
\bea
 \label{fluxes}
 M_A = N \Bigl(\frac{1}{n_+},\frac{1}{\sigma n_-}, -\fn_a \Bigr) \, ,
\eea
with
\bea
 \sum_{a = 1}^{d} \fn_a = \chi_\sigma \, , \qquad  \chi_\sigma \equiv \frac{1}{n_+} + \frac{\sigma}{n_-} \, ,
\eea
while the remaining components, $I = 2 , \ldots , n$, can be used to express the fibration vectors $w^\pm$ in terms of the fluxes.

\subsubsection{The gravitational extremal functional}
\label{Igrav}
The relation \eqref{FIN} implies
\bea
 \label{FIN:master}
 \cS ( \lambda_A,b_I) = 2\pi \frac{b_1}{b_0}\left( \cV_{2 n-1}(\lambda^+_a \, , b^+_i) -\cV_{2n-1}(\lambda^-_a \, , b^-_i ) \right) ,
\eea
where $\cV_{2 n-1} ( \lambda_a , b_i )$ is the master volume of CY$_n$, given by
\bea
 \label{fixedpoint2}
 \cV ( \lambda_a^\pm , b_i^\pm ) = \sum_{A=\{a_1 , \ldots , a_{m}\}} \frac{e^{\Phi_A^\pm}}{ d_A \prod_{i=1}^{m} \epsilon^{(A)\pm}_i} \, ,
\eea
with%
\footnote{We used the relation $\sum_{i=1}^n \epsilon^{(A)\pm}_i =b_1^\pm$, which follows from $\Phi \bigl|_{\lambda_a = 1} = - b_1$ (see \eqref{rell}).}
\bea
 \Phi_A^\pm = -\sum_{i=1}^n \epsilon^{(A)\pm}_i \lambda^\pm_{a_i}
 = - \sum_{i=1}^n \epsilon^{(A)\pm}_i \lambda_{a_i} -\frac{b_0}{V_0^\pm} \lambda_\pm \, ,
\eea

The supersymmetry conditions \eqref{BPS} then give
\begin{subequations}
 \begin{align}
  & \sum_{a=1}^d \frac{\pd \cV_{2n-1}( \lambda_a^\pm , b_i^\pm )}{\pd \lambda_a} = - \frac{b_1^\pm}{b_1} N \, , \label{BPS1} \\
  & \frac{b_1}{b_0}\left( \frac{\pd \cV_{2n-1}( \lambda_a^+ , b_i^+ )}{\pd \lambda_a} - \frac{\pd \cV_{2n-1}( \lambda_a^- , b_i^- )}{\pd \lambda_a} \right) = N \fn_a \, . \label{BPS2}
 \end{align}
\end{subequations}
We now introduce the dual variables
\bea
 \label{Rpm}
 \Delta_a^\pm =-\frac{2}{N}\frac{\pd \cV_{2n-1}( \lambda_a^\pm , b_i^\pm )}{\pd \lambda_a} \, ,
\eea
and replace $\cV_{2n-1}( \lambda_a^\pm , b_i^\pm )$ with their Legendre transform%
\footnote{The reader may worry that equations \eqref{BPS1} and \eqref{Rpm} are similar to \eqref{def:N} and \eqref{def:R-charge}, but differ by some factors of $b_1^\pm$ and $b_1$. It is easy to see, with a scaling argument, that all these factors combine nicely to give the formula in the text.}
\bea
 \cV_{2n-1}( \lambda_a^\pm , b_i^\pm ) = \cV_{2n-1}(N, \Delta^\pm, b_1) \, ,
\eea
where $\cV_{2n-1}(N, \Delta^\pm, b_1)$ is the free energy defined in section \eqref{freegrav}, with $\Delta_a$ replaced by $\Delta_a^\pm$.%
\footnote{For three-dimensional theories, the presence of square roots introduces sign ambiguities in the relative sign between blocks. In particular, the entropy functions in the literature feature a characteristic relative sign $\sigma$ between the two blocks, distinguishing between twist and anti-twist configurations \cite{Boido:2022iye,Boido:2022mbe}. This ambiguity is a consequence of the analytic continuation and can be fixed through a careful examination of orientation. We refer the reader to \cite{Boido:2022mbe} for a detailed discussion.}
The dual variables satisfy
\bea
 b_i^\pm = \frac{b_1}{2} \sum_{a=1}^d v_i^a \Delta_a^\pm \, , \quad \text{ for } \quad i = 1 , \ldots, n \, ,
\eea
as follows from combining \eqref{conb}, with $(b_i,\lambda_a)$ replaced by $(b_i^\pm,\lambda_a^\pm)$, and using the supersymmetry condition \eqref{BPS1}.
 
By introducing the normalized variable $\epsilon \equiv \frac{2 b_0}{b_1}$, we can write the supersymmetric action as
\bea
\boxed{
 \label{rot:susyaction}
 \cS ( \Delta_a , \fn_a , \epsilon) = \frac{ 4 \pi}{ \epsilon} \left( \cV_{2 n - 1} (N, \Delta_a^{+} ,b_1) - \cV_{2 n - 1} (N, \Delta_a^{-} , b_1) \right)\, .}
\eea
The remaining supersymmetry conditions \eqref{BPS2} can be written as
\bea
 \label{diffR}
 \boxed{\Delta_a^+ - \Delta_a^- = - \epsilon\,  \fn_a \, ,}
\eea
from which we see that the Legendre transform indeed acts as a gravitational block \cite{Hosseini:2019iad} for the extremal function. Notice that the variables $\Delta_a^\pm$ satisfy
\bea
 \sum_{a = 1}^{d} \Delta_a^{\pm} = 2 \frac{b_1^\pm}{b_1}= 2 - \frac{1}{V_0^\pm} \epsilon \, , \qquad
 \Delta_a^{+} - \Delta_a^{-} = - \epsilon \fn_a \, .
\eea
Sometimes it is useful to introduce canonically normalized R-charges $\Delta_a$
\bea
 \label{diffR2}
 \Delta_a^\pm \equiv \Delta_a \mp \frac{\epsilon}{2} \left( \fn_a \pm \frac{r_a}{2} \chi_{-\sigma} \right) \, ,
\eea
where
\bea
 \sum_{a = 1}^{d} \Delta_a = \sum_{a = 1}^{2} r_a = 2 \, , \qquad \sum_{a = 1}^{d} \fn_a = \chi_\sigma \, .
\eea
Notice that the relation between $\Delta_a^\pm$ and $\Delta_a$  depends on the choice of a reference R-symmetry $r_a$ in the field theory and is therefore not canonical. It becomes unambiguous only for the topological twist on a two-sphere, where $n_\pm = \sigma = 1$ and $\chi_{- \sigma} = 0$.

The extremal function $\cS ( \Delta_a , \fn_a )$ must be extremized with respect to all the variables $\Delta_a$, including those corresponding to baryonic directions. At first glimpse, this might appear to be in tension with the approach of \cite{Couzens:2018wnk,Gauntlett:2018dpc}, where the supersymmetric action is expressed as a function of $b_i$ and $\fn_a$ using the supersymmetry conditions, and extremization is performed only over the components $b_i$ of the Reeb vector. It turns out that extremizing $\cS ( \Delta_a , \fn_a )$ with respect to the baryonic symmetries is automatically enforced by the supersymmetry conditions and is, in fact, equivalent to them. Indeed, we have
\bea
\boxed{
 \label{barextr}
 \sum_{a = 1}^{d} B_a^{(r)} \frac{\pd \cS ( \Delta_a , \fn_a, \epsilon )}{\pd \Delta_a}= 0 \, , \quad \text{ for } \quad r = 1, \ldots , d - n \, .}
\eea
This relation readily follows by differentiating \eqref{rot:susyaction} and using \eqref{bar}
\bea
 \label{proof:barextr}
 \sum_{a = 1}^{d} B_a^{(r)} \frac{\pd \cS ( \Delta_a , \fn_a , \epsilon) }{\pd \Delta_a}
 & = \frac{4\pi}{\epsilon} \sum_{a = 1}^{d} B_a^{(r)} \left( \frac{\pd \cV_{2 m - 1} (N,\Delta_a^{+} ,b_1)}{\pd \Delta_a} - \frac{\pd \cV_{2 m - 1} (N, \Delta_a^{-} , b_1) }{\pd \Delta_a}\right)
 \\ & = - \frac{2\pi \alpha_n N}{\epsilon} \sum_{a = 1}^{d} B_a^{(r)} (\lambda_a^+ -\lambda_a^-)
 = - \frac{2\pi \alpha_n N}{\epsilon} \sum_{a = 1}^{d} B_a^{(r)} (\lambda_a -\lambda_a) =0 \, ,
\eea
since $\sum_{a = 1}^{d} B_a^{(r)} =0$.

The case of the topological twist on a two-sphere, where $n_\pm = \sigma = 1$, deserves a few words. From \eqref{diffR2}, we see that
\bea
 \Delta_a^\pm =\Delta_a \mp \frac{\epsilon}{2} \fn_a \, ,
\eea
and, by sending $\epsilon$ to zero in \eqref{rot:susyaction}, we find 
\bea
 \label{rot:susyaction2}
 \boxed{\cS ( \Delta_a , \fn_a ) = - 4 \pi \sum_{a = 1}^d \fn_a \frac{\pd \cV_{2 n - 1} (N, \Delta,b_1)}{\pd\Delta_a} \, ,}
\eea
where $\sum_{a=1}^d \Delta_a = \sum_{a=1}^d \fn_a = 2$. The condition \eqref{barextr} in this case can also be written as
\bea
 0 =
 \sum_{a , b = 1}^{d} B_a^{(r)} \fn_b
 \frac{\pd^2 \cV_{2 n - 1} ( \Delta_a )}{\pd \Delta_a \pd \Delta_b}
 =
- \frac{\alpha_n N}{2}
 \sum_{b = 1}^{d} \fn_b
 \frac{\pd X^{(r)} ( \Delta_a )}{\pd \Delta_b}
 \, , \quad \text{ for } \quad r = 1, \ldots , d - n \, .
\eea
In the static case, we can find solutions where the sphere is replaced with a Riemann surface of genus $\fg$. An inspection of the supersymmetry conditions written in \cite{Couzens:2018wnk,Gauntlett:2018dpc} shows that the genus-$\fg$ case can be obtained from the sphere case by a simple rescaling of parameters. The supersymmetric action for solutions where $\text{SE}_{2n-1}$ is fibered over a Riemann surface of genus $\fg$ still takes the form \eqref{rot:susyaction2}, but now the fluxes satisfy
\bea
 \sum_{a = 1}^{d}  \fn_a = 2 - 2 \fg \, .
\eea
The form \eqref{rot:susyaction2} is the hallmark of $\cI$-extremization for SCFTs dual to magnetically charged black holes in AdS$_4$  \cite{Benini:2015eyy,Benini:2016rke,Hosseini:2016tor}.

\subsubsection{Comparing with the quantum field theory results}
\label{entropyext}

The functional $\cS ( \Delta_a , \fn_a , \epsilon)$ for CY$_3$ provides, up to a normalization, the gravitational prediction the large-$N$ central charge of the $(0 , 2)$ two-dimensional CFT obtained by compactifying the four-dimensional CFT on the spindle \cite{Gauntlett:2019roi}. The precise relation is
\bea
 c_r  ( \Delta_a , \fn_a , \epsilon ) = \frac{24}{b_1} (2\pi)^2 \cS ( \Delta_a , \fn_a , \epsilon ) \, .
\eea
Using \eqref{rot:susyaction} and \eqref{4da}, we find
\bea
 c_r  ( \Delta_a , \fn_a , \epsilon ) = \frac{32}{9\epsilon} \left ( a_{(4)} (\Delta_a^+) - a_{(4)} (\Delta_a^-) \right ) \, ,
\eea
and, in the static case $(\epsilon = 0)$,
\bea
 c_r  ( \Delta_a , \fn_a ) = - \frac{32}{9} \sum_{a=1}^d  \fn_a \frac{\pd a_{(4)} (\Delta_a) }{\pd \Delta_a} \, ,
\eea
which matches the field theory computation of the two-dimensional central charge \cite{Hosseini:2019use,Hosseini:2020vgl,Boido:2022mbe,Martelli:2023oqk}. These relations imply the equivalence of the gravitational approach of \cite{Couzens:2018wnk,Gauntlett:2018dpc} and the field theory $c$-extremization of \cite{Benini:2013cda,Benini:2012cz}. This equivalence was originally proven in \cite{Hosseini:2019use} in the static case, and later extended in \cite{Boido:2022mbe} to the spindle compactification. Here, we have substantially simplified the proof and provided a geometric interpretation.

For CY$_4$, the functional  $\cS ( \Delta_a , \fn_a , \epsilon )$ provides the gravitational prediction for the entropy of AdS$_4$ rotating and accelerating black holes with magnetic fluxes $\fn_a$ \cite{Gauntlett:2019roi}. More precisely, it gives the logarithm of the grand-canonical partition function:
\bea
 \log \cZ ( \Delta_a , \fn_a , \epsilon ) = \frac{2}{\sqrt{b_1}} (2\pi)^2 \cS ( \Delta_a , \fn_a , \epsilon) \, .
\eea
The Bekenstein-Hawking entropy of a black hole with angular momentum $J$ and electric charges $\fq_a$ is then obtained via a Legendre transform, which is equivalent to the extremization of the entropy function
\bea
 \label{ef}
 \cI ( \Delta_a , \fn_a , \epsilon ) = \log \cZ ( \Delta_a , \fn_a , \epsilon ) + 2 \pi \ii \biggl( \epsilon J + \sum_{a=1}^d \Delta_a \fq_a \biggr) \, .
\eea
In the class of examples considered in \cite{Gauntlett:2019roi}, all electric charges $\fq_a$ vanish, but we expect the result to extend to the case of electrically charged black holes as well. Using \eqref{rot:susyaction} and \eqref{free3d}, we find
\bea
 \label{id1}
 \log \cZ  ( \Delta_a , \fn_a , \epsilon) = \frac{1}{2\epsilon} \left ( F_{S^3} (\Delta_a^+) - \sigma F_{S^3} (\Delta_a^-) \right ) \, ,
\eea
where the relative sign to take into account the sign ambiguities of the antitwist case \cite{Boido:2022mbe}, and, in the static case, which corresponds to $\sigma=n_\pm=1$,
\bea
 \label{id2}
 \log \cZ ( \Delta_a , \fn_a ) = - \frac{1}{2} \sum_{a=1}^d \fn_a  \frac{\pd F_{S^3} (\Delta_a) }{\pd \Delta_a} \, .
\eea

The main difficulty in the comparison with field theory is again the fact that large-$N$ computations are available only for non-chiral quivers and are independent of baryonic charges. Indeed, it turns out that one can find a large-$N$ saddle point for the topologically twisted index or the spindle index only for those quivers and R-charges for which there exists a saddle point of $S^3$ free energy $F_{S^3} (\Delta_a)$ \cite{Hosseini:2016tor,Hosseini:2022vho,Colombo:2024mts}. When field theory results are available, they confirm the identities \eqref{id1} and \eqref{id2}.
On the gravitational side, the restriction to mesonic R-charges is imposed by enforcing the mesonic twist, which amounts to requiring that the baryonic moduli vanish:
\bea
 X^{(r)} = \sum_{a=1}^d B^{(r)} \lambda_a =0 \, , \quad \text{ for } \quad r = 1 , \ldots , d - 4 \, .
\eea
Given \eqref{bar}, this is equivalent to demanding that the $S^3$ free energy is independent of the baryonic R-charges:
\bea
 \sum_{a = 1}^{d} B_a^{(r)} \frac{\pd F_{S^3} (N , \Delta  , b_1 )}{\pd \Delta_a} =0 \, .
\eea
The equivalence between the gravitational approach of \cite{Couzens:2018wnk,Gauntlett:2018dpc} and $\cI$-extremization, \emph{restricted to the mesonic twist}, has been checked in many examples in the static case \cite{Hosseini:2019ddy,Gauntlett:2019roi}. As discussed in section \ref{freeext}, the vanishing of $X^{(r)}$ implies the parameterization \eqref{BZ} of the R-charges, matching the results found in \cite{Hosseini:2019ddy}.
It should not come as a surprise that the mesonic twist also implies a constraint on the fluxes $\fn_a$, which are restricted to span mesonic directions only. However, this result is clearly incomplete. It is well known that there exist black holes with baryonic charges and nonzero entropy, in tension with the current field theory predictions. We have already speculated that the $S^3$ free energy should be extended to generalized form \eqref{free3d}. Here, we have shown that the gravitational approach, valid for arbitrary CY$_4$ and choices of magnetic fluxes, can always be recast in the form of an $\cI$-extremization, where the functional $\cI$ has the characteristic gravitational block form \eqref{id1}, based on the generalized free energy \eqref{free3d}. Once again, we see these results as predictions for the large-$N$ limit of many supersymmetric partition functions that remain to be found.

Finally, we note that the approach in \cite{Couzens:2018wnk,Gauntlett:2018dpc} is limited to a particular class of solutions. Nevertheless, based on the evidence provided by these examples, we are confident that the generalized free energy \eqref{free3d} is the basic building block for constructing entropy functions for \emph{all} asymptotically AdS$_4\times\text{SE}_7$ black holes, in the spirit of the gravitational block philosophy \cite{Hosseini:2019iad}.

\section{Examples}\label{sec:examples}

In this section, we compute the master volume and its Legendre transform---the generalized free energy \eqref{free3d}---for various examples of M2-branes at conical singularities, focusing on the toric case and considering the same set of examples discussed in \cite{Hosseini:2019ddy}. So far, we have not been able to find a general closed-form expression for the free energy in terms of the toric data, and therefore we proceed case-by-case. This is in contrast to the case of D3-branes, where the simple cubic expression \eqref{cubic} applies. Our results generalize the quartic expressions found in \cite{Amariti:2011uw, Amariti:2012tj}, which are valid only for the mesonic twist.

For all the examples considered, corresponding black hole solutions can also be studied. Their entropy can be computed using $\cI$-extremization, with the entropy function \eqref{ef} obtained by gluing together the generalized free energy as in \eqref{id1} and \eqref{id2}. A word of caution is in order here. Whenever the CY$_4$ fan has facets with worse-than-orbifold singularities---which translate into corresponding singularities on the link $\text{SE}_7$---the master volume depends on the choice of resolution. Many of the dual pairs studied in the literature \cite{Hanany:2008cd,Hanany:2008fj,Martelli:2008si,Benini:2009qs,Jafferis:2009th} fall into this category. To demonstrate our results using simple examples, we have chosen to consider these cases as well. It is an interesting question whether different families of black holes are associated with different resolutions, as suggested by our construction.
For the mesonic twist, the free energy reduces to the Sasakian volume (see \eqref{aFv}), which is resolution-independent.%
\footnote{In particular, the results in \cite{Hosseini:2019ddy,Gauntlett:2019roi} are not affected by this ambiguity.}
The ambiguity arises only for black holes carrying baryonic charges, where the resolution could affect the structure of the five-cycles. Of course no one has yet studied baryonic black holes in such complicated and singular geometries.

For all the examples, we refer to \cite{Hosseini:2019ddy} for a discussion of the matter content of the dual field theory and the assignment of R-charges to the fields in the quiver.

\label{sec:examples}

\subsection{The $\cC \times \bC$ geometry}
\label{ssec:conifold}
\begin{figure}[H]
\centering
\begin{tikzpicture}[scale=0.45]
  \draw[draw=none,fill=blue!20,opacity=0.6] (3,0) -- (1,-2) -- (0,2.5) -- cycle;
  \draw[draw=none,fill=blue!30,opacity=0.6] (0,0) -- (3,0) -- (0,2.5) -- cycle;
  \draw[draw=none,fill=blue!30,opacity=0.6] (0,0) -- (-2,-2) -- (0,2.5) -- cycle;
  \draw[draw=none,fill=blue!20,opacity=0.6] (-2,-2) -- (1,-2) -- (0,2.5) -- cycle;
  \draw[draw=none,fill=blue!30,opacity=0.6] (-2,-2) -- (3,0) -- (0,2.5) -- cycle;
  
  \draw[-,dashed] (0,0) -- (3,0) node[below] {};
  \draw[->,solid] (3,0) -- (4.5,0) node[below] {};
  
  \draw[-,dashed] (0,0) -- (0,2.5) node[below] {};
  \draw[->,solid] (0,2.5) -- (0,4.5) node[below] {};
  
  \draw[-,dashed] (0,0) -- (-2,-2) node[below] {};
  \draw[->,solid] (-2,-2) -- (-3.2,-3.2) node[below] {};
  
  \draw (-2.7,-3.6) node {$e_2$};
  \draw (4.5,0.5) node {$e_3$};
  \draw (0.6,4.5) node {$e_4$};
  
  \draw[-,solid] (3,0) -- (1,-2);
  \draw[-,solid] (1,-2) -- (-2,-2);
  \draw[-,solid] (-2,-2) -- (0,2.5);
  \draw[-,solid] (0,2.5) -- (1,-2);
  \draw[-,solid] (0,2.5) -- (3,0);

  \draw (0.2,-0.5) node {$v^1$};
  \draw (-2.3,-1.6) node {$v^2$};
  \draw (3.3,.65) node {$v^3$};
  \draw (0.5,2.9) node {$v^4$};
  \draw (1.3,-2.3) node {$v^5$};
\end{tikzpicture}
\caption{The toric diagram for $\cC \times \bC$.\label{toric:conifold}}
\end{figure}
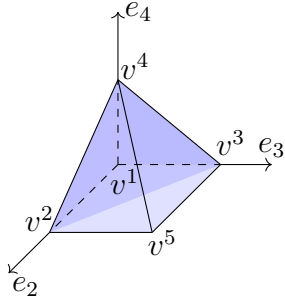

The geometry is specified by the vectors $v^a=(1,\vec{v}^{\, a})$ with
\bea
 \vec{v}^{\,1} = ( 0 , 0, 0 ) \, , \qquad \vec{v}^{\,2} = ( 1 , 0 , 0 ) \, , \qquad \vec{v}^{\,3} = ( 0, 1, 0 ) \, , \qquad \vec{v}^{\,4} = ( 0, 0, 1) \, , \qquad \vec{v}^{\,5} = ( 1, 1, 0) \, ,
\eea
and is the product of a conifold singularity $\cC$ with a copy of $\bC$.
The toric diagram is the polytope defined by the $\vec{v}^{\, a}$ and is given in figure \ref{toric:conifold}. The dual SCFT, identified in \cite{Benini:2009qs}, is a flavored ABJM quiver with Chern-Simons levels set to zero.  Note that the facet $(1253)$ is a pyramid in $\bR^4$---this introduces a singularity on the facet---which we resolve by blowing up the surface. This resolution can be done in two distinct ways, by adding an extra line that stretches from:
\begin{enumerate*}[label=(\roman*)]
    \item $ v^2 \to v^3$, or
    \item $v^1 \to v^5$.
\end{enumerate*}
In the following, we consider the first resolution. The result for the second resolution is obtained by applying the cyclic permutation $3 \to 1 \to 2 \to 5$. While the results  for the mesonic twist \cite{Hosseini:2019ddy,Gauntlett:2019roi} remains invariant under this permutation, the general resolved master volume $\cV ( \lambda_a , b_i )$ is \emph{not}, and the permutation maps one resolution onto the other.
\noindent There is a single baryonic symmetry given by $B = ( 1 , - 1 , - 1 , 0 , 1 )$. We compute the master volume using the localization formula \eqref{fixedpoint}. We can triangulate the toric polytope \ref{toric:conifold} by considering the tetrahedra $(1243)$ and $(2354)$. It is convenient to write the master volume in terms of the quantities $\phi_A$ that are invariant under the gauge transformations \eqref{gauge}.%
\footnote{Notice that the $\phi_A$ are not necessarily independent variables, since there are only $d - 3$ independent gauge invariant quantities, while the number of fixed points can be larger. In this example, however, there are only two fixed points, and $\phi_1$ and $\phi_2$ are independent.}
The gauge invariants $\phi_1$ and $\phi_2$ corresponding to the two tetrahedra are, respectively:
\bea
 \phi_1 ( \lambda_a , b_i ) & \equiv
 ( b_1 - b_2 - b_3 - b_4 ) \lambda_1 + b_2 \lambda_2 + b_3 \lambda_3 + b_4 \lambda_4 \, , \\
 \phi_2 ( \lambda_a , b_i ) & \equiv
 ( b_1 - b_3 - b_4 ) \lambda_2 + ( b_1 - b_2 - b_4 ) \lambda_3 + b_4 \lambda_4 - ( b_1 - b_2 - b_3 - b_4) \lambda_5 \, .
\eea
We then calculate the master volume \eqref{master},
\bea
 \label{master:conifold}
 \cV ( \lambda_a , b_i ) & = - \frac{8 \pi^4}{3 ( b_1 - b_2 - b_3 - b_4 ) b_4}
 \left( \frac{\phi_1^3 ( \lambda_a , b_i )}{b_2 b_3} - \frac{\phi_2^3 ( \lambda_a , b_i )}{( b_1 - b_2 - b_4 ) ( b_1 - b_3 - b_4 )} \right) \, ,
\eea
as a function of the Reeb vector $b$ and the K\"ahler parameters $\lambda_a$. This was computed in  \cite[(5.18)]{Hosseini:2019ddy} with a different method involving the dual polytope. We also note that
\bea
 \cV_{\text{res.\,(\romannum{1})}} ( \lambda_a , b_i ) - \cV_{\text{res.\,(\romannum{2})}} ( \lambda_a , b_i )
 = - \frac{8 \pi^4}{3} \frac{X^3}{b_4} \, ,
\eea
where $X=\sum_{a=1}^5 B_a \lambda_a$ is the baryonic K\"ahler modulus. Thus, for the mesonic twist $X = 0$, we find $\cV_{\text{res.\,(\romannum{1})}} ( \lambda_a , b_i ) \bigl|_{X = 0} = \cV_{\text{res.\,(\romannum{2})}} ( \lambda_a , b_i ) \bigl|_{X = 0}$.
The R-charges \eqref{def:R-charge} corresponding to each vertex are
\bea
 \label{R-charge:conifold}
 \Delta_1 & = \frac{16 \pi^4}{N} \frac{\phi_1^2 ( \lambda_a , b_i )}{b_2 b_3 b_4} \, ,
 \qquad \Delta_2 = \frac{16 \pi^4}{N b_4 ( b_1 - b_2 - b_3  - b_4 )} \biggl( \frac{\phi_1^2 ( \lambda_a , b_i )}{b_3}
 - \frac{\phi_2^2 ( \lambda_a , b_i )}{( b_1 - b_2 - b_4 )} \biggr) \, , \\
 \Delta_3 & = \frac{16 \pi^4}{N b_4 ( b_1 - b_2 - b_3 - b_4 )} \biggl( \frac{\phi_1^2 ( \lambda_a , b_i )}{b_2}
 - \frac{\phi_2^2 ( \lambda_a , b_i )}{b_1 - b_3 - b_4} \biggr) \, , \\
 \Delta_4 & = \frac{16 \pi^4}{N ( b_1 - b_2 - b_3 - b_4 )} \biggl( \frac{\phi_1^2 ( \lambda_a , b_i )}{b_2 b_3}
 - \frac{\phi_2^2 ( \lambda_a , b_i )}{(b_1 - b_2 - b_4) (b_1 - b_3 - b_4)} \biggr) \, , \\
 \Delta_5 & = \frac{16 \pi^4}{N (b_1 - b_2 - b_4 ) ( b_1 - b_3 - b_4 ) b_4} \phi_2^2 ( \lambda_a , b_i ) \, ,
\eea
and satisfy, see \eqref{reeb:id},
\bea
 \label{b:Delta:conifold}
 2 = \sum_{a = 1}^{5} \Delta_a \, , \qquad \frac{2 b_2}{b_1} = \Delta_2 + \Delta_5 \, , \qquad \frac{2 b_3}{b_1} = \Delta_3 + \Delta_5 \, , \qquad \frac{2 b_4}{b_1} = \Delta_4 \, .
\eea
The previous relations can be inverted to express the $\phi_l$ in terms of $(\Delta_a, N, b_1)$:
\bea
 \label{phi:conifold}
 \phi_1^2 & = \frac{b_1^3 N}{128 \pi^4} ( \Delta_2 + \Delta_5 ) ( \Delta_3 + \Delta_5 ) \Delta_1 \Delta_4 \, , \\
 \phi_2^2 & = \frac{b_1^3 N}{128 \pi^4} ( \Delta_1 + \Delta_2 ) ( \Delta_1 + \Delta_3 ) \Delta_4 \Delta_5 \, . 
\eea
By eliminating $\lambda_a$ and $b_i$ using \eqref{b:Delta:conifold} and \eqref{phi:conifold}, the master volume \eqref{master:conifold} takes the form
\bea
 \label{V:on-shell:Delta:conifold}
 \cV ( \Delta_a ) & = \pm \frac{N^{3/2} \sqrt{b_1}}{24 \sqrt{2} \pi^2}
 \frac{\Delta_1 \sqrt{( \Delta_2 + \Delta_5 ) ( \Delta_3 + \Delta_5 ) \Delta_1 \Delta_4}
 \pm \Delta_5 \sqrt{( \Delta_1 + \Delta_2 ) ( \Delta_1 + \Delta_3 ) \Delta_4 \Delta_5}}
 {\Delta_1 - \Delta_5}
 \, .
\eea
Note that, the signs in the two square-root terms are not correlated, reflecting the ambiguities in taking square roots. These ambiguities are sometimes fixed by physical arguments.  Moreover,
\bea
 \label{V:volS:on-shell:Delta:conifold}
 \cV ( \Delta_a ) & = \pm \frac{N^{3/2}}{12 b_1 \sqrt{\pi} ( \Delta_1 - \Delta_5 )}
 \left( \Delta_1 \sqrt{ \frac{\Delta_1}{\Vol_{S_1} ( \Delta_a )}} \pm \Delta_5 \sqrt{\frac{\Delta_5}{\Vol_{S_5} ( \Delta_a )}} \right) ,
\eea
where $S_a$ is the torus-invariant five-cycle associated with $v^a$. The geometric interpretation of this formula arises from the fact that the toric diagram can be viewed as a suspension over a triangle in the $(234)$ plane, as explained in appendix \ref{app:fp}. The baryonic K\"ahler modulus is given by
\bea
 \label{X:conifold}
 X 
 = \frac{2}{b_1} \frac{\phi_1 - \phi_2}{\Delta_1 - \Delta_5} \, .
\eea

By introducing the quartic function \cite{Amariti:2011uw}
\bea
 \label{a3d:conifold}
 a_{(3)} ( \Delta_a ) & \equiv \frac{1}{24} \sum_{a , b , c , e= 1}^{5} | ( v^a , v^b , v^c , v^e ) | \Delta_a \Delta_b \Delta_c \Delta_e \\
 & = \Delta_4 ( \Delta_1 \Delta_2 \Delta_3 + \Delta_1 \Delta_5 \Delta_3 + \Delta_2 \Delta_5 \Delta_3 + \Delta_1 \Delta_2 \Delta_5 ) \, ,
\eea
\eqref{V:on-shell:Delta:conifold} can also be rewritten as
\bea
 \cV ( \Delta_a ) = \pm \frac{N^{3/2} \sqrt{b_1}}{24 \sqrt{2} \pi^2}
 \sqrt{a_{(3)} ( \Delta_a ) + \frac{2 (2 \pi )^4}{b_1 N} \Delta_1 \Delta_5 X^2} \, .
\eea
For the mesonic twist, $X = 0$, and thus
\bea
 \cV_{\text{mes.}} ( \Delta_a ) = \pm \frac{N^{3/2} \sqrt{b_1}}{24 \sqrt{2} \pi^2} \sqrt{a_{(3)} ( \Delta_a )} \, ,
\eea
which is consistent with \cite[(5.25)]{Hosseini:2019ddy}, where by supersymmetry $b_1=1$. 

The supersymmetric action for compactifications with fluxes $\fn_a$ is then given by \eqref{rot:susyaction}, and it is automatically extremized with respect to the baryonic directions \eqref{barextr}. This action can be used to construct the entropy function \eqref{ef} for generic black holes with the same asymptotic geometry.

\subsubsection{A purely baryonic twist}

Magnetically charged static black holes with horizons given by Riemann surfaces of genus $\fg$ and a mesonic twist were studied in \cite{Hosseini:2019ddy}. Here, we focus on the case of a purely baryonic twist, which is obtained by setting the Reeb vector to its Sasaki-Einstein value,
\bea
 b_2 = b_3 = \frac{3}{2} b_4 = \frac{3}{8} b_1 \, ,
\eea
corresponding to a consistent truncation of the problem by symmetry.%
\footnote{In general, even without symmetries, we can consistently restrict the Reeb vector to its Sasaki-Einstein value. Indeed, we already proved that the Legendre transform of the master volume is extremized precisely at the Sasaki-Einstein value of $b$.}
Therefore, from \eqref{b:Delta:conifold}, we impose the following conditions:
\begin{equation}
 \label{frozen:conifold}
 \begin{alignedat}{3}
  \Delta_1 - \varsigma \big|_{\varsigma = 0}  & = \Delta_5 + \varsigma \big|_{\varsigma = 0} \equiv \delta_1 \, ,\qquad & \Delta_2 & = \Delta_3 \equiv \delta_2 \, ,\qquad & \Delta_4 & = \frac23 ( \delta_1 + \delta_2 ) \, ,\\[0.5em]
  \fn_1 &= \fn_5 \equiv \fp_1 \, , \qquad & \fn_2 & = \fn_3 \equiv \fp_2 \, , \qquad & \fn_4 & = \frac{2}{3} ( \fp_1 + \fp_2 ) \, ,
 \end{alignedat}
\end{equation}
where we introduce the parameter $\varsigma$, which will be taken to zero at the end of the computation. The R-charge constraint $\sum_{a = 1}^{5} \Delta_a = 2$ and the twisting condition $\sum_{a = 1}^{5} \fn_a = 2 - 2 \fg$ reduce to
\bea
 \delta_1 + \delta_2 = \frac34 \, , \qquad \fp_1 + \fp_2 = \frac34 ( 1 - \fg ) \, .
\eea
Observe that we must require $1 - \fg \in 4 \bZ$ in order to satisfy the quantization condition $\fn_a \in \bZ$. Taking the limit $\varsigma \to 0$ in \eqref{R-charge:conifold}, \eqref{b:Delta:conifold}, \eqref{X:conifold}, and \eqref{master:conifold}, we find that
\bea
 \phi_l & =  \pm \frac{b_1^{3/2} \sqrt{N}}{8 \sqrt{3} \pi^2} ( \delta_1 + \delta_2 )^{3/2} \sqrt{\delta_1} \, , \quad \text{ for } \quad l = 1, 2 \, , \\
  X & = \mp \frac{\sqrt{b_1 N}}{8 \sqrt{3} \pi^2} ( \delta_1 - \delta _2 ) \sqrt{\frac{\delta_1 + \delta_2}{\delta_1}}
  = - \frac{4}{N} \sum_{a = 1}^{2} B_a \frac{\pd \cV_B ( \delta_a )}{\pd \delta_a} \, ,
\eea
where the restriction of the master volume is given by
\bea
 \label{VB:conifold}
 \cV_{B} ( \delta_a ) = \mp \frac{\sqrt{b_1} N^{3/2}}{3 \sqrt{3} (4 \pi)^2}
 ( \delta_1 + 3 \delta_2 ) \sqrt{\delta_1 ( \delta_1 + \delta_2 )} \, .
\eea
Alternatively, one may directly take the limit $\varsigma \to 0 $ in \eqref{V:on-shell:Delta:conifold}, leading to the same result.

The master volume $\cV_B$ predicts the large-$N$ free energy on $S^3$ for purely baryonic R-charges, as given by \eqref{free3d}:
\bea 
\label{VB:conifold2}
 F_{S^3} ( \delta_a ) = \mp \frac{4 \pi N^{3/2}}{3 \sqrt{3}}
 ( \delta_1 + 3 \delta_2 ) \sqrt{\delta_1 ( \delta_1 + \delta_2 )} \, .
\eea
This expression is extremized at $\delta_1 = \delta_2 = \frac{3}{8}$, which corresponds to $\Delta_{1,2,3,5}=\frac{3}{8}$ and $\Delta_{4}=\frac12$. Using the dictionary in \cite[(5.8)]{Hosseini:2019ddy}, we can predict the R-charges of the fields and monopoles in the quiver
\bea 
 \Delta_{A_1}=\frac78 \, ,\qquad \Delta_{A_2}=\Delta_{B_1}=\Delta_{B_2} =\frac38 \, , \qquad \Delta_{m}=-\frac{1}{16} \, ,
\eea
in agreement with \cite[(6.13)]{Jafferis:2011zi} and the dimensions of baryonic operators in the theory.%
\footnote{This should not come as a total surprise, since we already proved that the R-charges take the form \eqref{def:R-charge}. Since the extremal values of $b$ correspond to the Sasaki-Einstein one, the exact R-charges are proportional to the volumes of the five-cycles. This also guarantees that such extremal values are independent of the resolution.}
Notice that this result was derived in \cite{Jafferis:2011zi} by a somehow indirect argument, since the large-$N$ free energy computed in that paper is independent of baryonic charges, and its extremization can predict only the dimensions of mesonic operators. Here, in contrast, it follows directly from a straightforward extremization of our generalized free energy. 

The entropy function for purely magnetic static black holes is obtained by taking the limit $\varsigma \to 0$ in \eqref{id2}:
\bea
 \label{SB:conifold}
 \cI_B ( \delta_a , \fp_a ) = \log \cZ_B ( \delta_a , \fp_a ) = \pm \frac{\pi \sqrt{b_1} N^{3/2}}{3 \sqrt{3}}
 \frac{(4 \delta_1^2 + 9 \delta_1 \delta_2 + 3 \delta_2^2 ) \fp_1 + \delta_1 ( 7 \delta_1 + 9 \delta_2 ) \fp_2}{\sqrt{\delta_1 ( \delta_1 + \delta_2 )}} \, ,
\eea
and can be used to predict their entropy. After imposing $\delta_2 = \frac34 - \delta_1$, the function \eqref{SB:conifold} has a critical point at
\bea
 \mathring{\delta}_1 ( \fp_a ) & = \frac{3}{16}  \frac{\fp_1 + 3 \fp_2 \pm \sqrt{\Theta ( \fp_a ) }}{\fp_1 + \fp_2} \, ,
\eea
where we define $\Theta ( \fp_a ) \equiv( \fp_2 - \fp_1 ) ( 7 \fp_1 + 9 \fp_2 )$.
At this critical value, the entropy function \eqref{SB:conifold} evaluates to the Bekenstein-Hawking entropy
\bea
 S_{\text{BH}} ( \fp_a ) \equiv \cI_B ( \delta_a , \fp_a ) \bigl|_{\text{crit.}} = - \frac{\pi \sqrt{b_1} N^{3/2}}{4 \sqrt{3}}
 \bigl( 2 \fp_1 + 6 \fp_2 \mp \sqrt{\Theta ( \fp_a )} \bigr)
 \sqrt{\frac{\fp_1 + 3 \fp_2 \pm \sqrt{ \Theta ( \fp_a )}}{\fp_1 + \fp_2 }} \, .
\eea

For dyonic static black holes with electric charges $\fq_a$, we extremize%
\footnote{We included a factor of $\sqrt{b_1}$ in the Legendre term for convenience, since $\log \cZ ( \delta_a , \fp_a )$ is proportional to $\sqrt{b_1}$. This factor must be set to one to match the Bekenstein-Hawking entropy.}
\bea
 \label{conifold:dyonic}
 \cI_B ( \delta_a , \fp_a ) \equiv \log \cZ_B ( \delta_a , \fp_a ) + 2 \pi \ii \sqrt{b_1} \sum_{a = 1}^{2} \delta_a \fq_a \, .
\eea

Finally, for rotating black holes with a spindle horizon and either a twist or anti-twist of the R-symmetry connection, parameterized by $\sigma = \pm 1$, we employ \eqref{ef} together with \eqref{id1}, restricted to baryonic charges and fluxes. Explicitly, we extremize
\be
 \label{conifold:rotating}
 \begin{aligned}
  \cI_B ( \delta_a , \fp_a , \epsilon) & \equiv \frac{1}{2 \epsilon} \left( F_{S^3} ( \delta_a^+ ) - \sigma F_{S^3} ( \delta_a^- ) \right)
  + 2 \pi \ii \sqrt{b_1} \biggl( \epsilon J + \sum_{a = 1}^{2} \delta_a \fq_a \biggr) \, ,
 \end{aligned}
\ee
subject to the constraint $\delta_1 + \delta_2 = \frac{3}{4}$, where $F_{S^3} ( \delta_a )$ is given by \eqref{VB:conifold2}, and
\bea
 \delta_a^\pm \equiv \delta_a \mp \frac{\epsilon}{2} \left( \fn_a \pm \frac{r_a}{2} \chi_{- \sigma} \right) \, ,
 \quad \text{ for } \quad a = 1, 2 \, .
\eea
Notice that
\bea
 \sum_{a = 1}^{2} \delta_a^+ = \frac38 \left( 2 - \frac{\epsilon}{n_+} \right) \, , \qquad
 \sum_{a = 1}^{2} \delta_a^- =  \frac38 \left( 2 + \frac{\sigma \epsilon}{n_-} \right) \, , \qquad
 \sum_{a = 1}^{2} \fn_a = \frac{3 \chi_\sigma}{8} \, .
\eea
The resulting expression for the entropy as a function of the charges is lengthy and offers little additional insight, so we do not report it here.

\subsection{Flavoring the $\bC^3$ quiver}
\label{ssec:FC3}

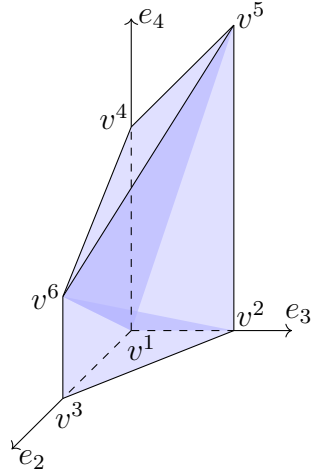
\begin{figure}[H]
\centering
\begin{tikzpicture}[scale=0.45]

  \draw[draw=none,fill=blue!20,opacity=0.6] (0,0) -- (3,0) -- (3,9) -- (0,6) -- cycle;
  \draw[draw=none,fill=blue!20,opacity=0.6] (0,0) -- (3,0)-- (-2,-2) -- cycle;
  \draw[draw=none,fill=blue!20,opacity=0.6] (0,0) -- (-2,-2) -- (-2,1) -- (0,6) -- cycle;
  \draw[draw=none,fill=blue!30,opacity=0.6] (0,0) -- (3,0)-- (-2,1) -- cycle;
  \draw[draw=none,fill=blue!30,opacity=0.6] (0,0) -- (3,9)-- (-2,1) -- cycle;
  
  \draw[-,dashed] (0,0) -- (3,0) node[below] {};
  \draw[->,solid] (3,0) -- (4.7,0) node[below] {};
  
  \draw[-,dashed] (0,0) -- (0,6) node[below] {};
  \draw[->,solid] (0,6) -- (0,9.2) node[below] {};
  
  \draw[-,dashed] (0,0) -- (-2,-2) node[below] {};
  \draw[->,solid] (-2,-2) -- (-3.5,-3.5) node[below] {};
  
  \draw (-2.9,-3.8) node {$e_2$};
  \draw (4.9,0.5) node {$e_3$};
  \draw (0.6,9.2) node {$e_4$};
  
  \draw[-,solid] (3,0) -- (-2,-2);
  \draw[-,solid] (-2,-2) -- (-2,1);
  \draw[-,solid] (3,0) -- (3,9);
  \draw[-,solid] (3,9) -- (-2,1);
  \draw[-,solid] (3,9) -- (0,6);
  \draw[-,solid] (0,6) -- (-2,1);

  \draw (0.3,-0.5) node {$v^1$};
  \draw (3.5,0.5) node {$v^2$};
  \draw (-1.8,-2.5) node {$v^3$};
  \draw (-0.5,6.3) node {$v^4$};
  \draw (3.5,9.3) node {$v^5$};
  \draw (-2.5,1) node {$v^6$};

\end{tikzpicture}
\caption{The toric diagram corresponding to an arbitrary flavoring of the $\bC^3$ quiver.\label{toric:N=8:SYM}}
\end{figure}

The geometry is specified by the vectors
\bea
 \vec{v}^{\,1} & = ( 0 , 0, 0 ) \, , ~~~ && \vec{v}^{\,2} = ( 0, 1, 0 ) \, , &&& \vec{v}^{\,3} = ( 1, 0, 0 ) \, , \\
 \vec{v}^{\,4} & = ( 0, 0, r_1) \, , && \vec{v}^{\,5} = ( 0, 1 , r_2) \, , &&& \vec{v}^{\,6} = ( 1, 0, r_3 ) \, ,
\eea
with the corresponding toric diagram shown in figure \ref{toric:N=8:SYM}.
\noindent There are two baryonic symmetries:
\bea
 B^{(1)} = \Bigl( \frac{r_3}{r_1} , 0 , -1 , - \frac{r_3}{r_1} , 0 , 1 \Bigr) \, , \qquad B^{(2)} = \Bigl( \frac{r_2}{r_1} , -1 , 0 , - \frac{r_2}{r_1} , 1 , 0 \Bigr) \, .
\eea
The corresponding dual SCFT, as identified in \cite{Benini:2009qs}, is a flavored version of $\cN = 8$ super Yang-Mills theory. The flavor symmetry group is given by $\U ( r_1 ) \times \U ( r_2 ) \times \U ( r_3 ) / \U (1)$. The presence of the square faces $(1254)$, $(3146)$, and $(3256)$ indicates that the geometry has worse-than-orbifold singularities, which we resolve by blowing up the surfaces. In the following discussion, we focus on the resolution achieved by introducing additional edges connecting the vertices $v^1 \to v^5$, $v^1 \to v^6$, and $v^2 \to v^6$. The toric polytope \ref{toric:N=8:SYM} can be triangulated using the tetrahedra $(1236)$, $(1265)$, and $(1564)$. The gauge invariants $\phi_l$, for $l=1,2,3$, corresponding to these tetrahedra, respectively, are given by
\bea
 \phi_1 ( \lambda_a , b_i ) & \equiv
 ( b_1 - b_2 - b_3 ) r_3 \lambda_1 + b_3 r_2 \lambda_2 + ( b_2 r_3 - b_4 ) \lambda_3 +  b_4 \lambda_6 \, , \\
 \phi_2 ( \lambda_a , b_i ) & \equiv
 ( b_1 - b_2 - b_3 ) r_2 \lambda_1 + ( b_3 r_2 + b_2 r_3 - b_4 ) \lambda_2 + ( b_4 - b_2 r_3 ) \lambda_5 + b_2 r_2 \lambda_6 \, , \\
 \phi_3 ( \lambda_a , b_i ) & \equiv
 \left( ( b_1 - b_2 - b_3 ) r_1 + b_3 r_2 + b_2 r_3 - b_4 \right) \lambda_1 - ( b_3 r_2 + b_2 r_3 - b_4 ) \lambda_4 + b_3 r_1 \lambda_5 + b_2 r_1 \lambda_6 \, .
\eea
Using \eqref{master}, we then calculate the master volume,
\bea
 \label{master:FC3}
 \cV ( \lambda_a , b_i ) & = - \frac{8 \pi^4}{3} \biggl[
 \frac{1}{( b_1 - b_2 - b_3 ) ( b_2 r_3 - b_4 )} \biggl( \frac{\phi_1^3 ( \lambda_a , b_i )}{r_3^2 b_3 b_4 }
 - \frac{\phi_2^3 ( \lambda_a , b_i )}{r_2^2 b_2 ( b_3 r_2 + b_2 r_3 - b_4 )} \biggr) \\
 & - \frac{\phi_3^3 ( \lambda_a , b_i )}{r_1^2 b_2 b_3 \left( ( b_1 - b_2 - b_3 ) r_1 + b_3 r_2 + b_2 r_3 - b_4 \right) ( b_3 r_2 + b_2 r_3 - b_4 )}
 \biggr] \, ,
\eea
as a function of the Reeb vector $b$ and the K\"ahler parameters $\lambda_a$. The master volume for the resolved space, achieved by introducing additional edges connecting the vertices $v^3 \to v^4$, $v^1 \to v^5$, and $v^2 \to v^6$, was obtained in \cite[(B.2)]{Hosseini:2019ddy} by computing the volume of the dual polytope. Observe that,
\bea
 \cV_{\text{here}} ( \lambda_a , b_i ) - \cV_{\text{there}} ( \lambda_a , b_i ) = \frac{8 \pi^4}{3} \frac{r_1 \bigl( X^{(1)} \bigr)^3}{b_3 r_3^2} \, ,
\eea
where $X^{(1)} = \sum_{a = 1}^{6} B_a^{(1)} \lambda_a$ is one of the baryonic K\"ahler moduli.
Thus, for the mesonic twist $X^{(1)} = 0$, we find $\cV_{\text{here}} ( \lambda_a , b_i ) \bigl|_{X^{(1)} = 0} = \cV_{\text{there}} ( \lambda_a , b_i ) \bigl|_{X^{(1)} = 0}$. The R-charges \eqref{def:R-charge} corresponding to each vertex are
\bea
 \label{R-charge:FC3}
 \Delta_1 & = \frac{16 \pi^4}{N} \left[
 \frac{\phi_1^2 ( \lambda_a , b_i )}{b_3 b_4 r_3 ( b_2 r_3 -  b_4 )}
 - \frac{1}{b_2 ( b_3 r_2 + b_2 r_3 - b_4 )}
 \left(
 \frac{\phi_3^2 ( \lambda_a , b_i )}{b_3 r_1^2}
 + \frac{\phi_2^2 ( \lambda_a , b_i )}{r_2 ( b_2 r_3 - b_4 )}
 \right)
 \right] \, , \\
 \Delta_2 & = \frac{16 \pi^4}{N ( b_1 - b_2 - b_3 ) ( b_2 r_3 - b_4 )}
 \left( \frac{\phi_1^2 ( \lambda_a , b_i )}{b_4 r_3}
 - \frac{\phi_2^2 ( \lambda_a , b_i )}{b_2 r_2^2} \right) \, , \\
 \Delta_3 & = \frac{16 \pi^4}{N b_3 b_4 r_3^2 ( b_1 - b_2 - b_3 )} \phi_1^2 ( \lambda_a , b_i )\, , \\
 \Delta_4 & = \frac{16 \pi^4}{N b_2 b_3 r_1^2 \left( ( b_1 - b_2 - b_3 ) r_1 + b_3 r_2 + b_2 r_3 - b_4 \right)} \phi_3^2 ( \lambda_a , b_i ) \, , \\
 \Delta_5 & = \frac{16 \pi^4}{N b_2 ( b_3 r_2 + b_2 r_3 - b_4 )} \left( \frac{\phi_2^2 ( \lambda_a , b_i ) }{r_2^2 ( b_1 - b_2 - b_3 )} - \frac{\phi_3^2 ( \lambda_a , b_i ) }{\left( ( b_1 - b_2 - b_3 ) r_1 + b_3 r_2 + b_2 r_3 - b_4 \right) r_1} \right) \, , \\
 \Delta_6 & = \frac{16 \pi^4}{N} \biggl[ \frac{1}{(b_1 - b_2 - b_3 ) ( b_2 r_3 - b_4 ) }
 \left(
 \frac{\phi_1^2 ( \lambda_a , b_i )}{b_3 r_3^2}
 - \frac{\phi_2^2 ( \lambda_a , b_i )}{r_2 ( b_3 r_2 + b_2 r_3 - b_4 )}
 \right) \\
 & - \frac{\phi_3^2 ( \lambda_a , b_i )}{b_3 r_1 ( b_3 r_2 + b_2 r_3 - b_4 ) \left( ( b_1 - b_2 - b_3 ) r_1 + b_3 r_2 + b_2 r_3 - b_4 \right)}
 \biggr] \, .
\eea
and satisfy, see \eqref{reeb:id},
\bea
 \label{b:Delta:FC3}
 2 = \sum_{a = 1}^{6} \Delta_a \, , \qquad \frac{2 b_2}{b_1} = \Delta_3 + \Delta_6 \, , \qquad \frac{2 b_3}{b_1} = \Delta_2 + \Delta_5 \, , \qquad \frac{2 b_4}{b_1} = r_1 \Delta_4 + r_2 \Delta_5 + r_3 \Delta_6 \, .
\eea
The previous relations can be inverted to express the $\phi_l$ in terms of $(\Delta_a, N, b_1)$:
\bea
 \label{phi:FC3}
 \phi_1^2  & = \frac{b_1^3 N r_3^2}{128 \pi^4} \Delta_3 ( \Delta_1 + \Delta_4 ) ( \Delta_2 + \Delta_5 ) ( r_1 \Delta_4 + r_2 \Delta_5 + r_3 \Delta_6 ) \, , \\
 \phi_2^2  & = \frac{b_1^3 N r_2^2}{128 \pi^4}  ( \Delta_1 + \Delta_4 ) ( \Delta_3 + \Delta_6 ) \left( r_1 \Delta_2 \Delta_4 + \Delta_5  ( r_2  \Delta_2 + r_3 \Delta_3 ) \right) \, , \\ 
 \phi_3^2 & = \frac{b_1^3 N r_1^2}{128 \pi^4} \Delta_4 ( \Delta_2 + \Delta_5 ) ( \Delta_3 + \Delta_6 ) ( r_1 \Delta_1 + r_2 \Delta _2 + r_3 \Delta_3 ) \, . 
\eea
By eliminating $\lambda_a$ and $b_i$ using \eqref{b:Delta:FC3} and \eqref{phi:FC3}, the master volume \eqref{master:FC3} takes the form
\bea
 \label{V:on-shell:Delta:FC3}
 \cV ( \Delta_a ) & = \pm \frac{N^{3/2} \sqrt{b_1}}{24 \sqrt{2} \pi^2}
 \biggl(
 \frac{r_1 \Delta_4 }{r_2 \Delta_2 + r_3 \Delta_3 - r_1 \Delta_4}
 \sqrt{\Delta_4 ( \Delta_2 + \Delta_5 ) ( \Delta_3 + \Delta_6 ) ( r_1 \Delta_1 + r_2 \Delta_2 + r_3 \Delta_3 )} \\
 & \pm \frac{r_3 \Delta_3}{r_1 \Delta_4 + r_2 \Delta_5 - r_3 \Delta_3}
 \sqrt{\Delta_3 ( \Delta_1 + \Delta_4 ) ( \Delta_2 + \Delta_5 ) ( r_1 \Delta_4 + r_2 \Delta_5 + r_3 \Delta_6 )} \\
 & \pm \frac{r_2 \left( r_1 \Delta_2 \Delta_4 + ( r_2 \Delta_2 + r_3 \Delta_3 ) \Delta_5 \right)}{( r_2 \Delta_2 + r_3 \Delta_3 - r_1 \Delta_4 ) ( r_1 \Delta _4 + r_2 \Delta_5 - r_3 \Delta_3 )} \\
 & \times \sqrt{( \Delta_1 + \Delta_4 ) ( \Delta_3 + \Delta_6 ) \left( r_1 \Delta_2 \Delta_4 + \Delta_5 ( r_2 \Delta_2 + r_3 \Delta_3 ) \right)}
 \biggr) \, .
\eea
Note that, the signs in the three square-root terms are not correlated. The baryonic K\"ahler moduli is given by
\bea
 \label{X:FC3}
 X^{(1)}
 & = \frac{2 \phi_1}{b_1 ( r_1 \Delta_4 + r_2 \Delta_5 - r_3 \Delta_3) }
 - \frac{2 r_3 \phi_3}{b_1 r_1 ( r_1 \Delta_4 - r_2 \Delta_2 - r_3 \Delta_3 )} \\
 & + \frac{2 ( \Delta_2 + \Delta_5 ) r_3 \phi_2}{b_1 ( r_1\Delta_4 - r_2 \Delta_2 - r_3 \Delta_3 ) ( r_1\Delta _4  + r_2 \Delta _5 - r_3 \Delta_3 )} \, , \\
 X^{(2)}
 & = - \frac{2 \left(r_1 \phi_2 - r_2 \phi_3\right)}{b_1 r_1 ( r_2 \Delta_2 + r_3 \Delta_3 - r_1 \Delta_4 )} \, .
\eea

By introducing the quartic function  \cite{Amariti:2011uw}
\bea
 \label{a3d:FC3}
 a_{(3)} ( \Delta_a ) & \equiv \frac{1}{24} \sum_{a , b , c , e= 1}^{6} | ( v^a , v^b , v^c , v^e ) | \Delta_a \Delta_b \Delta_c \Delta_e
\, ,
\eea
\eqref{V:on-shell:Delta:FC3} can also be rewritten as
\bea
 \cV ( \Delta_a ) = \pm \frac{N^{3/2} \sqrt{b_1}}{24 \sqrt{2} \pi^2}
 \sqrt{a_{(3)} ( \Delta_a ) + Y} \, ,
\eea
with
\bea
 Y \equiv  \frac{2 (2 \pi )^4}{b_1 N r_2 r_3} \left[ r_1 \Delta_4 \left( r_2 \Delta_3 ( X^{(1)} )^2 + r_3 \Delta_2 ( X^{(2)} )^2 \right) + \Delta_3 \Delta_5 \left( r_2 X^{(1)} - r_3 X^{(2)} \right)^2 \right] .
\eea
For the mesonic twist, $X^{(1)} = X^{(2)} = 0$, leading to
\bea
 \cV_{\text{mes.}} ( \Delta_a ) = \pm \frac{N^{3/2} \sqrt{b_1}}{24 \sqrt{2} \pi^2} \sqrt{a_{(3)} ( \Delta_a )} \, ,
\eea
which is consistent with \cite[(5.72)]{Hosseini:2019ddy}. 

\subsection{The cone over $Q^{1,1,1}$}
\label{ssec:Q111}

\begin{figure}[H]
\centering
\begin{tikzpicture}[scale=0.45]

  \draw[draw=none,fill=blue!20,opacity=0.6] (3,0) -- (3,3) -- (0,3.3) -- (-2,1) -- cycle;
  \draw[draw=none,fill=blue!20,opacity=0.6] (3,0) -- (1,-2) -- (-2,1) -- (3,3) -- cycle;
  \draw[draw=none,fill=blue!20,opacity=0.6] (3,0) -- (1,-2) -- (-2,-2) -- (-2,1) -- cycle;
  \draw[draw=none,fill=blue!20,opacity=0.6] (3,0) -- (-2,-2) -- (-2,1) -- (0,3.3) -- cycle;
  \draw[draw=none,fill=blue!20,opacity=0.6] (3,0) -- (3,3) -- (-2,1) -- (-2,-2) -- cycle;

  \draw[-,dashed] (0,0) -- (3,0) node[below] {};
  \draw[->,solid] (3,0) -- (4.5,0) node[below] {};
  
  \draw[-,dashed] (0,0) -- (0,3) node[below] {};
  \draw[->,solid] (0,3) -- (0,5) node[below] {};
  
  \draw[-,dashed] (0,0) -- (-1.6,-1.6) node[below] {};
  \draw[->,solid] (-1.6,-1.6) -- (-3.,-3.) node[below] {};
  
  \draw (-2.4,-3.4) node {$e_2$};
  \draw (4.7,0.5) node {$e_3$};
  \draw (0.6,5) node {$e_4$};
  
  \draw[-,solid] (-2,1) -- (1,-2);
  \draw[-,solid] (-2,1) -- (3,3);
  \draw[-,solid] (1,-2) -- (3,3);
  \draw[-,solid] (-2,-2) -- (1,-2);
  \draw[-,solid] (-2,-2) -- (-2,1);
  \draw[-,dashed] (-2,-2) -- (3,0);
  \draw[-,solid] (3,3) -- (3,0);
  \draw[-,solid] (3,0) -- (1,-2);
  \draw[-,solid] (0,3.3) -- (3,3);
  \draw[-,solid] (0,3.3) -- (-2,1);
  \draw[-,dashed] (-2,-2) -- (0,3.3);
  \draw[-,dashed] (3,0) -- (0,3.3);

  \draw (-1.7,-2.5) node {$v^1$};
  \draw (3.3,-0.5) node {$v^2$};
  \draw (0.5,3.85) node {$v^3$};
  \draw (-2.3,1.5) node {$v^4$};
  \draw (1.6,-2.) node {$v^5$};
  \draw (3.5,3.25) node {$v^6$};

\end{tikzpicture}
\caption{The toric diagram for $Q^{1,1,1}$.\label{toric:Q111}}
\end{figure}
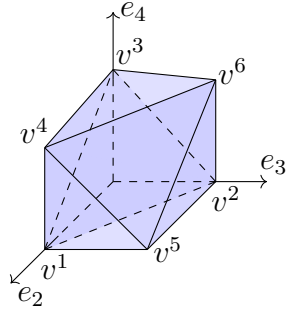

The geometry is specified by the vectors
\bea
 \vec{v}^{\,1} & = ( 1 , 0, 0 ) \, , ~~~ && \vec{v}^{\,2} = ( 0, 1, 0 ) \, , &&& \vec{v}^{\,3} = ( 0, 0, 1 ) \, , \\
 \vec{v}^{\,4} & = ( 1, 0, 1) \, , && \vec{v}^{\,5} = ( 1 , 1 , 0 ) \, , &&& \vec{v}^{\,6} = ( 0, 1, 1 ) \, .
\eea
with the corresponding toric diagram shown in figure \ref{toric:Q111}.
\noindent There are two baryonic symmetries: $B^{(1)} = ( 1 , - 1 , 0 , - 1 , 0 , 1 )$ and $B^{(2)} = ( 0 , -1 , 1 , - 1 , 1 , 0 )$.
For more details about the geometry, see \cite{Fabbri:1999hw}. The dual SCFT, identified in \cite{Benini:2009qs}, is a flavored ABJM quiver. The toric polytope \ref{toric:Q111} can be triangulated using the tetrahedra $(2436)$, $(2546)$, $(2514)$, and $(2341)$. All the faces of the toric diagram are triangles and the master volume is independent of the resolution. The gauge invariants $\phi_l$, for $l=1,\dots,4$, corresponding to these tetrahedra, respectively, are given by
\bea
 \phi_1 ( \lambda_a , b_i ) & \equiv
 ( b_1 - b_4 ) \lambda_2 + ( b_1 - b_2 - b_3 ) \lambda_3 + b_2 \lambda_4 - ( b_1 - b_3 - b_4 ) \lambda_6 \, , \\
 \phi_2 ( \lambda_a , b_i ) & \equiv
 ( 2 b_1 - b_2 - b_3 - b_4 ) \lambda_2 + ( b_1 - b_3 ) \lambda_4 - ( b_1 - b_2 - b_3) \lambda_5 - ( b_1 - b_3 - b_4) \lambda_6 \, , \\
 \phi_3 ( \lambda_a , b_i ) & \equiv
 ( b_1 - b_3 -b_4 ) \lambda_1 + ( b_1 - b_2 ) \lambda_2 + b_4 \lambda_4 - ( b_1 - b_2 - b_3) \lambda_5 \, , \\
 \phi_4 ( \lambda_a , b_i ) & \equiv
 ( b_1 - b_3 - b_4 ) \lambda_1 + b_3 \lambda_2 + ( b_1 - b_2 - b_3 ) \lambda_3 - ( b_1 - b_2 - b_3 - b_4) \lambda_4 \, .
\eea
We note that
\bea
 \label{phi:const:Q111}
 \phi_1 (\lambda_a , b_i) - \phi_2 ( \lambda_a , b_i) + \phi_3 (\lambda_a , b_i) - \phi_4 ( \lambda_a , b_i) = 0 \, .
\eea
In general, one can construct $d - 3$ independent gauge invariant variables, consisting of $d - 4$ baryonic K\"ahler moduli $X^{(r)}$ and one additional variable given by $\sum_{a = 1}^{d} \alpha^a ( b ) \lambda_a$, where $\alpha^a ( b )$ is a linear function of the Reeb vector $b$. Consequently, the ``minimal'' triangulation of the toric polytope \eqref{toric:Q111}, which consists of four tetrahedra, is subject to the constraint \eqref{phi:const:Q111}. The master volume can be computed from \eqref{master} as a sum of four contributions:%
\footnote{This was also computed by different methods in \cite{Hosseini:2019ddy, Cassia:2025aus}.}
\bea
 \label{master:Q111}
 \cV ( \lambda_a , b_i ) = \frac{8 \pi^4}{3 ( b_1 - b_2 - b_3 ) ( b_1 - b_3 - b_4 )} & \biggl[
 \frac{\phi_1^3 ( \lambda_a , b_i  )}{b_2 ( b_1 - b_4 )}
 - \frac{\phi_2^3 ( \lambda_a , b_i  )}{ ( b_1 - b_3 ) ( 2 b_1 - b_2 - b_3 - b_4 )} \\
 & + \frac{\phi_3^3 ( \lambda_a , b_i  )}{b_4 ( b_1 - b_2 )}
 + \frac{\phi_4^3 ( \lambda_a , b_i  )}{b_3 ( b_1 - b_2 - b_3 - b_4 )}
 \biggr] \, .
\eea
The R-charges \eqref{def:R-charge} corresponding to each vertex are
\begin{align}
 \label{R-charge:Q111}
 \Delta_1 & = - \frac{16 \pi^4}{N ( b_1 - b_2 - b_3 )}
 \left( \frac{\phi_3^2 ( \lambda_a , b_i )}{b_4 ( b_1 - b_2 )}
 + \frac{\phi_4^2 ( \lambda_a , b_i )}{b_3 ( b_1 - b_2 - b_3 - b_4 )} \right) , \nn \\
 \Delta_2 & = - \frac{16 \pi^4}{N (b_1 - b_2 - b_3 ) ( b_1 - b_3 - b_4 )}
 \left( \frac{\phi_1^2 ( \lambda_a , b_i )}{b_2}
 + \frac{\phi_3^2 ( \lambda_a , b_i )}{b_4}
 - \frac{\phi_2^2 ( \lambda_a , b_i )}{b_1 - b_3}
 + \frac{\phi_4^2}{ b_1 - b_2 - b_3 - b_4} \right) , \nn \\ \displaybreak
 \Delta_3 & = - \frac{16 \pi^4}{N ( b_1 - b_3 - b_4 )}
 \left( \frac{\phi_1^2 ( \lambda_a , b_i )}{b_2 ( b_1 - b_4 )}
 + \frac{\phi_4^2 ( \lambda_a , b_i )}{b_3 ( b_1 - b_2 - b_3 - b_4 )} \right) , \nn \\
 \Delta_4 & = - \frac{16 \pi^4}{N ( b_1 - b_2 - b_3 ) ( b_1 - b_3 - b_4 )}
 \left( \frac{\phi_1^2 ( \lambda_a , b_i )}{b_1 - b_4}
 - \frac{\phi_2^2 ( \lambda_a , b_i )}{2 b_1 - b_2 - b_3 - b_4}
 + \frac{\phi_3^2 ( \lambda_a , b_i )}{b_1 - b_2}
 - \frac{\phi_4^2 ( \lambda_a , b_i )}{b_3} \right) , \nn \\
 \Delta_5 & = - \frac{16 \pi^4}{N ( b_1 - b_3 - b_4)}
 \left( \frac{\phi_2^2 ( \lambda_a , b_i )}{( b_1 - b_3 ) ( 2 b_1 - b_2 - b_3 - b_4 )}
 - \frac{\phi_3^2 ( \lambda_a , b_i )}{b_4 ( b_1-b_2 )} \right) , \nn \\
 \Delta_6 & = \frac{16 \pi^4}{N ( b_1 - b_2 - b_3 )}
 \left( \frac{\phi_1^2 ( \lambda_a , b_i )}{b_2 ( b_1 - b_4 )}
 - \frac{\phi_2^2 ( \lambda_a , b_i )}{( b_1 - b_3 ) ( 2 b_1 - b_2 - b_3 - b_4 )} \right) ,
\end{align}
and satisfy, see \eqref{reeb:id},
\bea
 \label{b:Delta:Q111}
 2 = \sum_{a = 1}^{6} \Delta_a \, , \qquad \frac{2 b_2}{b_1} = \Delta_1 + \Delta_4 + \Delta_5 \, , \qquad \frac{2 b_3}{b_1} = \Delta_2 + \Delta_5 + \Delta_6 \, , \qquad \frac{2 b_4}{b_1} = \Delta_3 + \Delta_4 + \Delta_6 \, .
\eea
The previous relations can be inverted to express the $\phi_l$ in terms of $(\Delta_a, N, b_1)$:
\bea
 \label{phi:Q111}
 \phi_1^2 & = \frac{b_1^3 N}{128 \pi^4}
 ( \Delta_1 + \Delta_2 + \Delta_5 ) ( \Delta_1 + \Delta_4 + \Delta_5 )
 \left( \Delta_6 ( \Delta_3 - \Delta_5 ) + \cQ ( \Delta_a ) \right) \, , \\
 \phi_3^2 & = \frac{b_1^3 N}{128 \pi^4}
 ( \Delta_2 + \Delta_3 + \Delta_6 ) ( \Delta_3 + \Delta_4 + \Delta_6 )
 \left( \Delta_5 ( \Delta_1 - \Delta_6 ) + \cQ ( \Delta_a ) \right) \, , \\ 
 \phi_4^2  & = \frac{b_1^3 N}{128 \pi^4}
 ( \Delta_2 + \Delta_5 + \Delta_6 ) ( \Delta_4 + \Delta_5 + \Delta_6 )
 \left( \Delta_1 \Delta_3 - \Delta_5 \Delta_6 + \cQ ( \Delta_a ) \right) \, .
\eea
Here, the function
\bea
 \cQ ( \Delta_a) \equiv \frac{128 \pi^4}{b_1^3 N} \frac{\phi_2^2}{( \Delta_1 + \Delta_2+ \Delta_3 ) ( \Delta_1 + \Delta_3 + \Delta_4 )} \, ,
\eea
is determined by solving the constraint \eqref{phi:const:Q111}, which results in a quartic equation for $\phi_2$. While the solution to a quartic equation is well known, we omit its rather unwieldy explicit expression for brevity. By eliminating $\lambda_a$ and $b_i$ using \eqref{b:Delta:Q111} and \eqref{phi:Q111}, the master volume \eqref{master:Q111} simplifies to the form
\bea
 \label{V:on-shell:Delta:Q111}
 \cV ( \Delta_a ) & = \frac{N}{3 b_1}
 \frac{\Delta _5 \left( \Delta_1 \phi_3 - \Delta_6 \phi_2 \right)
 - \Delta_3 \left( \Delta_1 \phi_4 - \Delta_6 \phi_1 \right)}
 {( \Delta_3 - \Delta_5 ) ( \Delta_1 - \Delta_6 )} \, .
\eea
The baryonic K\"ahler moduli are given by
\bea
 \label{X:Q111}
 X^{(1)}
 = - \frac{2}{b_1}  \frac{\phi_2 - \phi_3}{\Delta_1 - \Delta_6} \, , \qquad
 X^{(2)}
 = - \frac{2}{b_1} \frac{\phi_3 - \phi_4}{\Delta_3 - \Delta_5} \, .
\eea

By introducing the quartic function  \cite{Amariti:2011uw}
\bea
 \label{a3d:def:Q111}
 a_{(3)} ( \Delta_a ) & \equiv \frac{1}{24} \sum_{a , b , c , e= 1}^{6} | ( v^a , v^b , v^c , v^e ) | \Delta_a \Delta_b \Delta_c \Delta_e
 + \frac{1}{2} ( \Delta_2 \Delta_4 + \Delta_3 \Delta_5 + \Delta_1 \Delta_6 )^2 \\
 & - ( \Delta_2 \Delta_4 )^2 - ( \Delta_3 \Delta_5)^2 - ( \Delta_1 \Delta_6 )^2 \, ,
\eea
\eqref{V:on-shell:Delta:Q111} can also be rewritten as
\bea
 \cV ( \Delta_a ) = \pm \frac{N^{3/2} \sqrt{b_1}}{24 \sqrt{2} \pi^2}
 \sqrt{a_{(3)} ( \Delta_a ) + Y} \, ,
\eea
with
\bea
 Y \equiv \frac{2 (2 \pi )^4}{b_1 N}
 \left[ \left( X^{(1)} - X^{(2)} \right) \left( \Delta_1 \Delta_6 X^{(1)} - \Delta_3 \Delta_5 X^{(2)} \right) + \Delta_2 \Delta_4 X^{(1)} X^{(2)} \right] .
\eea
For the mesonic twist, $X^{(1)} = X^{(2)} = 0$, leading to
\bea
 \cV_{\text{mes.}} ( \Delta_a ) = \pm \frac{N^{3/2} \sqrt{b_1}}{24 \sqrt{2} \pi^2} \sqrt{a_{(3)} ( \Delta_a )} \, ,
\eea
which agrees with \cite[(5.47)]{Hosseini:2019ddy}.

\subsubsection{Half-baryonic twist}

We define the \emph{half-baryonic twist} by setting either $X^{(1)}$ or $X^{(2)}$ equal to zero. First, consider $X^{(1)} = 0$. Thus, from \eqref{X:Q111} and \eqref{phi:const:Q111}, we obtain
\bea
 \label{case1:Q111}
 \textbf{Case \Romannum{1}:} \quad \phi_3 = \phi_2 \, , \qquad \phi_4 = \phi_1 \, .
\eea
Substituting \eqref{case1:Q111} into \eqref{R-charge:Q111}, and using \eqref{b:Delta:Q111}, we find that
\bea
 \label{hB:case1:phi:Q111}
 \phi_1^2 = \frac{b_1^3 N}{128 \pi^4}
 \frac{ \Delta_3 \Delta_{125} \Delta_{145} \Delta_{256} \Delta_{456}}{ \Delta_{124} + 2 \Delta_5 + \Delta_6} \, , \qquad
 \phi_2^2 = \frac{b_1^3 N}{128 \pi^4}
 \frac{\Delta_5 \Delta_{123} \Delta_{134} \Delta_{236} \Delta_{346}}{\Delta_{124} + 2 \Delta_3 + \Delta_6} \, ,
\eea
along with the following constraint among the R-charges
\bea
 \label{hB:case1:const:Q111}
 \sum_{a = 1}^{6} B_a^{(1)} \frac{\pd \cV ( \Delta_a )}{\pd \Delta_a} = 0 \, .
\eea
Here, we introduced the shorthand $\Delta_{abc} \equiv \Delta_a + \Delta_b + \Delta_c$ to improve readability.
Plugging back \eqref{hB:case1:phi:Q111} into \eqref{V:on-shell:Delta:Q111}, we obtain
\bea
 \label{hB:case1:master:Q111}
 \cV ( \Delta_a ) & = \pm \frac{N^{3/2} \sqrt{b_1}}{24 \sqrt{2} \pi^2}
 \frac{1} {\Delta_3 - \Delta_5}
 \Biggl(
 \Delta_3 \sqrt{ \frac{ \Delta_3 \Delta_{125} \Delta_{145} \Delta_{256} \Delta_{456}}{ \Delta_{124} + 2 \Delta_5 + \Delta_6}}
 \pm \Delta_5 \sqrt{\frac{\Delta_5 \Delta_{123} \Delta_{134} \Delta_{236} \Delta_{346}}{\Delta_{124} + 2 \Delta_3 + \Delta_6}}
 \Biggr) \, .
\eea
Note that the signs in the two square-root terms are not correlated. Moreover, \eqref{hB:case1:master:Q111} can be rewritten in terms of the volume of the torus-invariant five-cycle associated with $v^a$:
\bea
 \label{hB:case1:master:volS:Q111}
 \cV ( \Delta_a ) & = \pm \frac{N^{3/2}}{12 b_1 \sqrt{\pi} ( \Delta_3 - \Delta_5 )}
 \left( \Delta_3 \sqrt{ \frac{\Delta_3}{\Vol_{S_3} ( \Delta_a )}} \pm \Delta_5 \sqrt{\frac{\Delta_5}{\Vol_{S_5} ( \Delta_a )}} \right) \, ,
\eea
which is consistent with appendix \ref{app:fp}, since the toric diagram can be viewed as a suspension over a square in the $(1264)$ plane, and the corresponding $\phi$ are equal.

Next, consider $X^{(2)} = 0$. Thus, from \eqref{X:Q111} and \eqref{phi:const:Q111}, we obtain
\bea
 \label{case2:Q111}
 \textbf{Case \Romannum{2}:} \quad \phi_4 = \phi_3 \, , \quad \phi_2 = \phi_1 \, .
\eea
Substituting \eqref{case2:Q111} into \eqref{R-charge:Q111}, and using \eqref{b:Delta:Q111}, we find that
\bea
 \label{hB:case2:phi:Q111}
 \phi_1^2  = \frac{b_1^3 N}{128 \pi^4}
 \frac{\Delta_6 \Delta_{123}\Delta_{134} \Delta_{125} \Delta_{145}}{2 \Delta_1 + \Delta_{234} + \Delta _5} \, , \qquad
 \phi_3^2  = \frac{b_1^3 N}{128 \pi^4}
 \frac{ \Delta_1 \Delta_{236} \Delta_{346} \Delta_{256} \Delta_{456}}{\Delta_{234} + \Delta_5 + 2 \Delta_6} \, ,
\eea
along with the following constraint among the R-charges
\bea
 \label{hB:case2:const:Q111}
 \sum_{a = 1}^{6} B_a^{(2)} \frac{\pd \cV ( \Delta_a )}{\pd \Delta_a} = 0 \, .
\eea
Plugging back \eqref{hB:case2:phi:Q111} into \eqref{V:on-shell:Delta:Q111}, we obtain
\bea
 \cV ( \Delta_a ) & = \pm \frac{N^{3/2} \sqrt{b_1}}{24 \sqrt{2} \pi^2}
 \frac{1} {\Delta_1 - \Delta_6}
 \Biggl(
 \Delta_1 \sqrt{\frac{ \Delta_1 \Delta_{236} \Delta_{346} \Delta_{256} \Delta_{456}}{ \Delta_{324} + \Delta_5 + 2 \Delta_6}}
 \pm \Delta_6 \sqrt{\frac{\Delta_6 \Delta_{123} \Delta_{134} \Delta_{125} \Delta_{145}}{2 \Delta_1 + \Delta_{234} + \Delta _5}}
 \Biggr) \, .
\eea
Note that, the signs in the two square-root terms are not correlated. Moreover,
\bea
 \label{hB:case2:master:volS:Q111}
 \cV ( \Delta_a ) & = \pm \frac{N^{3/2}}{12 b_1 \sqrt{\pi} ( \Delta_1 - \Delta_6 )}
 \left( \Delta_1 \sqrt{ \frac{\Delta_1}{\Vol_{S_1} ( \Delta_a )}} \pm \Delta_6 \sqrt{\frac{\Delta_6}{\Vol_{S_6} ( \Delta_a )}} \right) \, ,
\eea
which is consistent with appendix \ref{app:fp}, since the toric diagram can also be viewed as a suspension over a square in the $(2345)$ plane, and the corresponding $\phi$ are equal.

\subsubsection{A purely baryonic twist}

Using the $\SU(2)^3$ symmetry of the model, it is also possible to consistently truncate to a purely baryonic twist by setting
\begin{equation}
 \label{SU2:Q111}
 \begin{alignedat}{3}
  \Delta_1 - \varsigma \big|_{\varsigma = 0} & = \Delta_6 + \varsigma \big|_{\varsigma = 0} \equiv \delta_1 \, ,\quad & \Delta_2 - \sigma \big|_{\varsigma = 0} & = \Delta_4 + \varsigma \big|_{\varsigma = 0} \equiv \delta_2 \, ,\quad & \Delta_3 - \varsigma \big|_{\varsigma = 0} & = \Delta_5 + \varsigma \big|_{\varsigma = 0} \equiv \delta_3 \, ,\\[0.5em]
  \fn_1 & = \fn_6 \equiv \fp_1 \, ,\quad & \fn_2 & = \fn_4 \equiv \fp_2 \, ,\quad & \fn_3 & = \fn_5 \equiv \fp_3 \, ,
 \end{alignedat}
\end{equation}
where we introduce the parameter $\varsigma$, which will be taken to zero at the end of the computation. The R-charge constraint $\sum_{a = 1}^{6} \Delta_a = 2$ and the twisting condition $\sum_{a = 1}^{6} \fn_a = 2 - 2 \fg$ reduce to
\bea
 \label{RB:const:Q111}
 \sum_{a = 1}^{3} \delta_a = 1 \, , \qquad \sum_{a = 1}^{3} \fp_a = 1 - \fg \, .
\eea
Using \eqref{b:Delta:Q111}, we find that the mesonic variables are frozen to their Sasaki-Einstein values:
\bea
 b_2 = b_3 = b_4 = \frac{1}{2} b_1 \, .
\eea
By taking the limit $\varsigma \to 0$ in \eqref{R-charge:Q111}, \eqref{b:Delta:Q111}, \eqref{X:Q111}, and \eqref{master:Q111}, we find that
\bea
 \phi_l & = \pm \frac{b_1^{3/2} \sqrt{N}}{(4 \pi )^2} ( \delta_1+ \delta_2 + \delta_3)^{3/2} \sqrt{\frac{ \delta_1 \delta_3}{ \delta_2}} \, , \quad \text{ for } \quad l = 1, \ldots, 4 \, , \\
 X^{(1)} & = \pm \frac{\sqrt{b_1 N}}{8 \pi ^2} \sqrt{\delta_3 (\delta_1 + \delta_2 + \delta_3 )} \biggl( \sqrt{\frac{\delta_2}{\delta_1}} - \sqrt{\frac{\delta_1}{\delta_2}} \biggr)
  = - \frac{4}{N} \sum_{a = 1}^{3} B^{(1)}_a \frac{\pd \cV_B ( \delta_a )}{\pd \delta_a} \, , \\
  X^{(2)} & = \pm \frac{\sqrt{b_1 N}}{8 \pi^2} ( \delta_2 - \delta_3 ) \sqrt{\frac{\delta_1 (\delta_1 + \delta_2 + \delta_3)}{\delta _2 \delta _3}}
  = - \frac{4}{N} \sum_{a = 1}^{3} B^{(2)}_a \frac{\pd \cV_B ( \delta_a )}{\pd \delta_a} \, ,
\eea
with
\bea
 \label{VB:Q111}
 \cV_{B} ( \delta_a ) = \mp \frac{\sqrt{b_1} N^{3/2}}{(4 \pi)^2} \sqrt{\delta_1 \delta_2 \delta_3 ( \delta_1 + \delta_2 + \delta_3 )} \, .
\eea
This expression is extremized at $\delta_1 = \delta_2 = \delta_3 = \frac{1}{3}$, which correctly leads to the known dimension of baryons realized as M5-branes wrapped on the five-cycles associated with $v^a$ \cite{Fabbri:1999hw,Hanany:2008fj}. 

As an example, we can write the entropy function for  purely magnetic static black holes. It is obtained by taking the limit $\varsigma \to 0$ in \eqref{id2}:
\bea
 \label{SB:Q111}
 \cI_B ( \delta_a , \fp_a ) &= \log \cZ_B ( \delta_a , \fp_a ) = - \frac{1}{2} ( 4 \pi )^3 \sum_{a = 1}^{2} \fp_a \frac{\pd_a \cV_B ( \delta_a )}{\pd \delta_a} \\ & = \pm \sqrt{b_1} \pi N^{3/2}
 \frac{\delta_1 \delta_2 ( \delta_1 + \delta_2 + 2 \delta_3 ) \fp_3 + \delta_2 \delta_3 ( 2 \delta_1 + \delta_2 + \delta_3 ) \fp_1 + \delta_1 \delta_3 ( \delta_1 + 2 \delta_2 + \delta_3 ) \fp_2}{\sqrt{\delta_1 \delta_2 \delta_3 ( \delta_1 + \delta_2 + \delta_3 )}} 
  \, .
\eea
For dyonic static black holes with electric charges $\fq_a$, we extremize
\bea
 \label{Q111:dyonic}
 \cI_B ( \delta_a , \fp_a ) \equiv \log \cZ_B ( \delta_a , \fp_a ) + 2 \pi \ii \sqrt{b_1} \sum_{a = 1}^{3} \delta_a \fq_a \, .
\eea

\subsection{The cone over $M^{1,1,1}$}
\label{ssec:M111}

\begin{figure}[H]
\centering
\begin{tikzpicture}[scale=0.45]
  \draw[-,solid] (-2,2) -- (2,-2) node[below] {};
  \draw[->,solid] (2,-2) -- (3,-3) node[below] {};
  
  \draw[-,solid] (-2,-2) -- (0,0) node[below] {};
  \draw[-,dashed] (0,0) -- (2,2) node[below] {};
  \draw[->,solid] (2,2) -- (3,3) node[below] {};
  
  \draw[-,solid] (0,0) -- (0,6.2) node[below] {};
  \draw[->,solid] (0,6.2) -- (0,9.5) node[below] {};
  
  \draw (3.5,-2.8) node {$e_2$};
  \draw (3.1,3.4) node {$e_3$};
  \draw (0.6,9.5) node {$e_4$};

  \draw[draw=none,fill=blue!20,opacity=0.6] (0,0) -- (2,-2) -- (2,2) -- (-2,9) -- cycle;
  \draw[draw=none,fill=blue!20,opacity=0.6] (2,-2) -- (2,2) -- (0,6.2) -- (-2,9) -- cycle;
  \draw[draw=none,fill=blue!30,opacity=0.6] (2,-2) -- (2,2) -- (-2,9) -- cycle;
  
  \draw[-,solid] (0,0) -- (2,-2);
  \draw[-,solid] (2,-2) -- (2,2);
  \draw[-,solid] (2,2) -- (0,6.2);
  \draw[-,solid] (0,6.2) -- (-2,9);
  \draw[-,solid] (-2,9) -- (0,0);
  \draw[-,dashed] (-2,9) -- (2,2);
  \draw[-,solid] (0,6.2) -- (2,-2);
  \draw[-,solid] (-2,9) -- (2,-2);
  
  \draw (-0.05,-0.7) node {$v^1$};
  \draw (2.6,-1.6) node {$v^2$};
  \draw (2.5,1.9) node {$v^3$};
  \draw (-2.1,9.4) node {$v^4$};
  \draw (0.5,6.6) node {$v^5$};
\end{tikzpicture}
\caption{The toric diagram for $M^{1,1,1}$.\label{toric:M111}}
\end{figure}
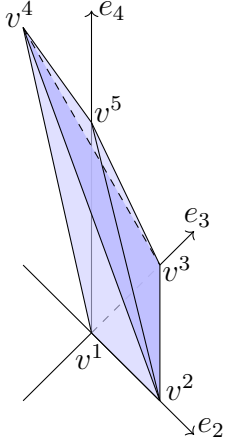

The geometry is specified by the vectors
\bea
 \vec{v}^{\,1} & = ( 0, 0, 0 ) \, , ~~~ && \vec{v}^{\,2} = ( 1, 0, 0 ) \, , &&& \vec{v}^{\,3} = ( 0, 1, 0 ) \, , \qquad \vec{v}^{\,4} = ( - 1, - 1, 3) \, ,  \qquad \vec{v}^{\,5} = ( 0, 0, 2) \, ,
\eea
with the corresponding toric diagram shown in figure \ref{toric:M111}.
\noindent There is a single baryonic symmetry given by $B = ( 3 , -2 , -2 , -2 , 3 )$. The dual SCFT was found in \cite{Martelli:2008si} and has a \emph{chiral} quiver (in the four-dimensional sense), which precludes a large-$N$ limit computation. However, a prediction for the large-$N$ free energy, restricted to mesonic variables, can be extracted from the known spectrum of gauge invariant mesonic operators \cite{Gulotta:2011aa,Kim:2019umc}. Here, we extend the prediction to include the full set of mesonic and baryonic variables. The toric polytope shown figure \ref{toric:M111} can be triangulated using the tetrahedra $(1243)$ and $(2354)$. Since the polytope is regular, the result is independent of the choice of triangulation. We define the gauge invariants $\phi_1$ and $\phi_2$ corresponding to the two tetrahedra, respectively,
\bea
 \phi_1 ( \lambda_a , b_i ) & \equiv
 3 ( b_1 - b_2 - b_3 - b_4 ) \lambda_1 + ( 3 b_2 + b_4 ) \lambda_2 + ( 3 b_3 + b_4 ) \lambda_3 + b_4 \lambda_4 \, , \\
 \phi_2 ( \lambda_a , b_i ) & \equiv
 ( 2 b_1 + b_2 - 2 b_3 - b_4 ) \lambda_2 + ( 2 b_1 - 2 b_2 + b_3 - b_4 ) \lambda_3 \\
 & + ( 2 b_1 - 2 b_2 - 2 b_3 - b_4) \lambda_4 - 3 ( b_1 - b_2 - b_3 - b_4 ) \lambda_5 \, .
\eea
Using \eqref{master}, we then calculate the master volume,
\bea
 \label{master:M111}
 \cV ( \lambda_a , b_i ) & = - \frac{8 \pi^4}{9 ( b_1 - b_2 - b_3 - b_4} \biggl(
 \frac{\phi_1^3 ( \lambda_a , b_i )}{b_4 ( 3 b_2 + b_4 ) ( 3 b_3 + b_4 )} \\
 & - \frac{\phi_2^3 ( \lambda_a , b_i )}{( 2 b_1 - 2 b_2 + b_3 - b_4 ) ( 2 b_1 + b_2 - 2 b_3 - b_4 ) \left( 2 ( b_1 - b_2 - b_3 ) - b_4 \right)}
 \biggr) \, .
\eea
as a function of the Reeb vector $b$ and the K\"ahler parameters $\lambda_a$. The R-charges \eqref{def:R-charge} corresponding to each vertex are
\bea
 \label{R-charge:M111}
 \Delta_1 & = \frac{16 \pi^4}{N} \frac{\phi_1 ( \lambda_a , b_i )^2}{b_4 ( 3 b_2 + b_4 ) ( 3 b_3 + b_4 )} \, , \\
 \Delta_2 & = \frac{16 \pi^4}{3 N ( b_1 - b_2 - b_3 - b_4 )} \biggl( \frac{\phi_1 ( \lambda_a , b_i )^2}{b_4 ( 3 b_3 + b_4 )}
 - \frac{\phi_2 ( \lambda_a , b_i )^2}{( 2 b_1 - 2 b_2 + b_3 - b_4 ) \left( 2 ( b_1 - b_2 - b_3 ) - b_4 \right)} \biggr) \, , \\
 \Delta_3 & = \frac{16 \pi^4}{3 N ( b_1 - b_2 - b_3 - b_4 )} \biggl( \frac{\phi_1 ( \lambda_a , b_i )^2}{b_4 ( 3 b_2 + b_4 )}
 - \frac{\phi_2 ( \lambda_a , b_i )^2}{( 2 b_1 + b_2 - 2 b_3 - b_4 ) \left( 2 ( b_1 - b_2 - b_3 ) - b_4 \right)} \biggr) \, , \\
 \Delta_4 & = \frac{16 \pi^4}{3 N ( b_1 - b_2 - b_3 - b_4 )} \biggl( \frac{\phi_1 ( \lambda_a , b_i )^2}{( 3 b_2 + b_4 ) ( 3 b_3 + b_4 )}
 - \frac{\phi_2 ( \lambda_a , b_i )^2}{( 2 b_1 + b_2 - 2 b_3 - b_4 ) ( 2 b_1 - 2 b_2 + b_3 - b_4 )} \biggr) \, , \\
 \Delta_5 & = \frac{16 \pi^4}{3 N} \frac{\phi_2 ( \lambda_a , b_i )^2}{( 2 b_1 + b_2 - 2 b_3 - b_4 ) ( 2 b_1 - 2 b_2 + b_3 - b_4 ) \left( 2 ( b_1 - b_2 - b_3 ) - b_4 \right)} \, ,
\eea
and satisfy, see \eqref{reeb:id},
\bea
 \label{b:Delta:M111}
 2 = \sum_{a = 1}^{5} \Delta_a \, , \qquad \frac{2 b_2}{b_1} = \Delta_2 - \Delta_4 \, , \qquad \frac{2 b_3}{b_1} = \Delta_3 - \Delta_4 \, , \qquad \frac{2 b_4}{b_1} = 3 \Delta_4 + 2 \Delta_5 \, .
\eea
The previous relations can be inverted to express the $\phi_l$ in terms of $(\Delta_a, N, b_1)$:
\bea
 \label{phi:M111}
 \phi_1^2 & = \frac{b_1^3 N}{128 \pi^4}
 \Delta _1 ( 3 \Delta_2 + 2 \Delta_5 ) ( 3 \Delta_3 + 2 \Delta_5 ) ( 3 \Delta_4 + 2 \Delta_5 ) \, , \\
 \phi_2^2 & = \frac{b_1^3 N}{128 \pi^4}
 ( 2 \Delta_1 + 3 \Delta_2 ) ( 2 \Delta_1 + 3 \Delta_3 ) ( 2 \Delta_1 + 3 \Delta_4 ) \Delta_5 \, . 
\eea
By eliminating $\lambda_a$ and $b_i$ using \eqref{b:Delta:M111} and \eqref{phi:M111}, the master volume \eqref{master:M111} takes the form
\bea
 \label{V:on-shell:Delta:M111}
 \cV ( \Delta_a ) & = \pm \frac{N^{3/2} \sqrt{b_1}}{72 \sqrt{2} \pi^2}
 \frac{1}{( \Delta_1 - \Delta_5 )}
 \bigl(
  \Delta _1 \sqrt{\Delta_1 ( 3 \Delta_2 + 2 \Delta_5 ) ( 3 \Delta_3 + 2 \Delta_5 ) ( 3 \Delta_4 + 2 \Delta_5 )} \\
 & \pm \Delta_5 \sqrt{ ( 2 \Delta_1 + 3 \Delta_2 ) ( 2 \Delta_1 + 3 \Delta_3 ) ( 2 \Delta_1 + 3 \Delta_4 ) \Delta _5}
 \bigr)
 \, .
\eea
Note that, the signs in the two square-root terms are not correlated. Furthermore,
\bea
 \label{V:volS:on-shell:Delta:M111}
 \cV ( \Delta_a ) & = \pm \frac{N^{3/2}}{12 b_1 \sqrt{\pi} ( \Delta_1 - \Delta_5 )}
 \left( \Delta_1 \sqrt{ \frac{\Delta_1}{\Vol_{S_1} ( b_i )}} \pm \Delta_5 \sqrt{\frac{\Delta_5}{\Vol_{S_5} ( b_i )}} \right) ,
\eea
which is consistent with appendix \ref{app:fp}, since the toric diagram can be seen as a suspension over a triangle in the $(234)$ plane. The baryonic K\"ahler modulus is given by
\bea
 \label{X:M111}
 X
 = \frac{2}{b_1} \frac{\phi_1 - \phi_2}{\Delta_1 - \Delta_5} \, .
\eea

By introducing the quartic function
\bea
 \label{a3d:M111}
 a_{(3)} ( \Delta_a ) & \equiv \frac{1}{24} \sum_{a , b , c , e= 1}^{5} | ( v^a , v^b , v^c , v^e ) | \Delta_a \Delta_b \Delta_c \Delta_e
 - \frac{8}{9} ( \Delta_1 \Delta_5 )^2 \, ,
\eea
as in \cite{Amariti:2011uw}, \eqref{V:on-shell:Delta:M111} can also be rewritten as
\bea
 \cV ( \Delta_a ) = \pm \frac{N^{3/2} \sqrt{b_1}}{24 \sqrt{2} \pi^2}
 \sqrt{a_{(3)} ( \Delta_a ) + \frac{2 (2 \pi )^4}{9 b_1 N} \Delta_1 \Delta_5 X^2} \, .
\eea
For the mesonic twist, $X = 0$, and thus
\bea
 \cV_{\text{mes.}} ( \Delta_a ) = \pm \frac{N^{3/2} \sqrt{b_1}}{24 \sqrt{2} \pi^2} \sqrt{a_{(3)} ( \Delta_a )} \, .
\eea

\subsubsection{A purely baryonic twist}

Due to the large isometry group $\SU(3) \times \SU(2)$ of $M^{1,1,1}$, we can choose the Reeb vector as
\bea
 b_2 = b_3 = 0 \, , \qquad b_4 = b_1 \, ,
\eea
which allows us to consistently consider a purely baryonic twist. From \eqref{b:Delta:M111}, we therefore impose the conditions
\begin{equation}
 \label{SU3xSU2}
 \begin{alignedat}{2}
  \Delta_1 - \varsigma \big|_{\varsigma = 0}  & = \Delta_5 + \varsigma \big|_{\varsigma = 0}  \equiv \delta_2 \, ,\qquad & \Delta_2 & = \Delta_3 = \Delta_4 \equiv \delta_1 \, ,\\[0.5em]
  \fn_1 &= \fn_5 \equiv \fp_2 \, , \qquad & \fn_2 & = \fn_3 = \fn_4 \equiv \fp_1 \, ,
 \end{alignedat}
\end{equation}
where we introduce the parameter $\varsigma$, which will be taken to zero at the end of the computation. The R-charge constraint $\sum_{a = 1}^{5} \Delta_a = 2$ and the twisting condition $\sum_{a = 1}^{5} \fn_a = 2 - 2 \fg$ reduce to
\bea
 \label{RB:const:M111}
 3 \delta_1 + 2 \delta_2 = 2 \, , \qquad 3 \fp_1 + 2 \fp_2 = 2 - 2 \fg \, .
\eea
By taking the limit $\varsigma \to 0$ in \eqref{R-charge:M111}, \eqref{b:Delta:M111}, \eqref{X:M111}, and \eqref{master:M111}, we find that
\bea
 \phi_l & =  \pm \frac{b_1^{3/2} \sqrt{N}}{8 \sqrt{2} \pi^2} ( 3 \delta_1 + 2 \delta_2 )^{3/2} \sqrt{\delta_2} \, , \quad \text{ for } \quad l = 1, 2 \, , \\
  X & = \pm \frac{\sqrt{b_1 N}}{8 \sqrt{2} \pi^2} ( 3 \delta_1 - 4 \delta _2 ) \sqrt{\frac{3 \delta_1 + 2 \delta_2}{\delta_2}}
  = - \frac{4}{N} \sum_{a = 1}^{2} B_a \frac{\pd \cV_B ( \delta_a )}{\pd \delta_a} \, ,
\eea
with
\bea
 \label{SU3xSU2:V}
 \cV_{B} ( \delta_a ) = \mp \frac{\sqrt{b_1} N^{3/2}}{\sqrt{2} (4 \pi)^2}
 \delta_1 \sqrt{\delta_2 ( 3 \delta_1 + 2 \delta_2 )} \, .
\eea
Alternatively, one may directly take the limit $\varsigma \to 0 $ in \eqref{V:on-shell:Delta:M111}, thereby obtaining the same result. The master volume is extremized at $\delta_1=\frac{4}{9}$ and $\delta_2=\frac{1}{3}$, which correctly leads to the known dimension of baryons realized as M5-branes wrapped on the five-cycles associated with $v^a$ \cite{Fabbri:1999hw,Hanany:2008fj}. From \eqref{VB:Q111}, we explicitly see that
\bea
 \cV_B^{Q^{1,1,1}} = \frac{3}{4} \cV_B^{M^{1,1,1}} \, ,
\eea
upon setting
\bea
 \delta_1^{Q^{1,1,1}} = \delta_3^{Q^{1,1,1}} = \frac{3}{4} \delta_1^{M^{1,1,1}} \, , \qquad \delta_2^{Q^{1,1,1}} = \delta_2^{M^{1,1,1}} \, .
\eea
Finally, the entropy function for purely magnetic static black holes is obtained by taking the limit $\varsigma \to 0$ in \eqref{id2}:
\bea
 \label{SB:M111}
  \cI_B ( \delta_a , \fp_a ) & = \log \cZ_B ( \delta_a , \fp_a ) = - \frac{1}{2} ( 4 \pi )^3 \sum_{a = 1}^{2} \fp_a \frac{\pd_a \cV_B ( \delta_a )}{\pd \delta_a}  \\& = \pm \frac{\pi \sqrt{b_1} N^{3/2}}{\sqrt{2}}
 \frac{\delta_2 ( 9 \delta_1 + 4 \delta_2 ) \fp_1 + \delta_1 ( 3 \delta_1 + 4 \delta_2 ) \fp_2}{\sqrt{\delta_2 ( 3 \delta_1 + 2 \delta_2 )}}
 \, .
\eea
After imposing $\delta_2 = 1 - \frac{3}{2} \delta_1$, the function \eqref{SB:M111} has a critical point at
\bea
 \mathring{\delta}_1 ( \fp_a ) & = \frac{9 \fp_2 \pm \sqrt{\Theta ( \fp_a )}}{6 ( 2 \fp_1 + 3 \fp_2 )} \, ,
\eea
where we define $\Theta ( \fp_a ) \equiv 3 ( 3 \fp_2 - 4 \fp_1 ) ( 4 \fp_1 + 9 \fp_2 )$.
One now easily computes the Bekenstein-Hawking entropy
\bea
 \label{SB:onshell:M111}
 S_{\text{BH}} ( \fp_a ) \equiv \cI_B ( \delta_a , \fp_a ) \bigl|_{\text{crit.}} = - \frac{2 \pi \sqrt{b_1} N^{3/2}}{9 \sqrt{6}}
 \bigl( 18 \fp_2 \mp \sqrt{\Theta ( \fp_a )} \bigr) \sqrt{\frac{9 \fp_2 \pm \sqrt{\Theta ( \fp_a ) }}{2 \fp_1 + 3 \fp_2}} \, ,
\eea
as also found in \cite{Kim:2019umc}.

For static dyonic black holes with electric charges $\fq_a$, we extremize
\bea
 \label{M111:dyonic}
 \cI_B ( \delta_a , \fp_a ) \equiv \log \cZ_B ( \delta_a , \fp_a ) + 2 \pi \ii \sqrt{b_1} \sum_{a = 1}^{2} \delta_a \fq_a \, .
\eea

\subsection{The complex cone over SPP}
\label{ssec:SPP}

The geometry is specified by the vectors
\bea
 \vec{v}^{\,1} & = ( 0 , 0, 0 ) \, , ~~~ && \vec{v}^{\,2} = ( 1 , - 1, 0 ) \, , &&& \vec{v}^{\,3} = ( 2 , 0 , 0 ) \, , \\
 \vec{v}^{\,4} & = ( 1, 1, 0) \, , && \vec{v}^{\,5} = ( 0 , 0 , 1 ) \, , &&& \vec{v}^{\,6} = ( 1 , 0 , 1 ) \, .
\eea
The toric diagram is given in Figure \ref{toric:conifold}. The dual SCFT describes the dynamics of $N$ M2-branes at a fibration over the suspended pinch point (SPP) singularity, as identified in \cite{Hanany:2008fj}.
\begin{figure}[H]
\centering
\begin{tikzpicture}[scale=0.45]

  \draw[draw=none,fill=blue!20,opacity=0.6] (-5.6,-2.3) -- (-4,-4) -- (-1.7,1) -- cycle;
  \draw[draw=none,fill=blue!20,opacity=0.6] (-5.6,-2.3) -- (0,0) -- (0,3) -- cycle;
  \draw[draw=none,fill=blue!20,opacity=0.6] (-5.6,-2.3) -- (-1.7,1) -- (0,3) -- cycle;
  \draw[draw=none,fill=blue!20,opacity=0.6] (0,0) -- (2,-2) -- (0,3) -- cycle;
  \draw[draw=none,fill=blue!20,opacity=0.6] (0,0) -- (2,-2) -- (-4,-4) -- cycle;
  \draw[draw=none,fill=blue!20,opacity=0.6] (-5.6,-2.3) -- (-4,-4) -- (0,0) -- cycle;

  \draw[draw=none,fill=blue!30,opacity=0.6] (0,0) -- (-5.6,-2.3) -- (0,3) -- (-1.7,1) -- cycle;
  \draw[draw=none,fill=blue!30,opacity=0.6]  (0,0) -- (-4,-4) -- (-5.6,-2.3) -- (-1.7,1) -- cycle;
  \draw[draw=none,fill=blue!40,opacity=0.6]  (0,0) -- (2,-2) -- (-4,-4) -- (-1.7,1) -- cycle;
  \draw[draw=none,fill=blue!40,opacity=0.6]  (0,0) -- (2,-2) -- (0,3) -- (-1.7,1) -- cycle;
  
  \draw[-,dashed] (0,0) -- (1.2,0) node[below] {};
  \draw[->,solid] (1.2,0) -- (3,0) node[below] {};
  
  \draw[-,dashed] (0,0) -- (0,3) node[below] {};
  \draw[->,solid] (0,3) -- (0,5.) node[below] {};
  
  \draw[-,dashed] (0,0) -- (-4,-4) node[below] {};
  \draw[->,solid] (-4,-4) -- (-5.3,-5.3) node[below] {};
  
  \draw (-4.6,-5.5) node {$e_2$};
  \draw (3.3,0.5) node {$e_3$};
  \draw (0.7,5.) node {$e_4$};
  
  \draw[-,solid] (-5.6,-2.3) -- (-1.7,1);
  \draw[-,solid] (-4,-4) -- (-1.7,1);
  \draw[-,solid] (2,-2) -- (-1.7,1);
  \draw[-,solid] (0,3) -- (-1.7,1);
  \draw[-,solid] (0,3) -- (2,-2);
  \draw[-,solid] (-4,-4) -- (2,-2);
  \draw[-,solid] (-4,-4) -- (-5.6,-2.3);
  \draw[-,dashed] (0,0) -- (-5.6,-2.3);
  \draw[-,dashed] (0,0) -- (2,-2);
  \draw[-,solid] (0,3) -- (-5.6,-2.3);
  
  \draw (0.5,0.5) node {$v^1$};
  \draw (-6,-2.1) node {$v^2$};
  \draw (-3.8,-4.4) node {$v^3$};
  \draw (2.5,-2.) node {$v^4$};
  \draw (0.5,3.4) node {$v^5$};
  \draw (-1.6,0.2) node {$v^6$};

\end{tikzpicture}
\caption{The toric diagram for SPP.\label{toric:SPP}}
\end{figure}
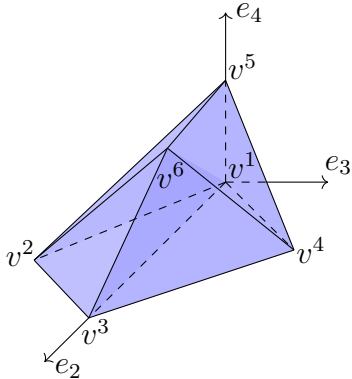
\noindent There are two baryonic symmetries: $B^{(1)} = ( 1 , 0 , - 1 , 0 , - 2 , 2 )$ and $B^{(2)} = ( - 1 , 1 , - 1 , 1 , 0 , 0 )$. The toric polytope \ref{toric:SPP} can be triangulated using the tetrahedra $(1256)$, $(3621)$, $(1364)$, and $(1654)$. Notice that there are worse-than-orbifold singularities and the master volume depends on the resolution. The gauge invariants $\phi_l$, for $l=1,\dots,4$, corresponding to the four tetrahedra, respectively, are given by
\bea
 \phi_1 ( \lambda_a , b_i ) & \equiv
 ( b_1 + b_3 - b_4 ) \lambda_1 - b_3 \lambda_2 - ( b_2 + b_3 - b_4 ) \lambda_5 + ( b_2 + b_3 ) \lambda_6 \, , \\
 \phi_2 ( \lambda_a , b_i ) & \equiv
 ( 2 b_1 - b_2 + b_3 - b_4 ) \lambda_1 - 2 b_3 \lambda_2 + ( b_2 + b_3 - b_4 ) \lambda_3 + 2 b_4 \lambda_6 \, , \\
 \phi_3 ( \lambda_a , b_i ) & \equiv
 ( 2 b_1 - b_2 - b_3 - b_4 ) \lambda_1 + ( b_2 - b_3 - b_4 ) \lambda_3 + 2 b_3 \lambda_4 + 2 b_4 \lambda_6 \, , \\
 \phi_4 ( \lambda_a , b_i ) & \equiv
 ( b_1 - b_3 - b_4 ) \lambda_1 + b_3 \lambda_4 - ( b_2 - b_3 - b_4 ) \lambda_5 + ( b_2 - b_3 ) \lambda_6 \, ,
\eea
subject to the constraint
\bea
 \label{phi:const:SPP}
 ( b_2 - b_3 - b_4 ) \left( 2 \phi_1 (\lambda_a , b_i) - \phi_2 (\lambda_a , b_i) \right)
 + ( b_2 + b_3 - b_4 ) \left( \phi_3 (\lambda_a , b_i) - 2 \phi_4 (\lambda_a , b_i) \right) = 0 \, ,
\eea
due to the fact that there are only $d-3=3$ independent gauge invariants in the model.
The master volume can be computed from \eqref{master} as a sum of four contributions:
\bea
 \label{master:SPP}
 \cV ( \lambda_a , b_i ) & = - \frac{8 \pi^4}{3 b_3} \biggl[
 \frac{1}{b_2 + b_3 - b_4} \left( \frac{\phi_1^3 ( \lambda_a , b_i  )}{( b_2 + b_3 ) ( b_1 + b_3 - b_4 )}
 - \frac{\phi_2^3 ( \lambda_a , b_i  )}{4 b_4 ( 2 b_1 - b_2 + b_3 - b_4 )} \right) \\
 & + \frac{1}{b_2 - b_3 - b_4} \left( \frac{\phi_3^3 ( \lambda_a , b_i  )}{4 b_4 ( 2 b_1 - b_2 - b_3 - b_4 )}
 - \frac{\phi_4^3 ( \lambda_a , b_i  )}{( b_2 - b_3 ) ( b_1 - b_3 - b_4 )} \right)
 \biggr] \, .
\eea
The R-charges \eqref{def:R-charge} corresponding to each vertex are
\bea
 \label{R-charge:SPP}
 \Delta_1 & = \frac{16 \pi^4}{N b_3}
 \left[ \frac{1}{b_2 + b_3 - b_4} \left( \frac{\phi_1^2 ( \lambda_a , b_i  )}{b_2 + b_3} - \frac{\phi_2^2 ( \lambda_a , b_i  )}{4 b_4} \right)
 + \frac{1}{b_2 - b_3 - b_4} \left( \frac{\phi_3^2 ( \lambda_a , b_i  )}{4 b_4} - \frac{\phi_4^2 ( \lambda_a , b_i  )}{b_2 - b_3} \right) \right] , \\
 \Delta_2 & = \frac{8 \pi^4}{N ( b_2 + b_3 - b_4 )}
 \left( \frac{\phi_2^2 ( \lambda_a , b_i  )}{b_4 ( 2 b_1 - b_2 + b_3 - b_4 )}
 - \frac{2 \phi_1^2 ( \lambda_a , b_i  )}{( b_2 + b_3 ) ( b_1 + b_3 - b_4 )} \right)  , \\
 \Delta_3 & = - \frac{16 \pi^4}{4 N b_3 b_4}
 \left( \frac{\phi_2^2 ( \lambda_a , b_i  )}{2 b_1 - b_2 + b_3 - b_4}
 - \frac{\phi_3^2 ( \lambda_a , b_i  )}{2 b_1 - b_2 - b_3 - b_4} \right)  , \\
 \Delta_4 & = - \frac{8 \pi^4}{N ( b_2 - b_3 - b_4 )}
 \left( \frac{2 \phi_4^2 ( \lambda_a , b_i  )}{( b_2 - b_3 ) ( b_1 - b_3 - b_4 )}
 - \frac{\phi_3^2 ( \lambda_a , b_i  )}{b_4 ( 2 b_1 - b_2 - b_3 - b_4 )} \right)  , \\
 \Delta_5 & = - \frac{16 \pi^4}{N b_3}
 \left( \frac{\phi_1^2 ( \lambda_a , b_i  )}{( b_2 + b_3 ) ( b_1 + b_3 - b_4 )}
 - \frac{\phi_4^2 ( \lambda_a , b_i  )}{( b_2 - b_3 ) ( b_1 - b_3 - b_4 )} \right)  , \\
 \Delta_6 & = \frac{8 \pi^4}{N b_3}
 \biggl[ \frac{1}{b_2 + b_3 - b_4} \left( \frac{2 \phi_1^2 ( \lambda_a , b_i  )}{b_1 + b_3 - b_4}
 - \frac{\phi_2^2 ( \lambda_a , b_i  )}{2 b_1 - b_2 + b_3 - b_4} \right) \\
 & + \frac{1}{b_2 - b_3 - b_4} \left( \frac{\phi_3^2 ( \lambda_a , b_i  )}{ 2 b_1 - b_2 - b_3 - b_4}
 - \frac{2 \phi_4^2 ( \lambda_a , b_i  )}{b_1 - b_3 - b_4} \right) \biggr] ,
\eea
and satisfy, see \eqref{reeb:id},
\bea
 \label{b:Delta:SPP}
 2 = \sum_{a = 1}^{6} \Delta_a \, , \qquad \frac{2 b_2}{b_1} = \Delta_2 + 2 \Delta_3 + \Delta_4 + \Delta_6 \, , \qquad \frac{2 b_3}{b_1} = \Delta_4 - \Delta_2 \, , \qquad \frac{2 b_4}{b_1} = \Delta_5 + \Delta_6 \, .
\eea
The previous relations can be inverted to express the $\phi_l$ in terms of $(\Delta_a, N, b_1)$:
\bea
 \label{phi:SPP}
 \phi_1^2 & = \frac{b_1^3 N}{128 \pi^4}
 ( \Delta_1 + \Delta_3 + 2 \Delta_4 ) \left( 2 ( \Delta_3 + \Delta_4 ) + \Delta_6 \right)
 \left( \Delta_5 ( \Delta_2 - \Delta_4 ) + \cQ ( \Delta_a) \right) \, , \\
 \phi_2^2 & = \frac{b_1^3 N}{64 \pi^4}
 \left( 2 ( \Delta_1 + \Delta_4 ) + \Delta_5 \right) ( \Delta_5 + \Delta_6 )
 \left( 2 \Delta_2 ( \Delta_3 + \Delta_4 ) - \Delta_4 \Delta_5 + \cQ ( \Delta_a) \right) \, , \\ 
 \phi_3^2 & = \frac{b_1^3 N}{64 \pi^4}
 \left( 2 ( \Delta_1 + \Delta_2 ) + \Delta_5 \right) ( \Delta_5 + \Delta_6 )
 \left( \Delta _4 \left( 2 ( \Delta_2 + \Delta_3 ) - \Delta_5 \right) + \cQ ( \Delta_a) \right) \, .
\eea
Here, the function
\bea
 \cQ ( \Delta_a) \equiv \frac{128 \pi^4}{b_1^3 N} \frac{\phi_4^2}{( \Delta_1 + 2 \Delta_2 + \Delta_3 ) \left( 2 ( \Delta_2 + \Delta_3 ) + \Delta_6 \right)} \, ,
\eea
is determined by solving the constraint \eqref{phi:const:SPP}, which leads to a quartic equation for $\phi_4$. Although the solution to a quartic equation is well known, we omit its rather unwieldy explicit expression for brevity. By eliminating $\lambda_a$ and $b_i$ using \eqref{b:Delta:SPP} and \eqref{phi:SPP}, the master volume \eqref{master:SPP} take the form
\bea
 \label{V:on-shell:Delta:SPP}
 \cV ( \Delta_a ) & = \frac{N}{3 b_1 ( \Delta_2 - \Delta_4 )}
 \left(
 \frac{\Delta_2 \left( \Delta_5 \phi_1 - ( \Delta_3 + \Delta_4 ) \phi_2 \right)}{2 ( \Delta_3 + \Delta_4 ) - \Delta_5}
 + \frac{ \Delta_4 \left( ( \Delta_2 + \Delta_3 ) \phi_3 - \Delta_5 \phi_4 \right)}{2 ( \Delta_2 + \Delta_3 ) - \Delta_5} \right) \, .
\eea
The baryonic K\"ahler moduli is given by
\bea
 \label{X:SPP}
 X^{(1)}
 = - \frac{2}{b_1} \frac{\phi_3 - 2 \phi_4}{2 ( \Delta_2 + \Delta_3 ) - \Delta_5} \, , \qquad
 X^{(2)}
 = \frac{1}{b_1} \frac{\phi_2 - \phi_3}{\Delta_2 - \Delta_4} \, .
\eea

By introducing the quartic function
\bea
 \label{a3d:def:SPP}
 a_{(3)} ( \Delta_a ) & \equiv \frac{1}{24} \sum_{a , b , c , e= 1}^{6} | ( v^a , v^b , v^c , v^e ) | \Delta_a \Delta_b \Delta_c \Delta_e
 - \frac{1}{2} ( \Delta_3 \Delta_5 - \Delta_1 \Delta_6 )^2 \, ,
\eea
as in \cite{Amariti:2011uw}, \eqref{V:on-shell:Delta:SPP} can also be rewritten as
\bea
 \cV ( \Delta_a ) = \pm \frac{N^{3/2} \sqrt{b_1}}{24 \sqrt{2} \pi^2}
 \sqrt{a_{(3)} ( \Delta_a ) + Y} \, ,
\eea
with
\bea
 Y \equiv \frac{(2 \pi )^4}{2 b_1 N}
 \left[ X^{(1)} \left( \Delta_3 \Delta_5 \left( X^{(1)} + 2 X^{(2)} \right)
 + \Delta_1 \Delta_6 \left( X^{(1)} - 2 X^{(2)} \right) \right) + 4 \Delta_2 \Delta_4 ( X^{(2)} )^2 \right] .
\eea
For the mesonic twist, $X^{(1)} = X^{(2)} = 0$, leading to
\bea
 \cV_{\text{mes.}} ( \Delta_a ) = \pm \frac{N^{3/2} \sqrt{b_1}}{24 \sqrt{2} \pi^2} \sqrt{a_{(3)} ( \Delta_a )} \, ,
\eea
which matches the result in \cite[(5.90)]{Hosseini:2019ddy}.

\subsubsection{Half-baryonic twist}

First, consider $X^{(1)} = 0$. Thus, from \eqref{X:SPP} and \eqref{phi:const:SPP}, we obtain
\bea
 \label{case1:SPP}
 \textbf{Case \Romannum{1}:} \quad \phi_3 = 2 \phi_4 \, , \qquad \phi_2 = 2 \phi_1 \, .
\eea
Substituting \eqref{case1:SPP} into \eqref{R-charge:SPP}, and using \eqref{b:Delta:SPP}, we find that
\bea
 \label{hB:case1:phi:SPP}
 \phi_1^2 ( \Delta_a ) & = \frac{b_1^3 N}{128 \pi^4}
 \frac{\Delta_2 ( \Delta_{13} + 2 \Delta_4 ) \left( 2 \Delta_{14} + \Delta_5 \right) \left( 2 \Delta_{34} + \Delta_6 \right) \Delta_{56}}{2 ( \Delta_{13} + 2 \Delta_4 ) + \Delta_{56}} \, , \\
 \phi_3^2 ( \Delta_a ) & = \frac{b_1^3 N}{32 \pi^4}
 \frac{\Delta_4 ( \Delta_{13} + 2 \Delta_2 ) \left( 2 \Delta_{12} + \Delta_5 \right) \left( 2 \Delta_{23} + \Delta_6 \right) \Delta_{56} }{2 ( \Delta_{13} + 2 \Delta_2 ) + \Delta_{56}} \, ,
\eea
along with the following constraint among the R-charges
\bea
 \label{hB:case1:const:SPP}
 \sum_{a = 1}^{6} B_a^{(1)} \frac{\pd \cV ( \Delta_a )}{\pd \Delta_a} = 0 \, .
\eea
Here, we introduced the shorthand $\Delta_{ab} \equiv \Delta_a + \Delta_b$ to improve readability.
Plugging back \eqref{hB:case1:phi:SPP} into \eqref{V:on-shell:Delta:SPP}, we obtain
\bea
 \label{hB:case1:master:SPP}
 \cV ( \Delta_a ) & = \pm \frac{N^{3/2} \sqrt{b_1}}{24 \sqrt{2} \pi^2}
 \frac{1} {\Delta_2 - \Delta_4}
 \Biggl(
 \Delta_2 \sqrt{\frac{ \Delta_2 ( \Delta_{13} + 2 \Delta_4 ) \left( 2 \Delta_{14} + \Delta_5 \right) \left( 2 \Delta_{34} + \Delta_6 \right) \Delta_{56}}{2 ( \Delta_{13} + 2 \Delta_4 ) + \Delta_{56}}} \\
 & \pm \Delta_4 \sqrt{\frac{ \Delta_4 ( \Delta_{13} + 2 \Delta_2 ) \left( 2 \Delta_{12} + \Delta_5 \right) \left( 2 \Delta_{23} + \Delta_6 \right) \Delta_{56} )}{2 ( \Delta_{13} + 2 \Delta_2 ) + \Delta_{56}}}
 \Biggr) \, .
\eea
Note that the signs in the two square-root terms are not correlated. Moreover, \eqref{hB:case1:master:SPP} can be rewritten in terms of the volume of the torus-invariant five-cycle associated with $v^a$:
\bea
 \label{hB:case1:master:volS:SPP}
 \cV ( \Delta_a ) & = \pm \frac{N^{3/2}}{12 b_1 \sqrt{\pi} ( \Delta_2 - \Delta_4 )}
 \left( \Delta_2 \sqrt{ \frac{\Delta_2}{\Vol_{S_2} ( \Delta_a )}} \pm \Delta_4 \sqrt{\frac{\Delta_4}{\Vol_{S_4} ( \Delta_a )}} \right) \, ,
\eea
which is consistent with appendix \ref{app:fp}, since the toric diagram can be seen as a suspension over a square in the $(1365)$ plane, and the corresponding $\Phi_{A\pm}$ are equal.
 
Next, consider $X^{(2)} = 0$. Thus, from \eqref{X:SPP} and \eqref{phi:const:SPP}, we obtain
\bea
 \label{case2:SPP}
 \textbf{Case \Romannum{2}:} \quad \phi_3 = \phi_2 \, , \quad \phi_1 = \frac{( \Delta_2 - \Delta_4 ) \phi_2 + \left( 2 \Delta_{34} - \Delta_5 \right) \phi_4}{2 \Delta_{23} - \Delta _5} \, .
\eea
Substituting \eqref{case2:SPP} into \eqref{R-charge:SPP}, and using \eqref{b:Delta:SPP}, we find that
\bea
 \label{hB:case2:phi:SPP}
 \phi_2^2 ( \Delta_a ) & = \frac{b_1^3 N}{64 \pi^4}
 \Delta_3 \left( 2 \Delta_{12} + \Delta_5 \right) \left( 2 \Delta_{14} + \Delta_5 \right) \Delta_{56} \, , \\
 \phi_4^2 ( \Delta_a ) & = \frac{b_1^3 N}{128 \pi^4}
 ( \Delta_{13} + 2 \Delta_2 ) \left( 2 ( \Delta_1 \Delta_3 - \Delta_2 \Delta_4 ) + \Delta_{34} \Delta_5 \right) \left( 2 \Delta_{23} + \Delta_6 \right) \, ,
\eea
along with the following constraint among the R-charges:
\bea
 \label{hB:case2:const:SPP}
  0 & = \frac{b_1^3 N}{256 \pi^4}
  \bigl[ \Delta_3 \left( 2 \Delta_{12} + \Delta_5 \right)
  \left( 2 \Delta_{14} + \Delta_5 \right) \left( 2 \Delta_{23} + \Delta_6 \right)
  - \Delta_1 \left(  2 \Delta_{23} - \Delta_5 \right)^2 \left( 2 \Delta_{34} + \Delta_6 \right) \bigr] \\
  & + \left( \frac{\Delta_{56}}{2 \Delta_{23} + \Delta_6} \phi_4 - \phi_2 \right) \phi_4 \, .
\eea
The expression for $\cV(\Delta_a)$ can be obtained straightforwardly using \eqref{hB:case2:phi:SPP} and \eqref{hB:case2:const:SPP}, although we do not present it here, as it is not particularly illuminating.

\section{Black hole entropy with baryonic fluxes}
\label{sec:BH}

In this section, we demonstrate that $\cI$-extremization for magnetically charged baryonic black holes is equivalent to the attractor mechanism in gauged supergravity and correctly predicts the black hole entropy. Moreover, we show that the master volume for the purely baryonic twist is precisely mapped to the supergravity prepotential. To this end, we focus on the class of black holes with baryonic charges in AdS$_4 \times Q^{1,1,1}$ and AdS$_4 \times M^{1,1,1}$ backgrounds, as found in \cite{Halmagyi:2013sla}. Since only static black holes have been discussed in the literature, we restrict ourselves to this case.

We start by considering dyonic static BPS black holes that can be embedded in M-theory and are asymptotic to AdS$_4 \times Q^{1,1,1}$ \cite{Halmagyi:2013sla}. These solutions are described within the framework of four-dimensional gauged $\cN = 2$ supergravity, coupled to $n_{\text{V}} = 3$ vector multiplets and $n_{\text{H}} = 1$ hypermultiplet, which is a consistent truncation of M-theory on $Q^{1,1,1}$ \cite{Cassani:2012pj}. The truncation includes, in addition to the graviphoton associated with the R-symmetry, two massless vector fields arising from the reduction of the M-theory four-form on three-cycles, known as Betti vector fields, which correspond to baryonic symmetries in the dual picture. The fourth vector field is massive. The dynamics of the theory are fully specified by the prepotential, the Killing vector, and the Killing prepotential. We use the notations of \cite{Halmagyi:2013sla} to which we refer for further details. The prepotential governing the vector multiplet sector is given by 
\bea
 \label{F:Q111}
 F ( X^{\Lambda} ) = 2 \sqrt{X^0 X^1 X^2 X^3} \, ,
\eea
where $X^{\Lambda}$, $\Lambda = 0, \ldots, 3$, are the projective coordinates encoding the special K\"ahler geometry of the scalar manifold $\cK \cM$. The gauging of the theory is determined by the Killing vector
\bea
  \label{k:Q111}
  k_{\Lambda}^a = \sqrt{2} ( e_0 , 2, 2, 2 ) \, ,
\eea
which acts on the hypermultiplet scalar fields $(a , \phi, \zeta , \tilde \zeta)$. Furthermore, the corresponding Killing prepotential is given by
\bea
 \label{P:Q111}
 P_{\Lambda}^{3} = \sqrt{2} \left( 4 - \frac{e_0}{2} e^{2\phi} , - e^{2\phi} , -e^{2\phi} , - e^{2\phi} \right) \, , \qquad P^{3,\;\!\Lambda} = 0 \, .
\eea
The constant $e_0$ determines the radius of AdS$_4$ via the relation
\bea
 R_{\text{AdS}_{4}} = \frac{1}{2} \left( \frac{e_0}{6} \right)^{3/4} \, .
\eea
The central charge of the black hole $\cZ$ and the superpotential $\cL$ are defined as
\bea
 \cZ \equiv e^{\cK/2} \left( q_\Lambda X^\Lambda - p^\Lambda F_\Lambda \right) \, , \qquad
 \cL \equiv e^{\cK/2} \left( P^3_\Lambda X^\Lambda - P^{3, \Lambda} F_\Lambda \right) \, ,
\eea
where $\cK$ denotes the K\"ahler potential, and $F_\Lambda \equiv \frac{\pd_\Lambda F ( X^\Lambda )}{\pd X^\Lambda}$. Then, the Bekenstein-Hawking entropy, expressed as a function of the magnetic and electric charges $(p^\Lambda, q_\Lambda)$, is given by the attractor equation
\bea
 \label{BH:S}
 \frac{\pd}{\pd z^i} \frac{\cZ}{\cL} & = 0 \, , \quad \text{for } \quad i = 1, 2, 3 \, , \\
 S_{\text{BH}} ( p^\Lambda , q_\Lambda ) & = \frac{\text{Area}}{4 G_{\text{N}}}
 = - \ii \frac{\cZ}{\cL} \frac{2 \pi \eta}{4 G_{\text{N}}} \, ,
\eea
where the derivatives are taken with respect to the physical scalars $z^i$. Here,
\bea
\eta =
\begin{cases}
2 \lvert \fg - 1\rvert, & \text{for } \fg \neq 1 \, , \\
1, & \text{for } \fg = 1 \, .
\end{cases}
\eea
The two equations in \eqref{BH:S} are equivalent to extremizing
\bea
 \label{attractor}
 \cI ( X^\Lambda) \equiv - \ii \frac{\pi \eta}{2 \alpha G_{\text{N}}} \left[ \left( q_\Lambda X^\Lambda - p^\Lambda F_\Lambda \right)
 - \mathsf{\Lambda} \left( P^3_\Lambda X^\Lambda - P^{3, \Lambda} F_\Lambda - \alpha \right) \right] \, ,
\eea
with respect to the projective coordinates $X^\Lambda$. However, the $X^\Lambda$ are only determined up to the ``gauge'' redundancy $X^\Lambda \mapsto e^{f} X^\Lambda$, which is related to K\"ahler transformations on $\cK \cM$. To eliminate this redundancy, we fix $\cL ( X^\Lambda )$ to a constant, which is enforced by including a Lagrange multiplier in \eqref{attractor}. Indeed, varying \eqref{attractor} with respect to $X^0$ yields a complex equation that determines both the dilaton and the `non-conserved charge' $q_0$ in terms of $X^\Lambda$ and $\mathsf{\Lambda}\;\! .$

The following Dirac quantization conditions, as given in \cite{Halmagyi:2013sla}, are imposed:
\bea
 p^\Lambda P^3_\Lambda = \mp 1 \, , \qquad
 p^\Lambda k^a_\Lambda = 0 \, ,
\eea
which consequently lead to the relations
\bea
 p^0 = \mp \frac{1}{4 \sqrt{2}} \, , \qquad \sum_{\Lambda = 1}^{3} p^\Lambda = \pm \frac{e_0}{8 \sqrt{2}} \, .
\eea
The attractor equation \eqref{attractor} needs to be supplemented by the hyperino BPS variation evaluated at the horizon,
\bea
 \label{hyperino:Q111}
 X^\Lambda k^a_\Lambda = 0 \quad \Longrightarrow \quad e_0 X^0 + 2 ( X^1 + X^2 + X^3 ) = 0 \, .
\eea
This condition can be used to eliminate $X^0$ (and the corresponding massive vector field) and to define an \emph{effective prepotential} for the massless vector fields,
\bea
 \label{F:eff:Q111}
 F_{\text{eff.}} ( X^{\Lambda} ) = 2 \sqrt{\frac{2}{e_0}} \sqrt{ - X^1 X^2 X^3 \left( X^1 + X^2 + X^3 \right)} \, ,
\eea
which can then be inserted into \eqref{attractor}. Notice that eliminating the field $A^0$ leads to a redefinition of the corresponding electric charges:
\bea
 \wh q_a = q_a - \frac{2}{e_0} q_0 \, , \quad \text{ for } \quad a = 1 , 2 , 3 \, .
\eea
The attractor equations in this form have been used extensively to compare gravity and field theory in various examples, including \cite{Benini:2016rke,Benini:2017oxt,Hosseini:2017fjo,Hosseini:2018uzp,Hosseini:2018usu,Bobev:2018uxk,Benini:2020gjh,Hosseini:2020wag}. 

\paragraph*{$M^{1,1,1}$ truncation.}
The truncation of M-theory on $M^{1,1,1}$ is derived from the truncation on $Q^{1,1,1}$ by eliminating a Betti vector multiplet. Specifically, this is achieved by imposing the conditions
\bea
 \label{HPZ:M111:trunc}
 X^3 = X^1 \, , \qquad A^3 = A^1 \, .
\eea
By substituting \eqref{HPZ:M111:trunc} into \eqref{F:Q111}, we find
\bea
 \label{F:M111}
 F_{M^{1,1,1}} ( X^{\Lambda} ) = 2 X^1 \sqrt{X^0 X^2} \, .
\eea
From \eqref{hyperino:Q111}, we obtain
\bea
 \label{hyperino:M111}
 X^0 = - \frac{2}{e_0} ( 2 X^1 + X^2 ) \, ,
\eea
which allows us to eliminate $X^0$ (along with its associated massive vector field). This leads to the definition of an effective prepotential for the remaining massless vector fields:
\bea
 \label{F:eff:M111}
 F_{\text{eff.}}^{M^{1,1,1}} ( X^{\Lambda} ) = 2 \sqrt{\frac{2}{e_0}} X^1 \sqrt{ - X^2 \left( 2 X^1 + X^2 \right)} \, ,
\eea
which can then be substituted into \eqref{attractor}. Eliminating the field $A^0$ also modifies the corresponding electric charges:
\bea
 \wh q_1 = q_1 - \frac{4}{e_0} q_0 \, , \quad \wh q_2 = q_2 - \frac{2}{e_0} q_0 \, .
\eea

\subsection{Dyonic black holes in AdS$_4 \times Q^{1,1,1}$}

We begin by identifying the index set $\Lambda = \{1, 2, 3\}$ with $a = \{1, 2, 3\}$. It is then necessary to construct a mapping between the geometric R-charges $\delta_a$, which satisfy the constraint \eqref{RB:const:Q111}, and the scalar fields $X^\Lambda$ belonging to the supergravity vector multiplets. A natural proposal for this mapping is
\bea
 \frac{X^a}{\sum_{b = 1}^{3} X^b} = \delta_a \, ,
\eea
which immediately ensures that the constraint \eqref{RB:const:Q111} is satisfied. With these identifications, we obtain the following relation between the master volume $\cV_B ( \delta_a )$ \eqref{VB:Q111} and the effective prepotential $F_{\text{eff.}}( X^\Lambda )$ \eqref{F:eff:Q111}:
\bea
  \cV_{B} ( \delta_a ) = \mp \frac{\sqrt{- 2 b_1 e_0} N^{3/2}}{( 8 \pi )^2 \left( \sum_{\Lambda = 1}^{3} X^\Lambda \right)^2} F_{\text{eff.}} ( X^{\Lambda} ) \, .
\eea
Furthermore, by identifying the quantized charges
\bea
 \frac{4 \sqrt{2} \eta}{e_0}\, p^a \in \bZ \, , \qquad \frac{e_0 \eta}{32 \sqrt{2} G_{N}} \, \wh q_a \in \bZ \, ,
\eea
with $\fp_a$ and $\fq_a$, respectively, we precisely recover the geometric extremization \eqref{Q111:dyonic}:
\bea
 \cI_B ( \delta_a , \fp_a ) \equiv \log \cZ ( \delta_a , \fp_a ) + 2 \pi \ii \sqrt{b_1} \sum_{a = 1}^{3} \delta_a \fq_a = \sqrt{b_1} \, S_{\text{BH}} ( p^\Lambda , q_\Lambda ) \, ,
\eea
where we used the holographic relation
\bea
 \frac{R_{\text{AdS}_{4}}^2}{G_{\text{N}}} = \frac{2 \sqrt{6} \pi^2}{9} \frac{N^{3/2}}{\sqrt{\Vol_{Q^{1,1,1}}}} \, ,
\eea
with $\Vol_{Q^{1,1,1}} = \frac{\pi^4}{8}$.

\subsection{Dyonic black holes in AdS$_4 \times M^{1,1,1}$}

The results in this section were previously obtained in \cite{Kim:2019umc} by explicitly solving the supersymmetry conditions in \cite{Gauntlett:2018dpc}. Here, we re-derive them using the master volume, which allows for a direct comparison with the supergravity prepotential and the attractor mechanism.
Let us first set $q_\Lambda = 0$. Then, \eqref{attractor} yields
\bea
 \label{M111:attractor}
 \cI ( X^\Lambda) & = \frac{\ii \pi \eta}{2 \alpha G_{\text{N}}}
 \biggl[
 \frac{2 p^1 X^2 ( e_0 X^0 - 2 X^1 ) + p^2 X^1 ( e_0 X^0 - 2 X^2 )}{e_0 \sqrt{X^0 X^2}} \\
 & + \frac{\mathsf{\Lambda}}{\sqrt{2}} \left( X^0 \left( 8 - e_0 e^{2 \phi } \right) - 2 ( 2 X^1 + X^2 ) e^{2 \phi} - \sqrt{2}\, \alpha \right)
 \biggr] \, .
\eea
By extremizing \eqref{M111:attractor} with respect to $X^0$ we obtain
\bea
 \label{e2phi:M111}
 e^{2 \phi} = \frac{8}{e_0} + \frac{1}{\sqrt{2} e_0^2 \mathsf{\Lambda}} \frac{2 p^1 X^2 ( e_0 X^0 + 2 X^1 ) + p^2 X^1 ( e_0 X^0 + 2 X^2 )}{\sqrt{( X^0 )^3 X^2}} \, .
\eea
At this stage we shall impose the hyperino BPS equation \eqref{hyperino:M111} to write down the attractor equation \eqref{M111:attractor} for the massless modes $X^1$ and $X^2$:
\bea
 \label{M111:massless:attractor}
 \cI ( X^\Lambda) & = \frac{\ii \pi \eta}{2 \alpha G_{\text{N}}}
 \biggl[
 \frac{2 \ii \sqrt{2} \left( p^2 X^1 ( X^1 + X^2 ) + p^1 X^2 ( 3 X^1 + X^2 ) \right)}{\sqrt{e_0 X^2 ( 2 X^1 + X^2 )}}
 -  \frac{8 \sqrt{2} \mathsf{\Lambda}}{e_0} \left( 2 X^1 + X^2 + \frac{e_0 \alpha}{8 \sqrt{2}} \right)
 \biggr] \, .
\eea
By extremizing \eqref{M111:massless:attractor} with respect to $X^1$, $X^2$, and the Lagrange multiplier $\mathsf{\Lambda}$, we find the critical points:
\bea
 \label{M111:I:crit}
 \mathring{\mathsf{\Lambda}} ( p^\Lambda ) & = \frac{\ii \sqrt{e_0}}{12 \sqrt{3}}
 \bigl( 6 p^1 \mp \sqrt{ \Theta ( p )} \bigr) \sqrt{\frac{ 3 p^1 \pm \sqrt{\Theta ( p ) }}{2 p^1 + p^2}} \, , \\
 \mathring{X}^1 ( p^\Lambda ) & = \frac{\alpha e_0}{48 \sqrt{2} } \frac{\pm \sqrt{ \Theta ( p ) } - 3 ( p^1 + p^2 )}{2 p^1+ p^2} \, , \qquad
 \mathring{X}^2 ( p^\Lambda ) = - \frac{\alpha e_0}{24 \sqrt{2}} \frac{ 3 p^1 \pm \sqrt{\Theta ( p ) }}{2 p^1 + p^2} \, ,
\eea
where we define $\Theta ( p ) \equiv 3 ( p^1 - p^2 ) ( 3 p^2 + p^2 )$. Moreover, substituting \eqref{M111:I:crit} into \eqref{e2phi:M111}, we obtain the value of the dilaton:
\bea
 e^{2 \phi} = \frac{24}{e_0} \frac{( 2 p^1 + p^2 )^2}{\left( 6 p^1 - \sqrt{ \Theta ( p )} \right) \left( 5 p^1 + p^2 + \sqrt{\Theta ( p ) } \right)} \, .
\eea
Finally, the Bekenstein-Hawking entropy \eqref{BH:S} reads%
\footnote{To compare with \cite{Halmagyi:2013sla}, one needs to set $e_0 = 6$ and $\mathring{X}^2 \equiv - \frac{\alpha v_1^2}{4 \sqrt{2}}$. Without loss of generality, we further choose $\alpha < 0$ and $1 < v_1 < \sqrt{3}$.}
\bea
 \label{mag:BH:M111}
 S_{\text{BH}} ( p^\Lambda )
 = - \frac{\ii \pi  \eta }{2 G_{\text{N}}} \mathring{\mathsf{\Lambda}} ( p^\Lambda )
 = \frac{\pi \sqrt{e_0} \eta}{24 \sqrt{3} G_{\text{N}}}
 \bigl( 6 p^1 \mp \sqrt{ \Theta ( p )} \bigr) \sqrt{\frac{ 3 p^1 \pm \sqrt{\Theta ( p ) }}{2 p^1 + p^2}} \, .
\eea
Upon identifying the quantized charges
\bea
 \frac{16 \sqrt{2} \eta}{3 e_0}\, p^1 \in \bZ \, , \qquad \frac{4 \sqrt{2} \eta}{e_0} \, p^2 \in \bZ \, ,
\eea
with $\fp_1$ and $\fp_2$, respectively, and using the holographic dictionary
\bea
 \label{holo:dict:M111}
 \frac{R_{\text{AdS}_{4}}^2}{G_{\text{N}}} = \frac{2 \sqrt{6} \pi^2}{9} \frac{N^{3/2}}{\sqrt{\Vol_{M^{1,1,1}}}} \, ,
\eea
with $\Vol_{M^{1,1,1}} = \frac{9 \pi^4}{128}$, we find that \eqref{mag:BH:M111} precisely matches \eqref{SB:onshell:M111}:
\bea
 S_{\text{BH}} ( p^\Lambda ) = S_{B} ( \fp_a ) \, .
\eea

In the presence of electric charges, explicitly solving the attractor equations \eqref{attractor} becomes more involved. Therefore, we shall compare the attractor equations, restricted to the fields $X^1$ and $X^2$, with the geometric extremization \eqref{M111:dyonic}. The geometric R-charges $\delta_a$, which satisfy the constraint \eqref{RB:const:M111}, are mapped to the scalar fields $X^\Lambda$ as
\bea
 \label{delta:X:M111}
 \delta _1 = \frac{4}{3} \frac{X^1}{2 X^1 + X^2} \, , \qquad \delta_2 = \frac{X^2}{2 X^1 + X^2} \, .
\eea
This mapping automatically ensures that the constraint \eqref{RB:const:M111} holds. With these identifications, we establish the following relation between the master volume $\cV_B ( \delta_a )$ \eqref{SU3xSU2:V} and the effective prepotential $F_{\text{eff.}}( X^\Lambda )$ \eqref{F:eff:M111}:
\bea
 \cV_B ( \delta_a ) = \mp \frac{\sqrt{- b_1 e_0} N^{3/2} }{24 \sqrt{2} \pi^2 ( 2 X^1 + X^2 )^2} F_{\text{eff.}} ( X^{\Lambda} ) \, .
\eea
Finally, using \eqref{holo:dict:M111} and \eqref{delta:X:M111}, and identifying the quantized charges
\begin{equation}
 \begin{alignedat}{2}
  \frac{16 \sqrt{2} \eta}{3 e_0}\, p^1 & \in \bZ \, , \qquad & \frac{4 \sqrt{2} \eta}{e_0} \, p^2 & \in \bZ \, , \\
  \frac{3 e_0 \eta}{128 \sqrt{2} G_{N}} \, \wh q_1 & \in \bZ \, , \qquad & \frac{e_0 \eta}{32 \sqrt{2} G_{N}} \, \wh q_2 & \in \bZ \, ,
 \end{alignedat}
\end{equation}
with $\fp_{1,2}$ and $\fq_{1,2}$, respectively, we precisely recover the geometric extremization \eqref{M111:dyonic}:
\bea
 \cI_B ( \delta_a , \fp_a ) \equiv \log \cZ ( \delta_a , \fp_a ) + 2 \pi \ii \sqrt{b_1} \sum_{a = 1}^{2} \delta_a \fq_a = \sqrt{b_1} \, S_{\text{BH}} ( p^\Lambda , q_\Lambda ) \, .
\eea

\section{Discussion and outlook}
\label{sec:discuss}

In this paper, we have defined a generalized free energy $\cF(\Delta_a)$ for three-dimensional $\cN=2$ SCFTs with an AdS$_4\times\text{SE}_7$ dual, as the constrained Legendre transform of the master volume of the internal geometry. We have shown that this function is the universal building block for constructing entropy functions for general black objects asymptotic to AdS$_4\times\text{SE}_7$. On the field theory side, it provides a prediction for the large-$N$ limit of several partition functions whose saddle points have yet to be found. There are many directions for future investigations.

It would be very interesting to obtain a general expression for the master volume $\cV(\Delta_a)$ purely in terms of toric data. While we have provided an algorithm to compute  $\cV(\Delta_a)$, we have not been able to find a closed-form, simple formula for the result. In comparison, for four-dimensional $\cN = 1$ theories, the corresponding $\cV(\Delta_a)$ is a simple cubic expression in terms of toric data \eqref{cubic}. There exist proposals based on a quartic expression \cite{Amariti:2011uw,Amariti:2012tj} which work well when restricted to mesonic directions. However, we have seen that the quartic formula must be corrected when baryonic moduli are introduced. It would be interesting to develop a general prescription for incorporating these corrections using only the toric data of the manifold.

We have also observed that when the CY$_4$ fan has facets with worse-than-orbifold singularities, the master volume depends on the choice of resolution, leading to an ambiguity in our prescription. Although many \emph{regular} four-fold singularities exist, curiously, a large number of dual pairs studied in the literature \cite{Hanany:2008cd,Hanany:2008fj,Martelli:2008si,Benini:2009qs,Jafferis:2009th} fall into the \emph{singular} category. This raises the question whether different families of black holes are associated with different resolutions, as suggested by our construction. Notably, this ambiguity affects only black holes with nonzero baryonic charges. Black hole solutions carrying baryonic charges have running scalars associated with baryonic moduli, which are typically reflected in the geometry by blown-up cycles. At present, there are no known examples of baryonic black holes in geometries with worse-than-orbifold singularities on the facets, and it would be interesting to investigate this issue in more detail.

Finally, it would be worthwhile to extend our analysis to other classes of AdS vacua. A natural next step is to relax the toric condition and consider solutions associated with D2-, D4-, or M5-branes in M-theory, or in massive type IIA. We still expect that the master volume---or more precisely, the equivariant volume in the spirit of \cite{Martelli:2023oqk,Colombo:2023fhu}---will continue to play a central role. We leave these interesting investigations to the future.

\section*{Acknowledgements}

SMH is supported by UK Research and Innovation (UKRI) under the UK government's Horizon Europe funding guarantee [grant number EP/Y027604/1]. AZ is partially supported by the INFN, and the MIUR-PRIN grant No. 2022NY2MXY [finanziato dall'Unione europea -- Next Generation EU, Missione 4 Componente 1 CUP H53D23001080006].

\appendix

\section{Manipulating the fixed point formula}
\label{app:fp}

In this appendix, we derive a formula for the Legendre transform of the master volume  for CY$_4$, whose toric diagram is the  suspension of a two-dimensional toric diagram. As in main text, we use notations where the vectors in the fan are written as  $v^a=(1,\vec{v}^{\, a})$, and the toric diagram is the three-dimensional polytope with vertices $\vec{v}^{\, a}$. We further assume that the $d$ vectors $\vec{v}^{\, a}$ can be split into $d-2$ vectors $\vec{v}^{\, i}$ \emph{lying on a plane}, and two vectors $\vec{v}^{\, \pm}$ located on opposite sides of this plane. We choose a triangulation of the two-dimensional toric diagram spanned by the vectors   $\vec{v}^{\, i}$, and decompose the fan into tetrahedra $( v^{\, \pm}\, , v^{\, i}\, , v^{\, j} \, , v^{\, k})$. The fixed point formula \eqref{fixedpoint} then gives
\bea
 \label{fixedpoint2}
 \bV ( \lambda_a , b_i ) = \sum_{A=\{i, j , k \}} \sum_{\pm} \frac{e^{\Phi_{A \pm}}}
 {\text{sign}( v^{\pm} , v^{i} , v^{j}, v^{k}) ( b , v^{i} , v^{j} , v^{k})
 \prod_{(j , k)} \frac{( b , v^{j} , v^{ k}  , v^{ \pm})}{(  v^{i} , v^{ j} , v^{ k} , v^{ \pm})}} \, ,
\eea
where
\bea
 \Phi_{A\pm} = - \frac{( b , v^{ i} , v^{ j}  , v^{ k}) }{( v^{ \pm} , v^{ i} , v^{ j} , v^{k})} \lambda_\pm  + \text{terms with } \lambda_i \, .
\eea

If we now further assume that \emph{all} the $\Phi_{A+}$ are equal to each other, and \emph{all} the $\Phi_{A-}$ are likewise equal, we obtain
\bea
 \label{Apm}
 \cV( \lambda_a , b_i )
 = (2\pi)^4 \bV^{(3)}( \lambda_a , b_i )
 = \frac{8\pi}{3} \sum_{\pm} \frac{( v^{ \pm} , v^{ i} , v^{ j} , v^{k})}{ ( b , v^{i} , v^{j}  , v^{ k})} \, \Phi^3_{A\pm} \Vol_{S_\pm}(b_i) \, ,
\eea
where $S_\pm$ are the five-cycles associated with $v^{ \pm}$, and we have used formula \eqref{Sa} to compute
\bea 
 \Vol_{S_\pm}(b_i) = \pi^3 \sum_{A=\{i , j , k \}}  \frac{1}{ |( v^{\pm}, v^{ i} , v^{ j} , v^{k})|
 \prod_{(j , k)} \frac{( b , v^{ j} , v^{k} , v^{ \pm})}{( v^{ i} , v^{ j} , v^{ k} , v^{\pm})}} \, .
\eea
Notice that in \eqref{Apm}, the quantity $\frac{( v^{\pm} , v^{ i} , v^{ j} , v^{k})}{ ( b , v^{ i} , v^{ j} , v^{k})}$ is actually independent of the choice of $(i , j , k)$ in the plane. Indeed, the vectors $v^i$ satisfy two linear conditions of the form $\sum_{k = 1}^4 \nu_k v_k^i = \text{constant}$---one being the CY condition, and the other reflecting the fact that they lie on a (different) plane. As a result, they also satisfy $\sum_{k = 1}^4 \wt \nu_k v_k^i = 0$ for some coefficients $\wt \nu_k$. This implies that $( v^{p} , v^{ i} , v^{ j} , v^{k}) = 0$ for all choices of $(i , j , k , p)$, meaning any four such vectors are linearly dependent in $\bR^4$. Then, by expanding $v^{p}$ as a linear combination of the other three vectors, we have
\bea
 \frac{( v^{ \pm} , v^{ i} , v^{ j} , v^{p})}{ ( b , v^{ i} , v^{ j} , v^{ p})}
 = \frac{( v^{ \pm} , v^{i} , v^{j} , v^{ k})}{ ( b , v^{i} , v^{j} , v^{ k})} \, .
\eea
The argument immediately extends to all possible triples of points lying in the plane, showing that the ratio is independent of the particular choice of $(i , j , k)$.

We then find
\bea 
 \Delta_{\pm} \equiv - \frac{2}{N} \frac{\pd \cV ( \lambda_a , b_i)}{\pd \lambda_\pm}
 = \frac{16 \pi}{N} \Phi^2_{A\pm} \Vol_{S_\pm}(b_i) \, ,
\eea
and, using
\bea
 ( b , v^{ i} , v^{j} , v^{k})
 & \overset{\eqref{reeb:id}}{=} \frac{b_1}{2} \sum_{a = 1}^{d} \Delta_a ( v^a , v^{i} , v^{ j} , v^{ k} ) \\
 & = \frac{b_1}{2} \left( \Delta_+ ( v^{\, +} , v^{ i} , v^{ j} , v^{ k}) + \Delta_- ( v^{ -} , v^{ i} , v^{ j} , v^{ k}) \right) ,
\eea
we finally obtain
\bea
 \cV ( \Delta ) = & \pm \frac{N^{3/2} }{12 b_1 \sqrt{\pi} \left (\Delta_+ ( v^{+} , v^{ i} , v^{ j}  , v^{ k}) + \Delta_- ( v^{-} , v^{ i} , v^{ j} , v^{ k}) \right)}
 \\
 & \times \biggl(( v^{+} , v^{ i} , v^{ j} , v^{k}) \frac{\Delta_+^{3/2}}{\sqrt{\Vol_{S_+}(\Delta)}}
 \pm ( v^{-} , v^{ i} , v^{ j} , v^{ k}) \frac{\Delta_-^{3/2}}{\sqrt{\Vol_{S_-}(\Delta)}} \biggr) \, .
\eea

\bibliographystyle{ytphys}
\bibliography{BarBH}

\end{document}